\documentclass[twocolumn,showpacs,preprintnumbers,amsmath,amssymb]{revtex4}
\usepackage{epsfig}
\usepackage{dcolumn}% Align table columns on decimal point
\usepackage{bm}% bold math
\newcommand{\remove}[1]{}
\newcommand{\dd}{\mathrm{d}}
\def\be{\begin{equation}}
\def\ee{\end{equation}}

\newcommand{\beq}{\begin{equation}}
\newcommand{\eeq}{\end{equation}}
\newcommand{\beqa}{\begin{eqnarray}}
\newcommand{\eeqa}{\end{eqnarray}}

\renewcommand{\pl}{\partial}

\newcommand{\ii}{{\rm i}}

\newcommand{\vv}{{\bf v}}

\newcommand{\vx}{{\bf x}}
\newcommand{\vk}{{\bf k}}
\newcommand{\vp}{{\bf p}}
\newcommand{\vq}{{\bf q}}
\renewcommand{\vr}{{\bf r}}

\newcommand{\cM}{{\cal M}}

\newcommand{\bea}{\begin{array}}
\newcommand{\ea}{\end{array}}

\newcommand{\XX}{{\cal X}}

\begin{document}

\title{Goldstone models of modified gravity}

\author{Philippe Brax}
\affiliation{Institut de Physique Th\'eorique,\\
CEA, IPhT, F-91191 Gif-sur-Yvette, C\'edex, France\\
CNRS, URA 2306, F-91191 Gif-sur-Yvette, C\'edex, France}
\author{Patrick Valageas}
\affiliation{Institut de Physique Th\'eorique,\\
CEA, IPhT, F-91191 Gif-sur-Yvette, C\'edex, France\\
CNRS, URA 2306, F-91191 Gif-sur-Yvette, C\'edex, France}
\vspace{.2 cm}

\date{\today}
\vspace{.2 cm}

\begin{abstract}

We investigate scalar-tensor theories where matter couples to the scalar field via
a kinetically dependent conformal coupling. These models can be seen as the
low-energy description of invariant field theories under a global Abelian symmetry.
The scalar field is then identified with the Goldstone mode of the broken symmetry.
It turns out that the  properties of these models are very similar to the ones of ultralocal
theories where the scalar-field value is directly determined by the local matter density.
This leads to a complete screening of the fifth force in the Solar System and between
compact objects, through the ultralocal screening mechanism.
On the other hand, the fifth force can have large effects in extended structures
with large-scale density gradients, such as galactic halos.
Interestingly, it can either amplify or damp Newtonian gravity, depending on the model
parameters.
We also study the background cosmology and the linear cosmological perturbations.
The background cosmology is hardly different from its $\Lambda$-CDM counterpart while
cosmological perturbations crucially depend on whether the coupling function is convex or
concave. For concave functions, growth is hindered by the repulsiveness of the fifth force
while it is enhanced in the convex case.
In both cases, the departures from the $\Lambda$-CDM cosmology increase on smaller
scales and peak for galactic structures.
For concave functions, the formation of structure is largely altered below some
characteristic mass, as smaller structures are delayed and would form later through
fragmentation, as in some warm dark matter scenarios.
For convex models, small structures form more easily than in the $\Lambda$-CDM scenario.
This could lead to an over-abundance of small clumps. We use a thermodynamic analysis
and show that although convex models have a phase transition between homogeneous and
inhomogeneous phases, on cosmological scales the system does not enter the
inhomogeneous phase. On the other hand, for galactic halos, the coexistence of small
and large substructures in their outer regions could lead to observational signatures
of these models.

\end{abstract}

 \vskip2pc

\maketitle

\section{Introduction}
\label{sec:Introduction}

The acceleration of the cosmic expansion \cite{Perlmutter:1998np,Riess:1998cb}
in the recent past of the Universe could be due
to a nearly massless field on large scales. One of the most natural ways of generating
such a field is via the breaking of a global symmetry. At energies below the symmetry breaking
scale, there are a number of Goldstone modes which are massless \cite{Goldstone:1962es}.
The potential energy of these modes after symmetry breaking is constant and could serve
as vacuum energy, i.e. potentially generating the cosmic acceleration
\cite{Copeland:2006wr}.
Another large-scale effect of such modes is to modify the way massive bodies interact
as the Goldstone modes propagate a new type of fifth force. The residual Abelian symmetry
at low energy translates into a shift symmetry of the scalar field which prevents
the existence of direct Yukawa interactions between the Goldstone modes and matter,
such as $\frac{\beta}{M_{\rm Pl}} m_{\psi} \varphi \bar\psi \psi$.
This implies that Goldstone fields evade the
stringent tests of extra gravitational forces in the Solar System \cite{Will:2001mx}.
On the other hand, effects in extended structures such as galaxy halos for instance
are present due to the derivative nature of the coupling of the Goldstone modes to matter.

From this particle physics point of view, the scale of symmetry breaking necessary
to induce the late time acceleration of the Universe must be very low, of order of a few
milli-eV, a scale which is far below most of particle physics scales (apart, possibly,
from the neutrino mass scales).
This requires the dark energy sector which we propose, i.e. the Goldstone modes generating
the late time acceleration, to be secluded and only very weakly coupled to the standard
model of particle physics. In a sense, we require this sector to be a new addition to the
standard model of particles, where the effects of its coupling to standard model matter
will appear only on very large scales and in astrophysics. This will be detailed
in this paper.

More generally, models of modified gravity have very different properties depending
on the type of couplings between the scalar field and matter. Conformal couplings,
i.e. leading to Yukawa interactions to matter, have been thoroughly investigated
and require a screening of fifth-force effects in our local environment\cite{Khoury:2010xi}.
They appear in four different guises, i.e. the  chameleon
\cite{Khoury:2003aq,Khoury:2003rn,Brax:2004qh}
and Damour-Polyakov \cite{Damour:1994zq} mechanisms, the K-mouflage
\cite{Babichev:2009ee,Brax:2014b}
and the Vainshtein \cite{Vainshtein:1972sx} effects.
No other mechanism is necessary for models with second-order equations of motion.
A fifth mechanism,  the ultralocal scenario \cite{Brax:2016vpd}, has been recently introduced
whereby the field does not propagate, i.e. there are no derivative terms in its equation
of motion therefore going beyond the previous classification.
Another coupling called disformal can also be relevant and plays a crucial role
in the Horndeski and beyond-Horndeski models
\cite{Zumalacarregui2014,Gleyzes:2014dya,Gleyzes:2014qga,Crisostomi:2016tcp},
i.e. the most general field
theories of one scalar with second-order equations of motion
\cite{Horndeski:1974wa,Deffayet:2011gz,Kobayashi2011}.
These two types of couplings cover almost all the
possibilities which have been unravelled by Bekenstein \cite{Bekenstein:1992pj}.
The most general coupling between matter and a scalar field that preserves Lorentz
invariance and causality depends on the Jordan frame metric
\be
g_{\mu\nu}= A^2(\varphi, X ) \tilde{g}_{\mu\nu}
+ B^2(\varphi, X) \partial_\mu\varphi \partial_\nu \varphi
\ee
where $\tilde{g}_{\mu\nu}$ appears in the Einstein-Hilbert term of gravity, i.e. when
the action is written in such a way that the Einstein-Hilbert action is normalized
with a constant Planck scale.
The disformal term $B^2(\varphi,X)$ and the conformal one $A^2(\varphi,X)$
are functions of the field $\varphi$ and of its kinetic term
$X=-\frac{1}{2} \tilde{g}^{\mu\nu} \partial_\mu \varphi \partial_\nu \varphi$ only.
In this paper we shall concentrate on the kinetic coupling where
\be
A(\varphi,X) \equiv A(X), \;\;\; B(\varphi,X) \equiv 0 ,
\label{coup}
\ee
and assume that the scalar field has no potential term. Hence the full dynamics of the models
are solely due to the coupling function $A(X)$. This is exactly the setting that can be
derived for Goldstone models, i.e. they are equivalent to modified gravity models with the
coupling (\ref{coup}) and a constant vacuum energy. As such they offer a relevant and
well-motivated scenario where consequences of both the cosmic acceleration and
modified gravity can be combined.

These Goldstone models are hardly different from $\Lambda$-CDM cosmology
at the background level (because the coupling function $A$ must remain very close
to unity, otherwise the dynamics of perturbations would strongly differ from the Newtonian
predictions).
At the linear perturbation level, models differ according to whether the function
$\ln[A(X)]$ is concave or convex.
For concave functions, the growth of structure is hampered by the
repulsiveness of the scalar interaction. This would lead to a scenario whereby, below
some characteristic mass, larger structures would form first and then smaller ones
would appear by fragmentation. On the other hand, for convex functions growth is largely
increased on shorter scales. This could lead to a high-redshift phase with
a catastrophic over-abundance of small clumps. We use a thermodynamical approach
to analyze this possibility and show that this is not the case for models of interest.

The paper is arranged as follows.
In section~\ref{sec:Derivative-conformal-coupling}, we define the models.
In section~\ref{sec:quantum}, we show that at low energy these models are best considered
as classical field theories for distances larger than a small cutoff scale.
In section~\ref{sec:Equations-of-motion-Einstein}, we study the models in the Einstein frame
whilst the same is done in the Jordan frame in section~\ref{sec:Equations-of-motion-Jordan}.
In section~\ref{sec:Hamilton-Jacobi}, we analyze the global solutions of the equations
of motion and the formation of defect surfaces in space.
In section~\ref{sec:models}, we introduce explicit models.
The stability and validity of the solution is studied in section~\ref{sec:stability},
the cosmological background and the linear perturbations in section~\ref{sec:linear},
and the spherical collapse in section~\ref{sec:Spherical-collapse}.
In section~\ref{sec:newtonian}, we consider the effects of the models on
extended dense structures, such as galactic and cluster halos, and the screening
of the fifth force in the Solar System.
Finally, in section~\ref{sec:nonlinear-analysis-convex} we perform an analysis
of the thermodynamics of the models that enhance the formation of large-scale structures
(i.e. with a convex coupling function).
We conclude then and have an appendix on the coupling to fermions

\section{Kinetic conformal coupling}
\label{sec:Derivative-conformal-coupling}

\subsection{The Models}
\label{sec:Models}

We consider scalar field models where the action has the form
\beqa
S & = & \int d^4 x \; \sqrt{-\tilde{g}} \left[ \frac{\tilde{M}_{\rm Pl}^2}{2} \tilde{R}
+ \tilde{\cal L}_{\varphi}(\varphi) \right]  \nonumber \\
&& + \int d^4 x \; \sqrt{-g} \, {\cal L}_{\rm m}(\psi^{(i)}_{\rm m},g_{\mu\nu}) ,
\label{S-def}
\eeqa
where $\tilde{g}$ is the determinant of the Einstein-frame metric tensor $\tilde{g}_{\mu\nu}$,
and $\psi^{(i)}_{\rm m}$ are various matter fields.
The additional scalar field $\varphi$ is explicitly coupled to matter through the
Jordan-frame metric $g_{\mu\nu}$, with determinant $g$,
which is given by the conformal rescaling
\beq
g_{\mu\nu} = A^2(\tilde\chi) \, \tilde{g}_{\mu\nu} , \;\;\; \mbox{with} \;\;\;
\tilde\chi = - \frac{1}{2\cM^4} \tilde{g}^{\mu\nu} \partial_{\mu}\varphi\partial_{\nu}\varphi .
\label{g-Jordan-def}
\eeq
Here $\tilde\chi = X/{\cal M}^4$ is the normalized kinetic term of the scalar field,
and $\cM^4$ is an energy scale, of the order of the
current cosmological energy density, that is a parameter of the model.
Throughout this paper we use the signature $(-,+,+,+)$ for
the metric and we denote Einstein-frame quantities with a tilde $\sim$, to distinguish them
from Jordan-frame quantities (without a tilde).

This model is similar to standard scalar-tensor models, except that the conformal mapping
(\ref{g-Jordan-def}) is taken as a function of the  derivative term $\tilde\chi$
instead of the scalar field value itself $\varphi$.
Since in this paper we are interested in the qualitative features associated with this
new coupling form, we do not consider the mixed dependence $A(\varphi,\tilde\chi)$
and we focus on the $\tilde\chi$-dependence of the coupling function $A$.
We also take the scalar field Lagrangian to be governed by its standard kinetic term,
\beq
\tilde{\cal L}_{\varphi}(\varphi)  =  - \frac{1}{2} \tilde{g}^{\mu\nu}
\partial_{\mu}\varphi \partial_{\nu}\varphi - {\cal M}^4 = \cM^4 ( \tilde\chi -1 ) ,
\label{Lphi-def}
\eeq
where the factor $-{\cal M}^4$ can be identified with a cosmological constant as in the
$\Lambda$-CDM scenario, or with the value of the potential $V(\varphi)$ that we assume
to be flat on the scales of interest.
Thus, in this paper we neglect the impact of possible variations of the potential
$V(\varphi)$.
This is to simplify the model, to avoid introducing a second arbitrary function $V(\varphi)$,
and to focus on the properties associated with the kinetic conformal coupling.

\subsection{Goldstone Models}
\label{sec:Goldstone}

\subsubsection{Shift symmetry}
\label{sec:shift}

This class of derivately coupled models is not artificial and in fact belongs
to a class of very well motivated ones: the Goldstone models. Indeed, let us consider
a global $U(1)$ symmetry broken at a scale $f$ by the vacuum expectation value (vev)
of a complex scalar field $\Phi$.
This should be understood as a new sector added to the standard model of particle physics.
The interactions with the standard model will be spelt out below.
Hence, let us consider the simple model
\be
S=\int d^4x \sqrt{-\tilde g} \, \left[ - \tilde{g}^{\mu\nu} \partial_\mu \bar \Phi \partial_\nu \Phi
- V(\vert \Phi\vert^2) \right]
\ee
in the Einstein frame. Let us assume that the potential has a minimum $f$ such that
\be
\partial_v V(v^2)=0 \;\;\;  \mbox{at} \;\;\; v = f .
\ee
Because of the global $U(1)$ invariance of the model
$\Phi \to e^{i\theta} \Phi$, at low energy, the breaking of the global $U(1)$ symmetry leads
to the presence of one Goldstone mode $\varphi$ such that
\be
\Phi= f e^{\ii \varphi/(\sqrt{ 2} f)} ,
\ee
whose low-energy Lagrangian follows from
\be
- \tilde{g}^{\mu\nu} \partial_\mu \bar \Phi \partial_\nu \Phi \to
- \frac{1}{2} \tilde{g}^{\mu\nu} \partial_\mu \varphi \partial_\nu \varphi
\ee
and represents a massless scalar field.
Moreover, at the minimum of the potential we have
\be
V(\vert \Phi\vert^2)\to V(f^2) ,
\ee
which acts as a cosmological constant.
Hence at energies below $f$, the model is described by a massless scalar field with
a cosmological constant term $V(f^2)$.
This would correspond to the constant factor ${\cal M}^4$ in the scalar-field
Lagrangian (\ref{Lphi-def}).
The scale $f$ should also be thought to be close to $\cM$.
Here we have in mind models such as $V(v^2) = -\mu^2 v^2 + \lambda v^4+ V_0$,
where we expect $\lambda \sim 1$ as it is dimensionless and $V_0$ is taken to be of order
$\mu^4$ as we assume that only one scale appears in the model.
The mass term is corrected by its interaction with other fields and the thermal mass becomes
$m^2(T)= \alpha T^2 -\mu^2$ where $\alpha$ is a model-dependent constant,
see \cite{Katz:2014bha} for instance. At high temperature, the minimum of the
potential is at the origin whereas at low temperature the symmetry is broken
below the critical temperature where the thermal mass vanishes.
This gives at low temperature $T\ll T_c$ the minimum $f=\mu/\sqrt{2\lambda}$, while the
transition occurs at the critical temperature
$T_c = \mu/\sqrt \alpha  = \sqrt{2\lambda} f/\sqrt \alpha$.
The minimum of the potential in the low-temperature phase is $
V(f^2) = V_0- \mu^4/(4\lambda)$, which is of order $\mu^4$.

As a result, we consider that there is a new sector whose role is to generate the
acceleration of the expansion of the Universe. It emerges from the breaking of a
$U(1)$ symmetry at very low energy, i.e. at an energy scale which is dictated
by the measurement of $\cM$. This is far below all particle physics scales, apart maybe
from the neutrino mass scales, and is motivated by the observation of the late-time cosmic
acceleration. This secluded sector is a viable description of the late-time acceleration
provided its coupling with the standard model of particle physics is not problematic.
We discuss this issue below.

\subsubsection{Coupling to matter and conformal gravitational coupling}
\label{sec:coupling-to-matter}

Let us now consider the coupling between the Goldstone mode $\varphi$ and matter.
First of all, we assume that the fermions of the standard model are not charged under
the global symmetry. This prevents the existence of new couplings at low energy between
the Goldstone mode and the gauge fields of the standard model arising from anomalous
triangle diagrams, i.e. couplings linear in $\varphi$ which could arise if the global symmetry
were anomalous. Moreover, this also implies that couplings of the type
\be
{\cal L}_{\rm linear}= \varphi \nabla_\mu J^\mu
\ee
are not present, as $J^\mu$ would be the Noether current built up from fermionic fields of
the standard model, which vanishes because of the vanishing charges.
This leaves us with higher-order interactions, which must respect the fact that
the existence of the global symmetry implies that the only couplings to matter must be
shift symmetric under $\varphi \to \varphi +c$.
As a result, a first class of interactions which could be considered at low energy
would arise from the Lagrangian
\be
{\cal L}_1= \alpha \partial_\mu \varphi \bar\psi \gamma^\mu \psi ,
\ee
where $\psi$ is a fermionic field of the standard model. Such interactions can be
unravelled by a change of fermionic field
\be
\psi= e^{\ii\alpha \varphi} \hat\psi ,
\ee
such that the kinetic terms
\be
{\cal L}_{\rm kin}= \ii \bar\psi \gamma^\mu \nabla_\mu \psi = \ii \bar {\hat \psi}
\gamma^\mu D_\mu \hat \psi - \alpha \partial_\mu \varphi \bar{\hat \psi} \gamma^\mu
\hat\psi ,
\ee
where $\nabla_\mu$ includes both the spin connection and the gauge fields, absorb the
interaction ${\cal L}_1$. Below the electro-weak scale, all the Dirac mass terms
$m_\psi \bar \psi \psi$ are invariant under this change of field implying that
the whole effective Lagrangian at very low energy below $\cM$ is invariant.

Another type of interactions arises from derivative interactions such as
\be
\frac{\tilde{g}^{\mu\nu}\partial_\mu \bar\Phi \partial_\nu \Phi}{{\cal M}^4} m_\psi
\bar{\tilde \psi} \tilde \psi ,
\ee
where the fermion $\tilde\psi$ of mass $m_{\psi}$ could represent dark matter fields.
At low energy this gives rise to the term
\be
\frac{\tilde{g}^{\mu\nu} \partial_\mu \varphi \partial_\nu \varphi}{2 \cM^4}
m_\psi \bar{\tilde \psi} \tilde\psi .
\ee
This interaction can equivalently be obtained from a derivative conformal coupling with
\be
A(X)= 1 + \tilde \chi ,
\ee
where $\tilde \chi$ was defined in Eq.(\ref{g-Jordan-def}).
More generally we have the correspondence
\be
A\left( -\frac{\tilde{g}^{\mu\nu}\partial_\mu \bar\Phi \partial_\nu \Phi}{{\cal M}^4}
\right) m_\psi \bar{\tilde \psi} \tilde \psi\to A(\tilde \chi) m_\psi \bar{\tilde\psi}
\tilde\psi ,
\ee
where $A(\tilde \chi)$ is the conformal derivative coupling in Bekenstein's metric with
\be
g_{\mu\nu}= A^2(\tilde \chi)\tilde g_{\mu\nu} ,
\ee
as in Eq.(\ref{g-Jordan-def}).
Hence these derivative coupling models are in one to one correspondence with Abelian
Goldstone models coupled to matter where the coupling function is directly related to the
interaction terms between fermions and the complex scalar $\Phi$.
Indeed, upon the identification between the Jordan- and Einstein-frame fermions
\be
\tilde \psi = A^{3/2}(\tilde\chi) \psi ,
\ee
we have the equality between the actions
\beqa
S_\psi & = & - \int d^4x \sqrt{-g} \, ( \ii \bar \psi \gamma^\mu \nabla_\mu \psi
+ m_\psi \bar \psi \psi) \nonumber \\
& = & - \int d^4x \sqrt{-\tilde g} \, ( \ii \bar{\tilde \psi} \tilde \gamma^\mu \tilde
\nabla_\mu \tilde \psi + A(\tilde\chi) m_\psi \bar{\tilde\psi}\tilde \psi) , \;\;\;
\label{S-psi-Einstein}
\eeqa
where $\nabla_\mu$ (respectively $\tilde \nabla_\mu$) are the covariant derivatives
including the spin connections in the two frames, see Appendix~\ref{sec:frames-fermions}.
In particular, the Einstein-frame mass can be identified with
$\tilde{m}_{\psi}= A m_{\psi}$.

These are not the only couplings that one could envisage. However, higher derivative
terms such as
\be
{\cal L}_{\rm higher}= \frac{(\Box \varphi)^2}{\cM^5} \bar \psi \psi,
\ee
which are invariant under the shift symmetry, are problematic.
Indeed they imply that, in the presence of matter, the
equation of motion for the scalar field $\varphi$ is of order higher than two.
This generically leads to the existence of instabilities.  For this reason, it is plausible, although a full proof is beyond
our work, that the low-energy Goldstone interactions for $\varphi$ obtained from a healthy,
e.g. ghost free, model for $\Phi$ at higher energy than $\cM$, should not lead to such
problematic interactions. This leaves the couplings in (\ref{S-psi-Einstein}) as the
remaining interactions at low energy.

\subsection{Total action and equation of motion for the scalar field}
\label{sec:total-action}

\subsubsection{Coupling to fermions}
\label{sec:fermions}

For the scalar field $\varphi$, associated for instance with the Goldstone mode in the
Goldstone models, the total action that includes both the scalar-field Lagrangian
(\ref{Lphi-def}) and the coupling to fermions through the fermionic action
(\ref{S-psi-Einstein}) reads in Einstein frame as
\beqa
S_{\varphi} & = & \int d^4 x \; \sqrt{-\tilde g} \; \left[ - \frac{1}{2} \tilde{g}^{\mu\nu}
\partial_\mu\varphi \partial_\nu \varphi - A(\tilde\chi) m_{\psi} \bar{\tilde\psi}\tilde\psi \right]
\nonumber \\
& = & \int d^4 x \; \sqrt{-\tilde g} \; \left[ {\cal M}^4 \tilde\chi - A(\tilde\chi) m_{\psi}
\bar{\tilde\psi}\tilde\psi \right] ,
\label{S-phi-eff}
\eeqa
where we did not include the cosmological constant, which does not contribute to the
equation of motion of the scalar field.
Using
\beq
\frac{\delta\tilde\chi(x)}{\delta\varphi(x')} = - \frac{1}{{\cal M}^4} \tilde{g}^{\mu\nu}(x)
\frac{\partial\varphi}{\partial x^{\nu}} \frac{\partial}{\partial x^{\mu}} \delta_D(x-x') ,
\eeq
the Klein-Gordon equation in the Einstein frame reads as
\beqa
&& \tilde\nabla_\mu \left[ \left( 1 - \frac{d\ln A}{d\tilde \chi} A
\frac{m_\psi \bar{\tilde\psi}\tilde \psi}{\cM^4} \right) \tilde\nabla^{\mu} \varphi \right]
\nonumber \\
&& = \tilde\nabla_\mu \left[ \left( 1 - \frac{d\ln A}{d\tilde \chi} A^4
\frac{m_\psi \bar{\psi} \psi}{\cM^4} \right) \tilde\nabla^\mu\varphi \right] = 0 .
\eeqa
Now in the Jordan frame, on-shell and for any fermion field, we have that the trace of the
energy-momentum tensor is given by
\be
T_\psi= -m_\psi \bar{\psi} \psi ,
\ee
implying that the Klein-Gordon equation becomes
\be
\tilde\nabla_\mu \left[ \left( 1 + \frac{d\ln A}{d\tilde\chi} A^4 \frac{T_\psi}{\cM^4} \right)
\tilde\nabla^\mu\varphi \right] = 0 .
\label{KG-T_psi}
\ee
We will see that this is the general Klein-Gordon equation for any derivatively coupled scalar
model.

Let us now assume that the fermions condense and acquire an occupation number at low
momentum
\be
n_\psi = \langle 0 \vert \bar \psi \psi\vert 0 \rangle .
\ee
We can identify the conserved matter density as
\be
\rho_\psi = m_\psi n_\psi ,
\ee
and $T_{\psi} = - \rho_{\psi}$.
This could be what happens for dark matter and therefore we have derived the
Klein-Gordon equation for the scalar $\varphi$ in the presence of Cold Dark Matter
\be
\tilde\nabla_\mu \left[ \left(1 - \frac{d\ln A}{d\tilde\chi} A^4 \frac{\rho_\psi}{\cM^4} \right)
\tilde\nabla^\mu\varphi \right] = 0 ,
\ee
whose solutions will be studied below.
In the following, we will only work in the low energy regime below the symmetry breaking
scale $f$. The Abelian model involving both fermions $\psi$ and the scalar $\Phi$
should be seen as the UV completion of the model.

Going back to the action (\ref{S-phi-eff}), we can see that the effective action for the
scalar field, which recovers the equation of motion of the scalar field, reads as
\beq
S_{\varphi}^{\rm eff} = \int d^4 x \; \sqrt{-\tilde g} \; \left[ - \frac{1}{2} (\tilde\partial \varphi)^2
+ A(\tilde\chi) {\rm \breve{T}}_{\psi} \right] ,
\label{S-phi-eff-1}
\eeq
with
\beq
{\rm \breve{T}}_{\psi} = A^3 T_{\psi} .
\eeq
In other words, the term associated with the coupling to matter that enters the effective
action (\ref{S-phi-eff-1}) is $A(\tilde\chi) {\rm \breve{T}}_{\psi}$, where the quantity
that is kept fixed when we look for the extremum with respect to $\varphi$ of the action
is neither the Jordan-frame $T_{\psi}$ nor the Einstein-frame-like quantity
$A^4 T_{\psi}$, but the intermediate quantity ${\rm \breve{T}}_{\psi}$.
The effective Lagrangian (\ref{S-phi-eff-1}) also applies to relativistic fermions,
as we did not assume the nonrelativistic limit.

\subsubsection{Coupling to classical point-particles}
\label{sec:point-particles}

Here we note that the same effective action (\ref{S-phi-eff-1}) holds when we have
a coupling to classical point-particles, described by the Jordan-frame action
\beq
S = - \int d^4x \; m \sum_{i=1}^N
\sqrt{ - g_{\mu\nu} \frac{dx^{\mu}}{d\tau} \frac{dx^{\nu}}{d\tau}}
\delta_D^3(\vec{x} - \vec{y}_i ) .
\label{S-point-Jordan}
\eeq
This gives for the trace of the energy-momentum tensor
\beq
T =  - \frac{m}{\sqrt{-g}} \sum_{i=1}^N \sqrt{ - g_{\mu\nu} \frac{dx^{\mu}}{d\tau}
\frac{dx^{\nu}}{d\tau} } \delta_D^3(\vec{x} - \vec{y}_i ) .
\label{T-point-masses}
\eeq
In the Einstein frame, we can write the action (\ref{S-point-Jordan}) as
\beq
S = - \int d^4x \; m A \sum_{i=1}^N
\sqrt{ - \tilde{g}_{\mu\nu} \frac{dx^{\mu}}{d\tau} \frac{dx^{\nu}}{d\tau}}
\delta_D^3(\vec{x} - \vec{y}_i ) ,
\label{S-point-Einstein}
\eeq
and as for the fermions we can identify $\tilde m = A m$.
The total action for the scalar field $\varphi$ is the sum of the scalar-field Lagrangian
(\ref{Lphi-def}) and of the action (\ref{S-point-Einstein}), which explicitly involves
the factor $A(\tilde\chi)$.
This gives again the Klein-Gordon equation (\ref{KG-T_psi}), where $T_{\psi}$
is replaced by the trace $T$ of Eq.(\ref{T-point-masses}).
Therefore, we recover the effective action (\ref{S-phi-eff-1}) for the scalar field,
where ${\rm \breve{T}}_{\psi}$ is replaced by ${\rm \breve{T}}$ with
\beqa
{\rm \breve{T}} & = & - \frac{m}{\sqrt{-\tilde{g}}} \sum_{i=1}^N \sqrt{ - \tilde{g}_{\mu\nu}
\frac{dx^{\mu}}{d\tau} \frac{dx^{\nu}}{d\tau} } \delta_D^3(\vec{x} - \vec{y}_i ) \nonumber \\
& = & A^3 T .
\label{Teff-point-masses}
\eeqa
Again, the coupling to matter is described by the simple term
$A(\tilde\chi) {\rm \breve{T}}$,
where ${\rm \breve{T}}$ is neither the Jordan-frame nor Einstein-frame
trace of the matter energy-momentum tensor, but the intermediate quantity
${\rm \breve{T}} = A^3 T$.

\subsection{Ultralocal models}
\label{sec:ultralocal-def}

The kinetic conformal coupling models defined by Eqs.(\ref{S-def})-(\ref{g-Jordan-def})
are closely related to the ultralocal models introduced in \cite{Brax:2016vpd}.
These ultralocal models are defined by the same action as Eq.(\ref{S-def})
but the conformal transformation involves the scalar field $\varphi$ instead of its kinetic term,
\beq
\mbox{ultralocal:} \;\;\; g_{\mu\nu} = A^2(\varphi) \tilde{g}_{\mu\nu} ,
\label{conformal-g-tg-ultralocal}
\eeq
and the scalar-field Lagrangian is dominated by its potential term, instead of its kinetic term,
\beq
\mbox{ultralocal:} \;\;\; \tilde{\cal L}_{\varphi}(\varphi) = - V(\varphi) .
\label{L-phi-def}
\eeq
Then, making the field redefinition
\beq
\tilde\chi \equiv - \frac{V(\varphi)}{{\cal M}^4} \;\;\; \mbox{and} \;\;\;
A(\tilde\chi) \equiv A(\varphi) ,
\label{tchi-def-ultralocal}
\eeq
the scalar-field Lagrangian becomes
\beq
\mbox{ultralocal:} \;\;\; \tilde{\cal L}_{\tilde\chi}(\tilde\chi) = {\cal M}^4 \tilde\chi ,
\label{L-chi-def-ultralocal}
\eeq
that is, the Lagrangian is only a linear potential.
Apart from the factor $-1$, we recover the Lagrangian (\ref{Lphi-def}) and the conformal
coupling $A(\tilde\chi)$.
However, for the ultralocal models $\tilde\chi$ is a standard scalar field, whereas for the
models (\ref{S-def}) that we study in this paper, it is the kinetic term (\ref{g-Jordan-def})
of an underlying scalar field $\varphi$.

The total action for the scalar field $\tilde\chi$ is again
\beq
\mbox{ultralocal:} \;\;\; S_{\tilde\chi}^{\rm eff} = \int d^4 x \; \sqrt{-\tilde g} \;
\left[ {\cal M}^4 \tilde\chi + A(\tilde\chi) {\rm \breve{T}} \right] ,
\label{S-chi-eff}
\eeq
with ${\rm \breve{T}} = A^3 T$, and the equation of motion reads as
\beq
\mbox{ultralocal:} \;\;\; 1 + \frac{d\ln A}{d\tilde\chi} A^4 \frac{T}{{\cal M}^4} = 0 .
\label{KG-ultralocal-0}
\eeq
We recover the expression within the inner brackets in Eq.(\ref{KG-T_psi}).
This means that, apart from the trivial solution $\varphi={\rm constant}$,
the other solution of the kinetic conformal coupling model associated with
the vanishing of the inner brackets in Eq.(\ref{KG-T_psi}) corresponds to the solution
of the ultralocal model with the same $A(\tilde\chi)$.
We shall see that we indeed recover the same equations of motion for both models
in the Jordan frame, both for the scalar field and matter.

\section{Quantum Properties}
\label{sec:quantum}

\subsection{Ultralocal models}
\label{sec:ultralocal-quantum}

We first consider whether the simpler ultralocal models can have a quantum description.
From the effective action (\ref{S-chi-eff}), working in Minkowski spacetime (i.e., we neglect
metric fluctuations) and performing a Wick rotation, the Euclidian generating functional of
these models reads as
\beq
Z[j] = \int {\cal D}\tilde\chi \; e^{\int d^4x [j \tilde\chi + {\cal M}^4 \tilde\chi
+ A(\tilde\chi) {\rm \breve{T}} ] } .
\label{Z-ultralocal-def}
\eeq
If we use a lattice regularization, we write Eq.(\ref{Z-ultralocal-def}) as
\beq
Z[j] = \lim_{\epsilon \to 0} \int \prod_k d\tilde\chi_k \; e^{\epsilon \sum_k [ j_k \tilde\chi_k + {\cal M}^4
\tilde\chi_k + A(\tilde\chi_k) {\rm \breve{T}}_k ] } ,
\label{Z-ultralocal-lattice}
\eeq
where $\epsilon$ is the spacetime volume of each lattice cell.
Because of the ultralocal character of the model, the lattice field variables $\tilde\chi_k$
are decoupled and we can write
\beq
Z[j] = \lim_{\epsilon \to 0} \left[ Z_{(\epsilon)}(\epsilon j) \right]^N
= \lim_{\epsilon \to 0}
e^{N \sum_{n=1}^{\infty} (\epsilon j)^n \langle \tilde\chi^n \rangle_{c(\epsilon)} /n! } ,
\label{Z-Z1-ultralocal}
\eeq
where we take the source $j$ to be constant over some finite spacetime volume $VT$
that covers $N$ lattice points, $VT=N \epsilon$, and zero elsewhere.
The one-cell generating function is
\beq
Z_{(\epsilon)}(j) = \frac{ \int d\tilde\chi \; e^{j \tilde\chi + \epsilon [ {\cal M}^4
\tilde\chi + A(\tilde\chi) {\rm \breve{T}} ] } }{ \int d\tilde\chi \; e^{\epsilon [ {\cal M}^4
\tilde\chi + A(\tilde\chi) {\rm \breve{T}} ] } }
= \langle e^{j \tilde\chi} \rangle_{(\epsilon)} ,
\label{Z1-ultralocal}
\eeq
where the subscript ``$(\epsilon)$'' denotes the dependence on the lattice discretization
$\epsilon$.
In the second equality in Eq.(\ref{Z-Z1-ultralocal})
we used the expansion over cumulants of the generating function $Z_{(\epsilon)}$.
In \cite{Brax:2016vpd} and \cite{Brax:2016a} we considered ultralocal models
where $\tilde \chi$ is restricted to a finite or half-bounded interval.
If the variable $\tilde\chi$ is restricted to a finite interval $[\tilde\chi_{\min},\tilde\chi_{\max}]$,
the cumulants $\langle \tilde\chi^n \rangle_{c(\epsilon)}$ are finite for all $n$ and $\epsilon$,
and in the limit $\epsilon \to 0$ they converge to the finite cumulants associated with the
uniform distribution over $[\tilde\chi_{\min},\tilde\chi_{\max}]$.
Then, there is no need for renormalization and in the continuum limit $\epsilon \to 0$
only the first term $n=1$ is nonzero in the sum in the exponential in Eq.(\ref{Z-Z1-ultralocal}),
which gives
\beq
Z[j] = e^{VT j \langle \tilde\chi \rangle_{(0)} } .
\eeq
For a nonconstant source $j(x)$, this yields
\beq
Z[j] = e^{\int d^4x \; j(x) \langle \tilde\chi \rangle_{(0)} } .
\eeq
This is the generating functional of a field $\tilde\chi(x)$ that has vanishing fluctuations,
with a Dirac distribution $\delta_D(\tilde\chi(x) - \langle \tilde\chi \rangle_{(0)} )$.
However, the mean $\langle \tilde\chi \rangle_{(0)}$ is not set by the saddle point
$\tilde\chi_c$ of the effective potential
$V^{\rm eff}(\tilde\chi) = - {\cal M}^4 \tilde\chi - A(\tilde\chi) {\rm \breve{T}}$,
but by the mean $(\tilde\chi_{\min}+\tilde\chi_{\max})/2$ of the uniform distribution
over the interval $[\tilde\chi_{\min},\tilde\chi_{\max}]$.
Thus, we find that in the continuum limit the ultralocal models with a finite range
for the field $\tilde\chi$ have a trivial limit that is independent of the coupling
function $A(\tilde\chi)$ (apart from the value of the boundaries
$[\tilde\chi_{\min},\tilde\chi_{\max}]$).

This is reminiscent of the case where the field $\tilde\chi$ is unbounded.
Then, the analysis is more complex and may involve a renormalization of the potential,
but one again finds a trivial limit, the Gaussian field \cite{Klauder2005,Rivers1990}.
In both cases, this is due to the central limit theorem (in our case the variance of the
Gaussian goes to zero because the lattice field was bounded).
In the unbounded case, one might obtain nontrivial results by introducing
a nonconventional quantization procedure, associated with a nonstandard
choice for the measure ${\cal D}\tilde\chi$ of the path integral \cite{Klauder2005}.
We do not consider such an approach here, as we do not wish to have different
path-integral measures for ordinary matter fields and the $\tilde\chi$ field.

We also note that because there is no kinetic term the action is very different
from the standard harmonic oscillator case. In particular, if the Hamiltonian reads at
quadratic order as $H = \int d\vx \frac{m^2}{2} \tilde\chi^2$, it reads as
$H= \int \frac{d\vk}{(2\pi)^3} \frac{m^2}{4\omega_{k}} [ a^{\dagger}_{\vk} a_{\vk}
+ a_{\vk} a_{\vk}^{\dagger} + a^{\dagger}_{-\vk} a^{\dagger}_{\vk} e^{2\ii\omega_k t}
+ a_{-\vk} a_{\vk} e^{-2\ii\omega_k t} ]$ in terms of the usual creation and annihilation
operators, instead of
$H= \int \frac{d\vk}{(2\pi)^3} \frac{\omega_{k}}{2} [a^{\dagger}_{\vk} a_{\vk}
+ a_{\vk} a_{\vk}^{\dagger}]$.
Besides, the plane waves $e^{\pm \ii k x}$ are not solutions of the equation of motion
of the scalar field, $\tilde\chi=0$, which is not a wave equation and does not propagate.
Thus we cannot expect ultralocal models to follow the standard quantization.

On the other hand, if the lattice cell $\epsilon$ is kept fixed,
as for standard quantum field theories we recover the usual classical limit
on macroscopic scales, where the action $S=N \epsilon S_{(\epsilon)}$ is much greater
than $1$, with $N \gg 1$.

In practice, we consider the ultralocal Lagrangian to be an effective theory,
which only applies on time and length scales greater than some UV cutoff $\epsilon^{1/4}$,
in a manner similar to the hydrodynamic approximation for fluids.
Then, we cannot use the ultralocal Lagrangian for quantum studies and it only makes
sense at the classical level, as a macroscopic theory.
Since the only dimensional scale that enters this effective Lagrangian is
${\cal M}^4 \sim \bar\rho_{\rm de 0}$, with ${\cal M} \sim 10^{-3} {\rm eV}$
and ${\cal M}^{-1} \sim 0.1 {\rm mm}$, it is natural to expect this cutoff to be around
the millimeter scale, which allows one to study cosmological and astrophysical
structures.
An alternative is to consider the ultralocal field $\tilde\chi$ to be a truly classical field,
without any quantum fluctuations.

\subsection{Kinetic conformal coupling models}
\label{sec:kinetic-quantum}

Going back to the kinetic conformal coupling models, which are the focus of this paper,
we can see that these models share some difficulties with the ultralocal models.
From the effective action (\ref{S-phi-eff-1}) the generating functional reads as
\beq
Z[j] = \int {\cal D}\varphi \; e^{\ii \int d^4x [j \varphi + {\cal M}^4 \tilde\chi
+ A(\tilde\chi) {\rm \breve{T}} ] } ,
\label{Z-kinetic-delf}
\eeq
where we consider Minkowski spacetime.
As we shall see below in section~\ref{sec:Hamilton-Jacobi}, it is natural to restrict
the kinetic term $\tilde\chi$ to the positive semiaxis, $\tilde\chi \geq 0$, and in
practice we shall consider finite intervals $[0,\gamma]$ with $\gamma>0$.
[The restriction to $\tilde\chi \geq 0$ is required to avoid multivalued
solutions to the equation of motion of the scalar field.]
This means that the coupling function $A(\tilde\chi)$ is not necessarily defined
on the negative semiaxis $\tilde\chi<0$, and may show a branch cut at $\tilde\chi=0$
as in the simple explicit model (\ref{lambda-concave-explicit})
(where $\XX=\tilde\chi/\gamma$).
In such cases we cannot perform a Wick rotation to Euclidian space, which transforms
$\tilde\chi$ to $\tilde\chi_{\rm E} < 0$.
Because of the constraint $\tilde\chi > 0$, and the fact that the equations of motion
(at the classical level) can be written in terms of $\tilde\chi$ only, as we shall see
in the following sections, it is tempting to change variable from $\varphi$ to $\tilde\chi$
in the path integral (\ref{Z-kinetic-delf}).
From the solution (\ref{phi-maximum}), which gives $\varphi$ in terms of $\tilde\chi$,
where $\psi=\sqrt{2{\cal M}^4\tilde\chi}$ from Eq.(\ref{psi-def})
(with $a=1$ in the Minkowski case),
we have
\beq
\frac{{\cal D}\varphi(x)}{{\cal D}\psi(x')} = - \int_0^{\tau} ds \sqrt{1-v^2(s)} \delta_D(x'-q(s)) .
\label{kinetic-Jacobian}
\eeq
Here we used that due to  the maximization in Eq.(\ref{phi-maximum}) the perturbations
$\delta\varphi$ are quadratic in the  perturbations $\delta q(s)$ with respect to the
maximizing path $q(s)$.
As the paths $q(s)$ that enter Eq.(\ref{kinetic-Jacobian}) are global functions
of the field $\psi$, the Jacobian $\vert {\cal D}\varphi / {\cal D}\psi \vert$ is not a constant
that can be absorbed in the normalization of the path integral.
However, if we choose to define the path integral (\ref{Z-kinetic-delf}) with
the measure $\int {\cal D} \psi$ instead of $\int {\cal D}\varphi$, we recover
ultralocal models (taking $j\psi$ for the source term) and the lack of standard
quantization.
This suggests that the kinetic conformal coupling models cannot be considered
at the quantum level.

A  simpler and rigorous argument comes from the study of the cosmological
background.
From Eq.(\ref{S-phi-eff-1}) the effective scalar-field Lagrangian is
\beq
{\cal L}_{\varphi}^{\rm eff} = {\cal M}^4 \tilde\chi - \breve\rho A(\tilde\chi) ,
\label{Leff-kinetic}
\eeq
with ${\rm \breve{T}} = - \breve\rho$.
Expanding around the cosmological background $\bar\varphi$,
$\bar{\tilde\chi}= (d\bar\varphi/d\tau)^2/(2{\cal M}^4a^2)$,
with $\varphi=\bar\varphi+\delta\varphi$, $\tilde\chi=\bar{\tilde\chi}+\delta\tilde\chi$,
we have
\beq
\delta\tilde\chi = \frac{1}{2{\cal M}^4a^2} \left[ 2 \frac{d\bar\varphi}{d\tau}
\frac{\partial\delta\varphi}{\partial\tau} + \left( \frac{\partial\delta\varphi}{\partial\tau}\right)^2
- (\nabla \delta\varphi)^2 \right] .
\eeq
If the background is given by the nontrivial solution of the equation of motion
(\ref{KG-T_psi}),
\beq
1 - \frac{\breve\rho}{{\cal M}^4} \frac{dA}{d\tilde\chi} = 0 ,
\label{chi-background-Leff}
\eeq
the Lagrangian (\ref{Leff-kinetic}) reads up to quadratic order as
\beq
\delta {\cal L}_{\delta\varphi}^{\rm eff} = - \frac{\breve\rho \bar{\tilde\chi}}{{\cal M}^4 a^2}
\frac{d^2 \bar{A}}{d\bar{\tilde\chi}^2}  \left( \frac{\partial\delta\varphi}{\partial\tau}\right)^2
+ \dots
\label{Leff-2-kinetic}
\eeq
This corresponds to a massless and vanishing-speed Gaussian field.
If we try to apply the canonical quantization to this field, adding a nonzero
velocity term $c_{\varphi}^2 (\nabla\delta\varphi)^2$ to the quadratic Lagrangian,
we find divergences in inverse powers of $c_{\varphi}$ in the limit $c_{\varphi}\to 0$
in the perturbative expansions.

For instance, let us consider scattering diagrams between photons, gravitons  and the scalar
field.
Higher-order contributions to the Lagrangian $\delta {\cal L}_{\varphi}^{\rm eff}$
include terms of the form $[(\partial \delta\varphi)^2]^{\ell}$, which give rise to a linear coupling
to the graviton of the form
\be
{\cal L}_{\delta\varphi}^{(2\ell)} \supset \frac{h_{\mu\nu}}{M_{\rm Pl}}
\frac{\partial^\mu\delta\varphi \partial^\nu\delta\varphi (\partial\delta\varphi)^{2(\ell-1)}}
{M_\ell^{4(\ell-1)}} ,
\ee
where $h_{\mu\nu} = M_{\rm Pl} \delta g_{\mu\nu}$ and $M_{\ell} \sim {\cal M}$.
Taking into account the vertex between one graviton and two photons of the type
(we only pick one part)
\be
{\cal L}_\gamma = \frac{h_{\mu\nu}}{M_{\rm Pl}} \partial^\mu A^\rho \partial^\nu A_\rho ,
\ee
we can draw a scattering diagram associated with the creation of $(2\ell)$ scalars
from two photons, with the exchange of a graviton.
The cross section reads as
\begin{eqnarray}
d\sigma_{(2\ell)} & = & \frac{1}{2 E_{\rm CM}^2}
\left( \prod_{i=1}^{2\ell} \frac{d^3p_i}{(2\pi)^3 2p_i c_\varphi} \right)
\vert {\cal M}_{(2\ell)}\vert^2 (2\pi)^4 \nonumber \\
&& \times \delta_D^{(3)}\left( \sum_i \vp_i \right) \, \delta_D\left(\sum_i E_i - E_{\rm CM}\right)
, \;\;\;
\end{eqnarray}
where $E_{\rm CM} = E_1' + E_2'$ is the total energy in the center-of-mass frame
of the two incoming photons $p_1'$ and $p_2'$.
The matrix element is symbolically
\be
{\cal M}_{(2\ell)} \sim \frac{p'_1 p'_2 (\prod_{i=1}^{2\ell} p_i)}
{M^2_{\rm Pl} M_\ell^{4(\ell-1)}p^2} ,
\ee
i.e., one momentum appears for each scalar and each photon, and there is one graviton
propagator, with $p=p_1'+p_2'$.
For each scalar we have the dispersion relation
\be
E_i = p_i c_\varphi .
\ee
We can integrate at once over $\vp_{2\ell}$ with the momentum Dirac function,
which gives $\vp_{2\ell} = - \sum_{i=1}^{2\ell-1} \vp_i$ and
\begin{eqnarray}
d\sigma_{(2\ell)} & = & \frac{\pi}{2 E_{\rm CM}^2 c_{\varphi}^{2\ell+1}}
\left( \prod_{i=1}^{2\ell-1} \frac{d^3p_i}{(2\pi)^3 2p_i} \right) \frac{1}{p_{2\ell}}
 \nonumber \\
&& \times \delta_D\left(\sum_{i=1}^{2\ell} p_i - \frac{E_{\rm CM}}{c_{\varphi}} \right)
\vert {\cal M}_{(2\ell)}\vert^2 .
\label{sigma-divergence}
\end{eqnarray}
If only one of the $p_i$ is of order $1/c_{\varphi}$ among $1 \leq i \leq 2\ell-1$
(and $p_{2\ell} \sim 1/c_{\varphi}$ as well), we have $\sigma \sim c_{\varphi}^{-2\ell-5}$.
If all momenta are of the same order $1/c_{\varphi}$, we have
$\sigma \sim c_{\varphi}^{-10\ell+3}$.
Thus, perturbative diagrams with $N$ scalars typically diverge as $1/c_{\varphi}^N$
or faster, when $c_{\varphi}\to 0$.
This is due to the increasingly large phase space volume when $c_{\varphi}\to 0$,
as greater numbers of modes fall in a given energy range.
This means that perturbative expansions are not well defined in this approach.

This can also be seen from the change of variable required to make the small-velocity
quadratic Lagrangian canonical. Indeed, writing the quadratic part of the action as
\beqa
S^{(2)} & = & \frac{1}{2} \int d^4x \left[ \left( \frac{\partial \delta\varphi}{\partial t} \right)^2
- c_{\varphi}^2 (\nabla\delta\varphi)^2 \right] \nonumber \\
& = &  \frac{1}{2} \int d^4 \tilde{x} \left[ \left( \frac{\partial \tilde\phi}{\partial \tilde{t}} \right)^2
- (\nabla\tilde\phi)^2 \right] ,
\eeqa
where we made the change of variables
\beq
t = \tilde{t} / c_{\varphi} , \;\;\; \delta\varphi = \tilde\phi/\sqrt{c_{\varphi}} ,
\label{rescaling-c-phi}
\eeq
we find that higher-order terms of the Lagrangian, such as $(\nabla\delta\varphi)^{2n}$,
diverge as $1/c_{\varphi}^n$ for $c_{\varphi} \to 0$
(higher-order gradients are not suppressed by $c_{\varphi}$).
Then, there is no perturbative regime and the limit $c_{\varphi} \to 0$ is singular.

Alternatively, if we only keep the quadratic part of the Lagrangian (\ref{Leff-2-kinetic}),
we recover a situation similar to the ultralocal models considered in the previous section.
As the Lagrangian only involves the time derivative
$(\partial\delta\varphi/\partial\tau)^2$, different space locations are decoupled.
Therefore, if we consider a lattice regularization, similar to Eq.(\ref{Z-ultralocal-lattice}),
the generating functional now factorizes as a product of 1D generating functionals,
associated with each spatial grid point. Again, for bounded fields the distribution
of each grid variable becomes flat in the continuum limit and we recover a Dirac
distribution for the field averaged over a finite spatial volume.
For unbounded fields one again recovers the Gaussian field in the continuum limit
\cite{Klauder2005}, in relation with the central limit theorem.
Again it is possible to obtain a nontrivial theory by using a nonconventional
quantization \cite{Klauder2005}, but we do not investigate this approach here.
Since at the quadratic level, around the cosmological background,
conventional quantization fails, we cannot develop the quantization of these models.

Then, as for the ultralocal models, we consider the kinetic conformal coupling models
to be effective theories, which only apply on time and length scales greater than some
UV cutoff, as the hydrodynamic approximation for fluids.
Since the dimensional scale that enters the Lagrangian (\ref{Lphi-def}) is again
${\cal M}^4 \sim \bar\rho_{\rm de 0}$, it is natural to expect this cutoff not to be far from
the millimeter scale, which allows us to study cosmological and astrophysical
structures.
Within the framework of the Goldstone models described in
section~\ref{sec:Goldstone}, we wish the phase transition to occur in this classical
regime. As the minimum of the potential $V(f^2)$ plays the role
of the cosmological constant ${\cal M}^4$, it is sufficient that the cutoff should be somewhat
above $10^{-3} {\rm eV}$, i.e. somewhat below $0.1$ mm.
The background scalar-field energy density reads as
$\bar\rho_{\varphi} \simeq {\cal M}^4 (\bar{\tilde\chi} + 1) \sim {\cal M}^4$
as we shall see in Eq.(\ref{rho-phi-p-phi-def}) below (for models
where $\tilde\chi$ is bounded as in this paper).
Therefore, at all redshifts the energy density of the scalar field remains below
the UV cutoff as soon as the latter is somewhat greater than ${\cal M}$.
An alternative is again to consider the scalar field $\varphi$ to be a truly classical field,
without any quantum fluctuations.

We note here that our Goldstone action is quite different from usual models.
Goldstone fields, associated for instance with a similar global Abelian $U(1)$ symmetry,
also appear in other contexts. For instance, the pion fields can be described as Goldstone
fields through chiral effective theory and a quantum description is possible,
around the lowest-order Lagrangian
${\rm Tr} [ \partial_{\mu}\Sigma^\dagger \partial^{\mu}\Sigma ]$.
However, in our case we expand the Lagrangian (\ref{Leff-kinetic})
around the background solution (\ref{chi-background-Leff}).
Then, the effective Lagrangian $\delta{\cal L}^{\rm eff}_{\delta\varphi}$
for the fluctuating field $\delta\varphi$ starts as $(\delta\tilde\chi)^2$,
which is very different from the usual kinetic term $\delta\tilde\chi$.
Moreover, the Minkowski limit is nonstandard.
First, the limit $\breve\rho \to 0$ is not the same as setting $\breve\rho=0$
in the Lagrangian (\ref{Leff-kinetic}). In the latter case, the Lagrangian reduces
to the free-field case ${\cal L}^{\rm eff}_{\varphi} \propto \tilde\chi$ and the
background is $\varphi=0$, with the usual result
$\delta{\cal L}_{\delta\varphi} \propto \delta\tilde\chi$.
In the former case, the background defined by Eq.(\ref{chi-background-Leff})
goes to the boundary $\tilde\chi_-$ where $dA/d\tilde\chi = +\infty$, and this value
is not necessarily zero.
If we consider a very small but nonzero uniform matter density $\breve\rho$,
which is constant with time, the background solution $\bar{\tilde\chi}$ is also
constant and nonzero. This corresponds to a background scalar field that
is a linear function of time, $\bar\varphi = - \sqrt{2{\cal M}^4 \bar{\tilde\chi}} t$.
This is thus similar to the cosmological background, with a simpler time dependence.
Then, the effective Lagrangian of the fluctuating field reads again as
$\delta{\cal L}^{\rm eff}_{\delta\varphi} = (\partial_t \delta\varphi)^2 + \dots$,
where the dots stand for higher-order terms, and there is no spatial gradient term,
i.e. a vanishing speed.
This vanishing speed, which breaks Lorentz invariance for the fluctuations,
is the source of the divergences found above in Eqs.(\ref{sigma-divergence})
and (\ref{rescaling-c-phi}), and the reason why standard perturbation theory and
canonical quantization do not apply.
Thus, the background solution breaks the Lorentz invariance of the Lagrangian, 
even in a time independent homogeneous matter environment.
In other contexts, such as chiral effective theory, the Lagrangian of the Goldstone
fields and their fluctuations is Lorentz invariant, with a nonzero velocity and a standard 
kinetic term, and one can apply the usual quantum field theory techniques.

\section{Equations of motion in the Einstein frame}
\label{sec:Equations-of-motion-Einstein}

Here and in the following sections we investigate the cosmological behavior of the kinetic
conformal coupling models defined by Eqs.(\ref{S-def})-(\ref{Lphi-def}).
We first derive in this section the equations of motion in the Einstein frame,
where gravity is easier to handle as it is given by the standard Einstein-Hilbert term.
We next express these results in the Jordan frame in the following
section~\ref{sec:Equations-of-motion-Jordan}, which is better suited to the analysis
of matter dynamics. Then, we work in the Jordan frame in the subsequent sections.

\subsection{Matter component}
\label{sec:Matter-component}

The Einstein-frame and Jordan-frame matter energy-momentum tensors are defined by
\beq
\tilde{T}_{\mu\nu} = \frac{-2}{\sqrt{-\tilde{g}}} \frac{\delta S_{\rm m}}{\delta \tilde{g}^{\mu\nu}} , \;\;\;
T_{\mu\nu} = \frac{-2}{\sqrt{-g}}
\frac{\delta S_{\rm m}}{\delta g^{\mu\nu}} ,
\label{T_munu-def}
\eeq
where we omit the subscript ``$\rm m$'' (for matter) in $\tilde{T}_{{\rm m};\mu\nu}$ and
$T_{{\rm m};\mu\nu}$ to simplify notations.
Because the conformal coupling function $A$ of Eq.(\ref{g-Jordan-def}) involves the
 derivative term $\tilde\chi$, it depends on both the scalar field $\varphi(x)$ and the
metric $\tilde{g}_{\mu\nu}(x)$. Then, in contrast with the standard scalar-tensor models, the relation
between the Einstein-frame and Jordan-frame matter energy-momentum tensors
is no longer $\tilde{T}_{\mu\nu} = A^2 T_{\mu\nu}$.
Indeed, the derivative conformal mapping (\ref{g-Jordan-def}) yields
$g^{\mu\nu} = A^{-2} \tilde{g}^{\mu\nu}$ and
\beq
\frac{\delta g^{\mu\nu}}{\delta \tilde{g}^{\alpha\beta}} = A^{-2} \left[ \delta^{\mu}_{\alpha}
\delta^{\nu}_{\beta} + \tilde{g}^{\mu\nu} \frac{d\ln A}{d\tilde\chi}
\frac{\partial_{\alpha}\varphi \partial_{\beta}\varphi}{\cM^4} \right] ,
\label{tg-g-A}
\eeq
where $\delta^{\mu}_{\alpha}$ is the Kronecker symbol, while $\sqrt{-g} = A^4 \sqrt{-\tilde{g}}$.
Then, the definition (\ref{T_munu-def}) gives
\beq
\tilde{T}_{\mu\nu} = A^2 T_{\mu\nu} + A^4 T \frac{d\ln A}{d\tilde\chi}
\frac{\partial_{\mu}\varphi \partial_{\nu}\varphi}{\cM^4} ,
\label{T_munu-Ttilde_munu}
\eeq
where we defined the traces of the Einstein-frame and Jordan-frame energy-momentum tensors
as
\beq
\tilde{T} = \tilde{T}^{\mu}_{\mu} = \tilde{g}^{\mu\nu} \tilde{T}_{\mu\nu} , \;\;\;
T = T^{\mu}_{\mu} = g^{\mu\nu} T_{\mu\nu} ,
\label{trace-T-def}
\eeq
i.e., operations on Einstein-frame (resp. Jordan-frame) tensors only involve the
Einstein-frame (resp. Jordan-frame) metric $\tilde{g}_{\mu\nu}$ (resp. $g_{\mu\nu}$).
The relation (\ref{T_munu-Ttilde_munu}) also yields
\beq
\tilde{T} = A^4 T \left[ 1 - 2 \tilde\chi  \frac{d\ln A}{d\tilde\chi} \right] ,
\label{T-Ttilde-def}
\eeq
so that Eq.(\ref{T_munu-Ttilde_munu}) can be inverted as
\beq
T_{\mu\nu} = A^{-2} \tilde{T}_{\mu\nu} - A^{-2} \tilde{T}
\frac{\frac{d\ln A}{d\tilde\chi}}{1 - 2 \tilde\chi \frac{d\ln A}{d\tilde\chi}}
\frac{\partial_{\mu}\varphi \partial_{\nu}\varphi}{\cM^4} .
\label{Ttilde_munu-T_munu}
\eeq

In the Jordan frame, the matter energy-momentum tensor satisfies the usual conservation
law, $\nabla_{\mu} T^{\mu}_{\nu} = 0$.
From the expression (\ref{T_munu-Ttilde_munu}) this implies for the Einstein-frame
energy-momentum tensor the ``nonconservation'' law
\beq
\tilde\nabla_{\mu} \tilde{T}^{\mu}_{\nu} = \partial_{\nu} \varphi \tilde\nabla_{\mu}
\left[ A^4 \frac{T}{\cM^4} \frac{d\ln A}{d\tilde\chi} \tilde\nabla^{\mu}\varphi \right] .
\label{non-conserv-Einstein}
\eeq
[It is the Jordan-frame trace $T$ that appears within the brackets but it can be expressed in
terms of $\tilde{T}$ through Eq.(\ref{T-Ttilde-def}).]

\subsection{Radiation component}
\label{sec:Radiation-component}

If we consider a radiation component with a pressure $p_{\rm rad}=\rho_{\rm rad}/3$, instead
of a matter component with a small pressure $p \ll \rho$, we have $T_{\rm rad}= 0$.
Then, $\tilde{T}_{\rm rad}=0$ and
$\tilde{T}_{\rm rad;\mu\nu} = A^2 T_{\rm rad;\mu\nu}$, and as in standard dilaton models,
both Einstein-frame and Jordan-frame radiation energy-momentum tensors satisfy the usual
conservation law, $\tilde{\nabla}_{\mu} \tilde{T}^{\mu}_{\rm rad;\nu} = 0$ and
$\nabla_{\mu} T^{\mu}_{\rm rad;\nu} = 0$.

\subsection{Scalar field}
\label{sec:Scalar-field}

The functional derivative of the Jordan-frame metric with respect to the scalar field
$\varphi$ reads as
\beqa
\frac{\delta g^{\mu\nu}(x')}{\delta\varphi(x)} & = &
\frac{2}{{\cal M}^4} A^{-2}(x') \frac{d\ln A}{d\tilde\chi}(x') \tilde{g}^{\mu\nu}(x')
\tilde{g}^{\alpha\beta}(x') \nonumber \\
&& \times \; \partial_{\alpha}\varphi(x') \frac{\partial}{\partial x'^{\beta}} \delta_D(x-x') ,
\eeqa
and the Klein-Gordon equation of motion of the scalar field $\varphi$, obtained from the
variation of the action (\ref{S-def}) with respect to $\varphi$, can be written as
\beq
\tilde\nabla_{\mu} \left[ \left( 1 + A^4 \frac{T}{\cM^4} \frac{d\ln A}{d\tilde\chi} \right)
\tilde\nabla^{\mu} \varphi \right] = 0 .
\label{K-G-1-def}
\eeq
We recover the equation of motion of the scalar field that we obtained in Eq.(\ref{KG-T_psi})
for the explicit case of fermionic matter and in section~\ref{sec:point-particles} for matter
classical point-particles.
The result (\ref{K-G-1-def}) is general and $T$ is the sum of the traces of the
energy-momentum tensors of all matter components.

The Einstein-frame energy-momentum tensor of the scalar field reads as
\beqa
\tilde{T}_{\varphi;\mu\nu} & = & \frac{-2}{\sqrt{-\tilde{g}}}
\frac{\delta S_{\varphi}}{\delta \tilde{g}^{\mu\nu}} \nonumber \\
& = & {\cal M}^4 (\tilde\chi-1) \tilde{g}_{\mu\nu} + \partial_{\mu}\varphi \partial_{\nu}\varphi .
\label{T-phi-def}
\eeqa
Using the Klein-Gordon equation (\ref{K-G-1-def}) we obtain the nonconservation law
\beq
\tilde\nabla_{\mu} \tilde{T}^{\mu}_{\varphi;\nu} = \partial_{\nu} \varphi \; \tilde\nabla_{\mu}
\tilde\nabla^{\mu} \varphi ,
\label{non-conserv-Einstein-phi}
\eeq
while the matter nonconservation law (\ref{non-conserv-Einstein}) simplifies as
\beq
\tilde\nabla_{\mu} \tilde{T}^{\mu}_{\nu} = - \partial_{\nu}  \varphi \; \tilde\nabla_{\mu}
\tilde\nabla^{\mu} \varphi .
\label{non-conserv-Einstein-1}
\eeq
Then, we can check that the full Einstein-frame energy-momentum tensor obeys the
usual conservation law,
\beq
\tilde\nabla_{\mu} \left[ \tilde{T}^{\mu}_{\nu} + \tilde{T}^{\mu}_{\rm rad;\nu}
+ \tilde{T}^{\mu}_{\varphi;\nu} \right] = 0 ,
\label{conserv-total}
\eeq
which ensures consistency with the Bianchi identity for the Einstein tensor,
$\tilde\nabla_{\mu} \tilde{G}^{\mu}_{\nu} =0$.

A constant scalar field, $\varphi = {\rm constant}$, is always a solution of the Klein-Gordon
equation (\ref{K-G-1-def}). Then, the scalar field plays no role; its kinetic energy vanishes,
the conformal transformation (\ref{g-Jordan-def}) is an irrelevant rescaling of coordinates
by a constant factor, and we recover the $\Lambda$-CDM scenario.
In the following, we focus on the nontrivial solution of the Klein-Gordon equation
(\ref{K-G-1-def}),
\beq
1 + A^4 \frac{T}{\cM^4} \frac{d\ln A}{d\tilde\chi} = 0 .
\label{K-G-2-def}
\eeq
This is a constraint equation for the kinetic term $\tilde\chi$, which becomes a function of the
Jordan-frame matter density when $T=-\rho$, and we have $\tilde\chi(x) = \tilde\chi[\rho(x)]$.

Then, the relation (\ref{T_munu-Ttilde_munu}) between the Einstein- and Jordan-frame matter
energy-momentum tensors simplifies as
\beq
\tilde{T}_{\mu\nu} = A^2 T_{\mu\nu} - \partial_{\mu}\varphi \partial_{\nu}\varphi .
\label{Ttilde-T-phi}
\eeq
Combining with Eq.(\ref{T-phi-def}) we find that the sum of the matter and scalar-field tensors is
\beq
\tilde{T}_{\mu\nu} + \tilde{T}_{\varphi;\mu\nu} = A^2 T_{\mu\nu}
+ {\cal M}^4 (\tilde\chi-1) \tilde{g}_{\mu\nu} .
\label{Tmatter-Tphi}
\eeq

\subsection{Background dynamics}
\label{sec:background-Einstein}

For the cosmological background, which is homogeneous and isotropic, both the
Einstein-frame and Jordan-frame metrics are of the Friedman-Lemaitre-Robertson-Walker
type, with
\beq
d \tilde{s}^2 = \tilde{g}_{\mu\nu} d x^{\mu} d x^{\nu} = \tilde{a}^2 [ - d\tau^2 + d \vx^2 ] ,
\label{FLRW-Einstein}
\eeq
\beq
d s^2 = g_{\mu\nu} d x^{\mu} d x^{\nu} = a^2 [ - d\tau^2 + d \vx^2 ] ,
\label{FLRW-Jordan}
\eeq
where $\tau$ is the conformal time.
The conformal transformation (\ref{g-Jordan-def}) means that the line elements transform
as $ds^2=A^2 d\tilde{s}^2$. At the background level, this means that the scale factor, physical
time and distance transform as
\beq
a = \bar{A} \tilde{a}, \;\; dt = \bar{A} d\tilde{t} , \;\; \vr = \bar{A} \tilde{\vr} .
\label{aE-aJ}
\eeq
Throughout this paper we denote with an overbar background quantities.

As we found in Eq.(\ref{T_munu-Ttilde_munu}), in contrast with models where the
conformal transformation (\ref{g-Jordan-def}) only depends on the value of the scalar
field $A(\varphi)$, the Einstein and Jordan matter energy-momentum tensors are not
proportional through a factor $A^2$. They also differ by an additive factor that explicitly
depends on the scalar field kinetic factor $\partial_{\mu}\varphi \partial_{\nu}\varphi$.
As the matter Lagrangian ${\cal L}_{\rm m}$ is given in the Jordan frame,
in the action (\ref{S-def}), it is the Jordan matter energy-momentum tensor that takes
the standard form, e.g. the perfect fluid form with an equation of state between matter
pressure and density, while the Einstein energy-momentum tensor takes a nonstandard
form with contributions from both the matter and scalar field sectors.
Moreover, the matter particles follow the geodesics defined by the Jordan-frame metric
$g_{\mu\nu}$ so that they obey standard continuity equations in the fluid limit.
At the background level, we write the Jordan- and Einstein-frame energy-momentum tensors
as
$\bar{T}^{\mu}_{\nu} = {\rm diag}(-\bar\rho,\bar{p},\bar{p},\bar{p})$ and
$\bar{\tilde{T}}^{\mu}_{\nu} =
{\rm diag}(-\bar{\tilde\rho},\bar{\tilde{p}},\bar{\tilde{p}},\bar{\tilde{p}})$.
As usual, we work in the nonrelativistic limit and we neglect the matter pressure, $p \ll \rho$.
Then, Eq.(\ref{T_munu-Ttilde_munu}) gives
\beq
\bar{\tilde\rho} = \bar{A}^4 \bar\rho \left[ 1 - 2 \frac{d\ln \bar{A}}{d\ln\bar{\tilde\chi}} \right] , \;\;\;\;
\bar{\tilde{p}} = \bar{A}^4 \bar{p} = 0 ,
\label{rho-E-J-mean}
\eeq
where the scalar field kinetic term is
\beq
\bar{\tilde\chi} = \frac{1}{2{\cal M}^4 \tilde{a}^2} \left( \frac{d\bar\varphi}{d\tau} \right)^2 \geq 0 .
\label{bar-chi}
\eeq
The radiation energy-momentum tensors are proportional, with
\beq
\bar{\tilde\rho}_{\rm rad} = \bar{A}^4 \bar\rho_{\rm rad} , \;\;\;\;
\bar{\tilde{p}}_{\rm rad} = \bar{A}^4 \bar{p}_{\rm rad} .
\label{rho-rad-E-J-mean}
\eeq
The Einstein-frame scalar-field energy-momentum tensor (\ref{T-phi-def}) takes its standard
form, with
\beq
\bar{\tilde\rho}_{\varphi} = {\cal M}^4 ( \bar{\tilde\chi} + 1 ), \;\;\;
\bar{\tilde{p}}_{\varphi} = {\cal M}^4 ( \bar{\tilde\chi} - 1 ) .
\label{rho-phi-mean}
\eeq

In the Einstein frame, we recover the usual Friedmann equation,
\beq
3 \tilde{M}_{\rm Pl}^2 \tilde{\cal H}^2 = \tilde{a}^2 \left( \bar{\tilde\rho} + \bar{\tilde\rho}_{\rm rad}
+ \bar{\tilde\rho}_{\varphi} \right) ,
\label{Friedmann-1}
\eeq
where $\tilde{\cal H}=d\ln\tilde{a}/d\tau$ is the Einstein-frame conformal expansion rate.

The Klein-Gordon equation (\ref{K-G-2-def}) reads as
\beq
\frac{d\ln \bar{A}}{d\bar{\tilde\chi}} = \frac{{\cal M}^4}{\bar{A}^4 \bar\rho} ,
\label{K-G-E-1}
\eeq
which can be written in terms of the Einstein-frame matter density as
\beq
\frac{\frac{d\ln \bar{A}}{d\bar{\tilde\chi}}}{1-2 \frac{d\ln \bar{A}}{d\ln\bar{\tilde\chi}}}
= \frac{{\cal M}^4}{\bar{\tilde\rho}} .
\label{K-G-E-2}
\eeq
This equation determines the scalar field kinetic term as a function of the Einstein-frame
background matter density, $\bar{\tilde\chi}(\tau) = \bar{\tilde\chi}[\bar{\tilde\rho}(\tau)]$.
This gives the value of the background scalar field by integrating Eq.(\ref{bar-chi}),
\beq
\bar\varphi(\tau) = - \int_0^{\tau} d\tau \sqrt{2 {\cal M}^4 \tilde{a}^2 \bar{\tilde\chi}} .
\label{phi-chi-integral}
\eeq
Here we choose without a loss of generality the boundary condition $\bar\varphi(\tau=0)=0$
and the negative sign for the square-root of Eq.(\ref{bar-chi}), because the action (\ref{S-def})
is invariant through the transformations $\varphi \to \varphi+{\rm constant}$ and
$\varphi \to - \varphi$.
[We choose more specifically the negative sign as it will correspond to the positive
convex Hamiltonian (\ref{Hamiltonian-def}) when we obtain $\varphi$ through the
Hamilton-Jacobi equation (\ref{phi-chi-minus-sqrt}) below.
Choosing the positive solution for $\varphi$ would lead to nonconventional signs in
this context.]
The scalar-field energy density evolves as
\beq
\frac{d\bar{\tilde\rho}_{\varphi}}{d\tau} = {\cal M}^4 \frac{d\bar{\tilde\chi}}{d\tau} .
\label{rhophi-evol-E}
\eeq

The matter nonconservation equation (\ref{non-conserv-Einstein-1}) yields
\beq
\frac{d\bar{\tilde\rho}}{d\tau} = - 3 \tilde{\cal H} \bar{\tilde\rho} - 6 \tilde{\cal H}
{\cal M}^4 \bar{\tilde\chi} - {\cal M}^4 \frac{d\bar{\tilde\chi}}{d\tau} .
\label{rho-evol-E}
\eeq
Combining with Eq.(\ref{rhophi-evol-E}) we recover the usual conservation law for the
sum of the matter and scalar-field components,
$d(\bar{\tilde\rho} + \bar{\tilde\rho}_{\varphi})/d\tau = - 3 \tilde{\cal H}
( \bar{\tilde\rho} + \bar{\tilde\rho}_{\varphi} + \bar{\tilde{p}}_{\varphi})$.
The radiation density obeys the standard evolution equation
\beq
\frac{d\bar{\tilde\rho}_{\rm rad}}{d\tau} = - 4 \tilde{\cal H} \bar{\tilde\rho}_{\rm rad} .
\label{rho-radiation-evol-E}
\eeq

\subsection{Perturbations}
\label{sec:perturbations-Einstein}

In the Einstein frame we write the perturbed metric in the Newtonian gauge as
\beq
d \tilde{s}^2= \tilde{a}^2 [ - (1+2\tilde\Phi) d\tau^2 + (1-2\tilde\Psi) d\vx^2 ] ,
\label{metric-E}
\eeq
At the linear order in the metric potentials and for scales much below the Hubble radius,
the $(0,0)$-component of the Einstein equations,
$\tilde{M}_{\rm Pl}^2 \tilde{G}^{\mu}_{\nu}=\tilde{T}^{\mu}_{\nu}+\tilde{T}^{\mu}_{\rm rad;\nu}
+\tilde{T}^{\mu}_{\varphi;\nu}$, gives
\beq
2\tilde{M}_{\rm Pl}^2 \frac{\nabla^2 \tilde\Psi}{\tilde{a}^2} = \delta( A^4 \rho )
- {\cal M}^4 \delta\tilde\chi ,
\label{Psi-E}
\eeq
where we used the nonrelativistic limit, $v^2 \ll c^2$, we neglected fluctuations of
the radiation component, and $\rho$ is again the Jordan-frame density.
Hereafter we denote with a $\delta$ perturbed quantities, such as
$\delta\tilde\chi \equiv \tilde\chi - \bar{\tilde\chi}$.
The nondiagonal $(i,j)$-components of the Einstein equations give
\beq
\frac{\tilde{M}_{\rm Pl}^2}{\tilde{a}^2} \partial_i\partial_j ( \tilde\Psi - \tilde\Phi )
= \bar{A}^4 \rho v_i v_j ,
\eeq
at linear order over $\delta A$, and in the nonrelativistic limit we obtain
\beq
\tilde\Phi = \tilde\Psi .
\label{Phi-Psi-E}
\eeq

\section{Equations of motion in the Jordan frame}
\label{sec:Equations-of-motion-Jordan}

We now express in the Jordan frame the results obtained in the previous section.
We study both the background dynamics and the growth of linear perturbations.

\subsection{Background dynamics}
\label{sec:background-Jordan}

From Eq.(\ref{aE-aJ}) the conformal expansion rates in the Jordan and Einstein frames
are related by
\beq
\tilde{\cal H} = (1-\epsilon_2) {\cal H} \;\;\; \mbox{with} \;\;\;
\epsilon_2(\tau) = \frac{d\ln\bar{A}}{d\ln a} .
\label{H-J-E}
\eeq
From Eq.(\ref{Ttilde-T-phi}) the Einstein- and Jordan-frame background matter densities
are related by
\beq
\bar{\tilde\rho} = \bar{A}^4 \bar\rho - 2 {\cal M}^4 \bar{\tilde\chi} .
\label{rhoJ-rhoE-mean}
\eeq
We also define the effective Jordan-frame scalar-field density and pressure by a simple
rescaling by $\bar{A}^4$,
\beq
\bar\rho_{\varphi} \equiv \frac{\bar{\tilde\rho}_{\varphi}}{\bar{A}^4} = {\cal M}^4
\frac{\bar{\tilde\chi}+1}{\bar{A}^4} , \;\;\;
\bar{p}_{\varphi} \equiv \frac{\bar{\tilde{p}}_{\varphi}}{\bar{A}^4} = {\cal M}^4
\frac{\bar{\tilde\chi}-1}{\bar{A}^4} .
\label{rho-phi-p-phi-def}
\eeq
Then, the Friedmann equation (\ref{Friedmann-1}) yields
\beq
3 M_{\rm Pl}^2 {\cal H}^2 = (1-\epsilon_2)^{-2} a^2 ( \bar{\rho} + \bar{\rho}_{\rm rad}
- \bar{p}_{\varphi} ) ,
\label{Friedmann-J}
\eeq
where we defined the time-dependent Jordan-frame Planck mass as
\beq
M_{\rm Pl}^2(\tau) \equiv \bar{A}^{-2} \, \tilde{M}_{\rm Pl}^2 .
\label{Planck-J}
\eeq
It is the scalar-field pressure with a negative sign, $- \bar{p}_{\varphi}$, instead of the density
$\bar\rho_{\varphi}$, that enters the Friedmann equation (\ref{Friedmann-J}), because of the
second term in Eq.(\ref{rhoJ-rhoE-mean}).
To recover the Friedmann equation in its standard form, we can define an effective dark-energy
density by
\beq
3 M_{\rm Pl}^2 {\cal H}^2 = a^2 ( \bar{\rho} + \bar{\rho}_{\rm rad} + \bar{\rho}_{\rm de} ) ,
\label{Friedmann-J-de}
\eeq
which gives
\beq
\bar{\rho}_{\rm de} \equiv - \bar{p}_{\varphi} + \frac{2\epsilon_2-\epsilon_2^2}{(1-\epsilon_2)^2}
( \bar{\rho} + \bar{\rho}_{\rm rad} - \bar{p}_{\varphi} ) .
\label{rho-de-def}
\eeq

In the Jordan frame the matter obeys the standard conservation equations,
$\nabla_{\mu} T^{\mu}_{\nu} =0$, and the background matter and radiation densities
evolve as
\beq
\bar\rho = \frac{\bar\rho_0}{a^3} , \;\,\,\,
\bar\rho_{\rm rad} = \frac{\bar\rho_{\rm rad 0}}{a^4} .
\label{rho-a-J}
\eeq
The scalar field is given by Eq.(\ref{K-G-E-1}), which reads as
\beq
\bar{A}^4 \frac{d\ln \bar{A}}{d\bar{\tilde\chi}} = \frac{{\cal M}^4}{\bar\rho} .
\label{K-G-J-1}
\eeq
This determines the scalar field kinetic term as a function of the Jordan-frame background
matter density, $\bar{\tilde\chi}(\tau) = \bar{\tilde\chi}[\bar\rho(\tau)]$, and the integrated
equation (\ref{phi-chi-integral}) becomes
\beq
\bar\varphi = - \int_0^{\tau} d\tau \frac{\sqrt{2 {\cal M}^4 a^2 \bar{\tilde\chi}}}{\bar{A}} .
\label{phi-chi-integral-J}
\eeq
We also have from Eq.(\ref{K-G-J-1})
\beq
\frac{d\bar{\tilde\chi}}{d\tau} = \epsilon_2 {\cal H} \frac{\bar{A}^4 \bar\rho}{{\cal M}^4} .
\label{chi-tau-epsilon2}
\eeq

\subsection{Perturbations}
\label{sec:Perturbations-J}

In the Jordan frame we write the Newtonian gauge metric as
\beq
d s^2= a^2 [ - (1+2\Phi) d\tau^2 + (1-2\Psi) d\vx^2 ] .
\label{metric-J}
\eeq
Using $ds^2 = A^2 d\tilde{s}^2$, the comparison with Eq.(\ref{metric-E}) leads to
\beq
\Phi = \tilde\Phi + \delta\ln A , \;\;\; \Psi = \tilde\Psi - \delta\ln A ,
\label{Phi-J-Phi-E}
\eeq
at linear order in $\delta A$, while the Einstein-frame Poisson equation (\ref{Psi-E})
also reads as
\beq
2 M_{\rm Pl}^2 \frac{\nabla^2 \tilde\Psi}{a^2} = \frac{\delta( A^4 \rho )
- {\cal M}^4 \delta\tilde\chi}{\bar{A}^4} .
\label{Psi-E-J}
\eeq
Since we wish the deviations from General Relativity and the $\Lambda$-CDM
cosmology to be small, at most of the order of ten percent,
the potentials $\Phi$ and $\Psi$ cannot deviate too much from the Jordan-frame
Newtonian potential, $\Phi \simeq \Psi \simeq \Psi_{\rm N}$.
From Eq.(\ref{Phi-J-Phi-E}) this implies that $\delta\ln A$ is smaller than the typical value of the
Newtonian potential.
Since $|\Psi_{\rm N}|$ is typically of order $10^{-5}$, over cosmological and astrophysical
scales, we can indeed linearize in $\delta A$ and we must have
\beq
| \delta \ln A | \lesssim 10^{-6}  , \;\;\; \mbox{hence} \;\;\; | A - 1 | \lesssim 10^{-6} ,
\label{dlnA-small}
\eeq
where we choose unity as the reference value of $A$.
Then, we can simplify the Poisson equation (\ref{Psi-E-J}) as,
\beq
\frac{\nabla^2 \Psi_{\rm N}}{a^2} = \frac{\delta\rho - {\cal M}^4 \delta\tilde\chi}{2 M_{\rm Pl}^2} ,
\label{PsiN-J}
\eeq
which defines a modified Newtonian potential, and the Jordan metric potentials read from
Eq.(\ref{Phi-J-Phi-E}) as
\beq
\Phi = \Psi_{\rm N} + \delta\ln A , \;\;\; \Psi = \Psi_{\rm N} - \delta\ln A .
\label{Phi-J-Psi-J-PsiN}
\eeq

In the Jordan frame, both the matter and radiation components obey the standard equations
of motion. This gives for the matter component the continuity and Euler equations
\beq
\frac{\partial\rho}{\partial\tau} + (\vv\cdot\nabla) \rho + ( 3 {\cal H}+\nabla\cdot\vv) \rho = 0 ,
\label{continuity-J}
\eeq
and
\beq
\frac{\partial\vv}{\partial\tau} + (\vv\cdot\nabla)\vv + {\cal H} \vv = -\nabla\Phi .
\label{Euler-J}
\eeq

The scalar field equation of motion (\ref{K-G-2-def}) simplifies as
\beq
\frac{d\ln A}{d\tilde\chi} = \frac{{\cal M}^4}{\rho} ,
\label{K-G-J-pert}
\eeq
which also gives
\beq
\nabla \ln A = \frac{d\ln A}{d\tilde\chi} \nabla\tilde\chi = \frac{{\cal M}^4}{\rho} \nabla\tilde\chi ,
\eeq
and the Euler equation also writes as
\beq
\frac{\partial\vv}{\partial\tau} + (\vv\cdot\nabla)\vv + {\cal H} \vv = -\nabla\Psi_{\rm N}
- \frac{\nabla p_A}{\rho}  ,
\label{Euler-1-J}
\eeq
with
\beq
p_A = \cM^4 c^2 \tilde\chi ,
\label{pA-def}
\eeq
where we explicitly wrote the factor $c^2$.

Thus, in terms of the matter dynamics, the modified-gravity
effects appear through two factors, a) the modification of the Poisson equation
(\ref{PsiN-J}), due to the additional source associated with the scalar-field
energy density fluctuations, but this effect will turn out to be negligible, and
b) the new pressure-like term $p_A$ in the Euler equation (\ref{Euler-1-J}).
This pressure $p_A$ corresponds to a polytropic equation of state, as it only
depends on the matter density (the sum of cold dark matter and baryons),
through Eq.(\ref{K-G-J-pert}) which implicitly determines $\tilde\chi$ as a function
of $\rho$.

\subsection{Linear growing mode}
\label{sec:linear-J}

On large scales or at early times we may linearize the equations of motion.
Expanding the coupling function $A(\tilde\chi)$ as
\beq
\ln A(\tilde\chi) = \ln\bar{A} + \sum_{n=1}^{\infty} \frac{\beta_n(\tau)}{n!} (\delta\tilde\chi)^n ,
\label{beta-n-def}
\eeq
the scalar field equation (\ref{K-G-J-pert}) gives at the background and linear
orders
\beq
\beta_1 = \frac{{\cal M}^4}{\bar\rho} > 0 , \;\;\;
\delta\tilde\chi = - \frac{\beta_1}{\beta_2} \delta ,
\label{dchi-linear}
\eeq
where we note $\delta\equiv \delta\rho/\bar\rho$ the matter density contrast.
The continuity equation (\ref{continuity-J}) reads as
$\pl_{\tau}\delta+\nabla\cdot[(1+\delta)\vv]=0$ in terms of the density contrast.
Combining with the Euler equation at linear order, this gives
\beq
\frac{\partial^2\delta}{\partial\tau^2} + {\cal H} \frac{\partial\delta}{\partial\tau}
+ \epsilon_1 c^2 \nabla^2\delta =
\frac{\bar\rho a^2}{2 M_{\rm Pl}^2} ( 1 + \epsilon_1 ) \delta ,
\label{delta-evol}
\eeq
where we introduced
\beq
\epsilon_1(\tau) \equiv \frac{\beta_1}{\beta_2} \frac{\cM^4}{\bar\rho}
= \frac{\beta_1^2}{\beta_2} .
\label{eps1-def}
\eeq

As compared with the $\Lambda$-CDM cosmology, the pressure-like term $\nabla^2\delta$
introduces an explicit scale dependence. Going to Fourier space, the linear
modes $D(k,\tau)$ now depend on the wave number $k$ and obey the evolution equation
\beq
\frac{\pl^2 D}{\pl (\ln a)^2} + \left( 2 + \frac{1}{H^2} \frac{d H}{d t} \right)
\frac{\pl D}{\pl\ln a} - \frac{3\Omega_{\rm m}}{2} (1+\epsilon) D = 0 ,
\label{DL}
\eeq
where $H=d\ln a/d t$ is the Jordan-frame expansion rate (with respect to the
Jordan-frame cosmic time $t$) and the factor $\epsilon(k,t)$, which describes the
deviation from the $\Lambda$-CDM cosmology, is given by
\beq
\epsilon(k,\tau) = \epsilon_1(\tau) \left( 1 + \frac{2}{3\Omega_{\rm m}}
\frac{c^2k^2}{a^2H^2} \right) .
\label{eps-def}
\eeq
Thus, the two effects of the scalar field, the contribution to the gravitational potential
from $\delta\rho_{\tilde\chi}$ and the pressure-like term associated with the conformal
transformation (\ref{g-Jordan-def}), modify the growth of structures in the same
direction, given by the sign of $\epsilon_1$. A positive $\epsilon_1$ gives a scale-dependent
amplification of the gravitational force and an acceleration of gravitational clustering.
The $k$-dependent pressure-like term dominates when $ck/aH>1$, that is, on subhorizon
scales. Moreover, we have $(ck/aH)^2 \sim 10^7$ today at scales of about
$1 \, h^{-1}$Mpc. Therefore, we must have
\beq
| \epsilon_1 | \lesssim 10^{-7}
\label{epsilon1-bound1}
\eeq
to ensure that the growth of large-scale structures is not too significantly modified.
Moreover, the fluctuations of the scalar field energy density
in the Poisson equations are negligible, as the factor $\epsilon_1$ is negligible with respect
to unity in the right-hand side in Eq.(\ref{delta-evol}).

From the definitions of $\beta_1$ and $\beta_2$, taking the derivative with respect to
the scale factor $a$ of the first Eq.(\ref{dchi-linear}) gives
\beq
\epsilon_2 = \frac{3\beta_1^2}{\beta_2} = 3 \epsilon_1 ,
\label{eps2-eps1}
\eeq
where $\epsilon_2$ was defined in Eq.(\ref{H-J-E}).
Therefore, we find from the condition (\ref{epsilon1-bound1}) that $\epsilon_2$ is also
very small, $|\epsilon_2| \lesssim 10^{-7}$,
\beq
| \epsilon_2 | = \left | \frac{d\ln \bar{A}}{d\ln a} \right| \lesssim 10^{-7} .
\label{epsilon2-small}
\eeq
We recover the condition (\ref{dlnA-small}) that $A$ remains very close to unity.
This also implies that the Einstein and Jordan frames are very close.

\subsection{Comparison with ultralocal models}
\label{sec:ultralocal}

The equations of motion that we have obtained in section~\ref{sec:Equations-of-motion-Jordan}
in the Jordan frame are identical to those associated with the ultralocal models introduced
in \cite{Brax:2016vpd}, if we make the change $\tilde\chi -1 \to \tilde\chi$.
The factor $-1$ comes from the explicit introduction of a cosmological constant in the
Lagrangian (\ref{Lphi-def}), which was not needed for the ultralocal models as any constant
shift could be absorbed within the definition of $\tilde\chi$.

We have recalled the definition of the ultralocal models in section~\ref{sec:ultralocal-def}.
In particular, we noticed that apart from the factor $-1$, we recover the Lagrangian
(\ref{Lphi-def}), $\tilde{\cal L}_{\tilde\chi}(\tilde\chi) = {\cal M}^4 \tilde\chi$,
and the conformal coupling $A(\tilde\chi)$.
However, whereas for the kinetic conformal coupling models that we study in this paper
$\tilde\chi$ is the kinetic term (\ref{g-Jordan-def}) of an underlying scalar field $\varphi$,
for the ultralocal models $\tilde\chi$ is a standard scalar field on its own.
These two different meanings of $\tilde\chi$ lead to some differences in the detailed behavior
of some quantities. First, for the ultralocal models the relation between the Einstein- and
Jordan-frame matter energy-momentum tensors takes the standard form,
\beq
\mbox{ultralocal:} \;\;\; \tilde{T}_{\mu\nu} = A^2 T_{\mu\nu} ,
\eeq
and the nonconservation equation for the Einstein-frame matter energy-momentum tensor
also takes the standard form
\beq
\mbox{ultralocal:} \;\;\; \tilde\nabla_{\mu} \tilde{T}^{\mu}_{\nu} = \tilde{T} \partial_{\nu} \ln A ,
\eeq
in contrast with Eqs.(\ref{T_munu-Ttilde_munu}) and (\ref{non-conserv-Einstein}).
Second, the Einstein-frame scalar-field energy-momentum tensor reads as
\beq
\mbox{ultralocal:} \;\;\; \tilde{T}_{\varphi;\mu\nu} = {\cal M}^4 \tilde\chi \tilde{g}_{\mu\nu} ,
\eeq
instead of Eq.(\ref{T-phi-def}), where we can see a sign of the change
$\tilde\chi -1 \to \tilde\chi$ between both models, and the equation of motion of the scalar
field reads as
\beq
\mbox{ultralocal:} \;\;\; 1 + \frac{\tilde{T}}{{\cal M}^4} \frac{d\ln A}{d\tilde\chi} = 0 ,
\label{KG-ultralocal}
\eeq
instead of Eq.(\ref{K-G-1-def}). The latter equation admits the new solution
$\varphi = {\rm constant}$,
and the other solution (\ref{K-G-2-def}) is only identical to Eq.(\ref{KG-ultralocal}) if we replace
in the latter $\tilde{T}$ by $A^4 T$ [and not if we express $T$ in terms of $\tilde{T}$ in
Eq.(\ref{K-G-2-def})].
On the other hand, this remark already shows that the two models are more closely related
if we work in Jordan frame.
Third, in the Einstein-frame Friedmann equation (\ref{Friedmann-1}) it is $-{\cal M}^4\tilde\chi$
instead of ${\cal M}^4\tilde(\chi+1)$ that enters the right-hand side.
However, the Poisson equation (\ref{Psi-E}) takes the same form, because of a cancellation
between the changes of the matter and scalar-field energy densities between the two models,
related to the simplification (\ref{Tmatter-Tphi}).

Despite these differences between both models in the Einstein frame, we recover almost the
same equations of motion in the Jordan frame. For the ultralocal models we recover
the Friedmann equation
(\ref{rho-de-def}), where $-\bar{p}_{\varphi} = {\cal M}^4 (1-\bar{\tilde\chi})$ is replaced by
$-{\cal M}^4 \bar{\tilde\chi}$ and the background scalar-field equation of motion is again
given by Eq.(\ref{K-G-J-1}).
At the perturbative level we obtain the same Poisson equation (\ref{PsiN-J}), which implies that
the equations of motion of matter particles and of fluids are the same because they follow
the geodesics of the Jordan metric $g_{\mu\nu}$ in both models.

Therefore, if we work in the Jordan frame, the only difference between the two models is that
for the kinetic conformal coupling case studied in this paper we have the additional constraint
that $\tilde\chi$ must be interpreted as the kinetic term of an underlying scalar field $\varphi$.
Neglecting the metric fluctuations this reads as
\beq
\mbox{kinetic coupling:} \;\;\;
\tilde\chi = \frac{1}{2{\cal M}^4 a^2} \left[ \left(\frac{\partial\varphi}{\partial\tau}\right)^2
- ( \nabla\varphi)^2 \right] ,
\label{chi-phi-def}
\eeq
where we used again $\bar{A} \simeq 1$.
Even though the field $\varphi$ does not appear in the Jordan-frame equations of motion,
it cannot be ignored as the requirement that $\tilde\chi$ can be written in the form
(\ref{chi-phi-def}) entails new constraints on the model.
Indeed, Eq.(\ref{chi-phi-def}) implies at once that at the background level we have
$\bar{\tilde\chi} \geq 0$, as in Eq.(\ref{bar-chi}).
This is actually the reason why we had to introduce an explicit cosmological constant in
the Lagrangian (\ref{Lphi-def}) through the factor $-{\cal M}^4$.
Indeed, the scalar-field contribution to the Friedmann equation
(\ref{Friedmann-J}) is $-\bar{p}_{\varphi} = {\cal M}^4 (1-\bar{\tilde\chi})$.
Without the unit factor we obtain the contribution $- {\cal M}^4 \bar{\tilde\chi} \leq 0$,
which can never give rise to an accelerated expansion as it is negative.
In contrast, in the ultralocal models there is no {\it a priori} restriction on the sign of $\tilde\chi$
[any constant shift can actually be absorbed by a change of the coupling function
$A(\tilde\chi)$], and we could obtain an accelerated expansion by building models where
$\bar{\tilde\chi}$ is negative in the late Universe.
For the kinetic conformal coupling model studied in this paper we need the unit factor
in Eq.(\ref{Lphi-def}), which can be interpreted as a standard cosmological constant or as
the value of the potential $V(\varphi)$, which is approximated as a constant on the scales
of interest, or is exactly constant at low energy within the Goldstone models
described in section~\ref{sec:shift}.

\section{The scalar field $\varphi$ as a function of its kinetic term $\tilde\chi$}
\label{sec:Hamilton-Jacobi}

We have seen above that the behavior of the system can be written in terms of $\tilde\chi$
only. This also allows us to recover the dynamics of the ultralocal models.
However, to be considered as a solution associated with the kinetic conformal coupling action
(\ref{S-def}), we must check that the field $\tilde\chi(\tau,\vx)$ obtained from these equations
of motion can also be written as the kinetic term of a scalar field $\varphi(\tau,\vx)$.
Therefore, we must integrate Eq.(\ref{chi-phi-def}) for $\varphi$.
This was done in Eq.(\ref{phi-chi-integral-J}) at the background level, but we must generalize
this result to the perturbed Universe.

We recall that at the background level we have from Eq.(\ref{bar-chi}) the condition
$\bar{\tilde\chi} \geq 0$, while the equation of motion of the background scalar field
(\ref{K-G-J-1}) simplifies as
\beq
\frac{d\ln \bar{A}}{d\bar{\tilde\chi}} = \frac{{\cal M}^4}{\bar\rho} ,
\label{K-G-J-back-1}
\eeq
where we used $\bar{A} \simeq 1$ from Eq.(\ref{dlnA-small}).
We require that this equation always has a solution, for any density $\bar\rho$.
This means that $d\ln A/d\tilde\chi$ covers the full positive semiaxis
over some range $] \tilde\chi_- , \tilde\chi_+ [$ of $\tilde\chi$ with $\tilde\chi_- \geq 0$,
\beq
\frac{d\ln A}{d\tilde\chi}( ] \tilde\chi_- , \tilde\chi_+ [ )  =  ] 0 , +\infty [   \;\;\ \mbox{with} \;\;\;
\tilde\chi_- \geq 0 .
\label{positive-axis}
\eeq
Then, to avoid multiple solutions and discontinuities, we restrict the general solution
$\tilde\chi(\rho)$ of the scalar-field equation of motion (\ref{K-G-J-pert}) to this interval
$]\tilde\chi_- , \tilde\chi_+ [$.
Thus, we require that $\tilde\chi(\rho)$ is an invertible function of $\rho$ from $] 0 , +\infty [$
to $]\tilde\chi_- , \tilde\chi_+ [$ with $\tilde\chi_- \geq 0$.
Then, the constraint $\tilde\chi \geq 0$ implies from Eq.(\ref{chi-phi-def})
\beq
\tilde\chi \geq 0 \;\;\; \mbox{and} \;\;\; \left| \frac{\partial\varphi}{\partial\tau}\right| \geq
| \nabla\varphi | .
\label{no-static}
\eeq
Therefore, in contrast with most scalar-field modified-gravity models,
the scalar field can never reach a quasi-static limit
(defined as $| \partial\varphi / \partial\tau | \ll | \nabla\varphi |$),
even on small astrophysical scales such as the Solar System.

We can write the solution of equation (\ref{chi-phi-def}) as
\beq
\frac{\partial\varphi}{\partial\tau} = - \sqrt{ 2{\cal M}^4 a^2 \tilde\chi + (\nabla\varphi)^2 } .
\label{phi-chi-minus-sqrt}
\eeq
Here we can choose the negative sign because Eq.(\ref{chi-phi-def}) is invariant under
the transformation $\varphi \to - \varphi$.
We can also choose the boundary condition,
\beq
\varphi(\tau=0,\vx) = 0 ,
\label{phi-boundary}
\eeq
which is consistent with the background solution (\ref{phi-chi-integral-J}) that applies at
early times when perturbations can be neglected.
Thus, we obtain the well-known Hamilton-Jacobi equation
\beq
\frac{\partial\varphi}{\partial\tau} + H( \tau,\vx , \nabla\varphi ) = 0 ,
\label{Hamilton-Jacobi}
\eeq
where we introduced the time-dependent Hamiltonian
\beq
H(\tau,\vx,\vp) = \sqrt{ \psi^2(\tau,\vx) + \vp^2 } ,
\label{Hamiltonian-def}
\eeq
with
\beq
\psi(\tau,\vx) = \sqrt{ 2{\cal M}^4 a^2 \tilde\chi } .
\label{psi-def}
\eeq

\subsection{Method of characteristics}
\label{sec:characteristics}

The nonlinear partial differential equation (\ref{Hamilton-Jacobi}) can be solved by the
method of characteristics, see \cite{Evans2010}.
This allows us to compute $\varphi$ along curves $x(s)=(\tau(s),\vx(s))$, with the parametric
coordinate $s$. Defining
\beq
p_0(s) = \frac{\partial\varphi}{\partial\tau}[x(s)] , \;\;\;  \vp(s) = \nabla\varphi[x(s)] ,
\eeq
the characteristic curves of Eq.(\ref{Hamilton-Jacobi}) are given by
\beq
\frac{d \tau}{ds} = 1 , \;\;\; \frac{d\vx}{ds} = \frac{\partial H}{\partial \vp}
\label{charact-1}
\eeq
and
\beq
\frac{d p_0}{ds} = - \frac{\partial H}{\partial\tau} , \;\;\; \frac{d\vp}{ds} =
- \frac{\partial H}{\partial \vx} .
\label{charact-2}
\eeq
Thanks to the first equation (\ref{charact-1}) we can identify $s$ with the time $\tau$,
$\tau(s)=s$. Then, the characteristic curves are given by the Hamilton equations,
\beq
\frac{d\vx}{d\tau} = \frac{\partial H}{\partial \vp} , \;\;\; \frac{d\vp}{d\tau} =
- \frac{\partial H}{\partial \vx} ,
\label{Hamilton-eqs}
\eeq
while the scalar field along the curve is given by the integration of
\beq
\frac{d\varphi}{d\tau} = p_0 + \frac{\partial H}{\partial \vp} \cdot \vp = - H
+  \frac{\partial H}{\partial \vp} \cdot \vp ,
\eeq
with the boundary condition (\ref{phi-boundary}).
For the Hamiltonian (\ref{Hamiltonian-def}) the characteristic curves read as
\beq
\frac{d\vx}{d\tau} = \frac{\vp}{\sqrt{\psi^2+\vp^2}} , \;\;\;
\frac{d\vp}{d\tau} = - \frac{\psi}{\sqrt{\psi^2+\vp^2}} \nabla\psi ,
\eeq
\beq
\frac{d\varphi}{d\tau} = - \frac{\psi^2}{\sqrt{\psi^2+\vp^2}} .
\label{phi-characteristics}
\eeq
For any initial point $\vx_{(0)}$, with $\vp_{(0)}=0$, at $\tau=0$, this method allows us
to obtain a local solution $\varphi(\tau,\vx)$ around the starting point until some finite time
$\tau>0$.
In particular, in the homogeneous case we recover the background solution
(\ref{phi-chi-integral-J}) (where we took $\bar{A} \simeq 1$), with $\vx(\tau)=\vx_{(0)}$
and $\vp(\tau)=0$.
However, it is well-known that in the general case this procedure cannot
extend to all space, as different characteristics may intersect, which would lead to different
values of $\varphi$ at those points.
This means that the initial-value problem (\ref{Hamilton-Jacobi}) does not in general have
a smooth solution, existing for all times $\tau>0$ \cite{Evans2010}.

\subsection{Viscosity solution}
\label{sec:viscosity}

Fortunately, it is possible to extend the local solutions obtained by the method of characteristics
to a unique generalized or weak solution, also called ``viscosity'' solution
\cite{Crandall1983,Crandall1984}.
Indeed, the Hamiltonian (\ref{Hamiltonian-def}) is bounded from below (by $0$), we can
also assume that $H(\tau,\vx,0)$ is bounded from above (e.g., if $\tilde\chi_+$ is finite)
and that $|\partial \psi/\partial x|$ is bounded.
Then, as $H(\tau,\vx,\vp)$ is convex in $\vp$, this ensures that
there exists a unique viscosity solution of the Hamilton-Jacobi equation (\ref{Hamilton-Jacobi}),
which is given by the explicit expression
\cite{Cannarsa2004,Tonon2011}
\beq
\varphi(\tau,\vx) = \min_{\vq(s)} \left\{ \int_0^{\tau} ds \, L(s,\vq(s),\dot{\vq}(s)) \;
\biggl \vert \; \vq(\tau)=\vx \right\} ,
\label{phi-minimum}
\eeq
where we used the boundary condition (\ref{phi-boundary}), $\varphi(0,\vx)=0$.
Here we take the minimum over all paths $\vq(s)$ that are of class $C^2$, start at any point
$\vx_{(0)}$ on the boundary surface $\tau=0$, and end at the point $(\tau,\vx)$.
We also introduced the Lagrangian $L(\tau,\vx,\vv)$ defined by
\beq
L(\tau,\vx,\vv) = \sup_{\vp} \left[ \vv\cdot\vp - H(\tau,\vx,\vp) \right] .
\label{Lagrangian-def}
\eeq
Thus, as could be expected the solution of the Hamilton-Jacobi equation
(\ref{Hamilton-Jacobi}) is the action, defined by the Hamiltonian $H$ or the Lagrangian $L$.
The Hamiltonian and the Lagrangian are related by the usual Legendre transform,
with
\beq
\vv = \frac{\partial H}{\partial\vp} , \;\;\; \vp = \frac{\partial L}{\partial\vv} .
\eeq
Thanks to the convexity of $H$ in $\vp$, the Lagrangian is also convex in $\vv$, both
the Hamiltonian and the Lagrangian are actually Legendre transforms of each other,
and from Eq.(\ref{Hamiltonian-def}) we obtain the explicit expressions
\beq
| \vv | < 1 , \;\; \vv= \frac{\vp}{\sqrt{\psi^2+\vp^2}} , \;\; \vp= \frac{\psi \vv}{\sqrt{1-\vv^2}} ,
\eeq
\beq
L(\tau,\vx,\vv) = - \psi(\tau,\vx) \sqrt{1-\vv^2} .
\label{Lagrangian-psi-v}
\eeq
Whereas $|\vp|$ is unbounded we have $|\vv| < 1$.
This ensures that causality is satisfied as the paths in the minimization (\ref{phi-minimum})
cannot travel faster than light.
The paths that minimize Eq.(\ref{phi-minimum}) also satisfy the Euler-Lagrange equations
defined by the Lagrangian $L$.
Since $\psi \geq 0$ we can also write the solution (\ref{phi-minimum}) as
\beq
\varphi(\tau,\vx) = - \max_{\vq(s)} \left\{ \int_0^{\tau} ds \, \psi(s) \sqrt{1 - \vv^2(s)}
\; \biggl \vert \; \vq(\tau)=\vx \right\} ,
\label{phi-maximum}
\eeq
where we wrote $\psi(s) = \psi(s,\vq(s))$ and $\vv(s) = \dot\vq(s)$ for each path.
We can intuitively figure out the behavior of $\varphi(\vx)$ from the explicit maximization
(\ref{phi-maximum}). Neglecting the time dependence of $\psi$, we can see that for a given
location $\vx$ the maximization selects the paths that spend most time over the maximum
$\psi_{\max}$ within the horizon. Local maxima $\psi_{\max}^{(i)}$ (i.e., within the
horizon scale) thus define attraction
basins, or cells such as Voronoi diagrams. Within each cell $\varphi(\vx)$ is regular, and at
the boundary between two cells we change of attractor $\psi_{\max}^{(i)}$.
Then, $\varphi$ remains continuous at the boundary
[this defines the boundary; if there were a jump, we could get a higher maximum in
Eq.(\ref{phi-maximum}) on the lower side by extending paths that come from the upper side]
but its gradient is discontinuous.
Taking into account the finite velocity and the time dependence makes the solution more
intricate than standard Voronoi diagrams, but theorems ensure that the viscosity solution is
Lipschitz continuous, with a piecewise smooth gradient that shows jump discontinuities
along surfaces of dimension three (in spacetime) \cite{Cannarsa2004,Tonon2011}.
These are rather mild singularities, in particular the gradient does not show Dirac-type
singularities.
Moreover, the Lorentz kinetic term (\ref{chi-phi-def}), from which we obtain $\varphi$,
is as smooth as the density field and does not show jumps if we take the density
to be continuous.

For the background case, we again recover the solution
(\ref{phi-chi-integral-J}) and the path obtained below Eq.(\ref{phi-characteristics}) by the
method of characteristics.
Indeed, since $\bar\psi(\tau)$ only depends on time and is positive, the maximization of
Eq.(\ref{phi-maximum}) corresponds to $\vv=0$ (to maximize the square-root factor at all
intermediate times $s$). This gives again the motionless path $\vx(s)=\vx(\tau)=\vx(0)$,
$\vv=\vp=0$, and the integral (\ref{phi-chi-integral-J}).

The concept of viscosity solutions is more general than the explicit expression
(\ref{phi-minimum}), which relies on the convexity of the Hamiltonian $H$.
In the general case, a bounded and continuous function $\varphi$ is a viscosity
subsolution (supersolution) of the Hamilton-Jacobi equation (\ref{Hamilton-Jacobi}) if,
for any $C^1$ test function $\Phi$ such that $\varphi-\Phi$ has a maximum (minimum)
at $(\tau,\vx)$, then $\partial_{\tau}\Phi+H(\tau,\vx,\nabla\Phi) \leq 0$
($\partial_{\tau}\Phi+H(\tau,\vx,\nabla\Phi) \geq 0$).
A viscosity solution is both a viscous subsolution and supersolution.
This concept ensures the existence, stability and uniqueness of the solution.
Thus, for a Lipschitz continuous Hamiltonian there is at most one viscous solution, and
for the convex case it exists and is given by the explicit expression (\ref{phi-minimum}).
Moreover, if a classical $C^1$ solution exists, then it is also a viscous solution (i.e.,
this concept contains the standard solutions and provides a generalization for the
case when the latter do not exist, when the method of characteristic develops singularities).
The viscosity solution can also be obtained as the limit $\epsilon\to 0^+$
when we add a small viscous term $\epsilon\nabla^2\varphi$ to the Hamilton-Jacobi
equation (\ref{Hamilton-Jacobi}).
It also satisfies the following stability property.
If $H_{\epsilon}$ is a sequence of continuous Hamiltonians, and $\varphi_{\epsilon}$ the
associated sequence of viscous solutions, then if $H_{\epsilon} \to H$ and
$\varphi_{\epsilon} \to \varphi$ for $\epsilon \to 0$, the limit function $\varphi$ is also
the viscous solution of $H$.
We refer the reader to the mathematical literature, e.g.
\cite{Crandall1983,Crandall1984,Cannarsa2004,Tonon2011,Barles2013},
for more detailed statements and further properties.

The many good properties of the viscous solutions explain why they are the relevant
solutions in usual applications.
In our case, we can choose to define $\varphi$ by Eq.(\ref{phi-maximum}).
Then, the stability theorem described above ensures that this definition remains valid
for small perturbations of $\tilde\chi$, and $\varphi$ is a continuous function of $\tilde\chi$.
Then, because the equations of motion only depend on $\tilde\chi$, the dynamical stability
can be analyzed in terms of $\tilde\chi$ only, as in section~\ref{sec:stability} below.
In practice, it may happen that the shift symmetry described in
section~\ref{sec:shift} is not exact and broken by small corrections, that is,
the action (\ref{S-def}) contains small terms that depend on the scalar field value $\varphi$,
in addition to the derivative $\tilde\chi$ terms studied in this paper.
Then, one can perform a perturbative analysis around the viscosity solution.
The fact that the latter is continuous means that these corrections to the equation of motion
are also well behaved, being continuous and small.

\section{Explicit models}
\label{sec:models}

We have seen in section~\ref{sec:ultralocal} that we recover the equations of motion
of the ultralocal models studied in \cite{Brax:2016vpd}. Therefore, we can use their results
and follow the same approach.

\subsection{Characteristic density $\rho_{\alpha}$ and redshift $z_{\alpha}$}
\label{sec:rhoc}

To satisfy the constraints (\ref{dlnA-small}) we write the coupling function as
\beq
\ln A(\tilde\chi) = \alpha \lambda(\XX)  \;\;\; \mbox{with} \;\;\; \alpha \lesssim 10^{-6} ,
\;\;\; 0 \leq \lambda \leq 1 ,
\label{lambda-def}
\ee
and
\beq
\tilde\chi = \gamma \XX  \;\;\; \mbox{with} \;\;\; \gamma > 0, \;\;\; 0 \leq \XX \leq 1 .
\label{gamma-chi-def}
\eeq
Thus, $\alpha$ is a small parameter that sets the magnitude of $\ln A$,
while $\gamma$ sets the magnitude of $\tilde\chi$.
This corresponds to models where $\tilde\chi$ ranges from $0$ to a finite value
$\gamma > 0$.
We could also consider models where $\tilde\chi$ goes up to infinity
(see \cite{Brax:2016vpd} for such a study in the case of ultralocal models), but most
of qualitative behaviors would not be modified.
Then, the equation of motion (\ref{K-G-J-pert}) reads as
\beq
\frac{d\lambda}{d\XX} = \frac{1}{\hat\rho} \;\;\; \mbox{with} \;\;\;
\hat\rho = \frac{\alpha\rho}{\gamma{\cal M}^4} .
\label{rho-hat-def}
\eeq
This implicitly defines the functions $\lambda(\hat\rho)$ and $\XX(\hat\rho)$,
from the value of $\XX$ that solves Eq.(\ref{rho-hat-def}) for a given density.
The changes of variables $\ln A \to \lambda$, $\tilde\chi \to \XX$, and $\rho \to \hat\rho$
have removed the explicit parameters
${\cal M}^4 \sim \bar\rho_{\rm de0}$, $\alpha \lesssim 10^{-6}$, and $\gamma$, so that
the functions $\lambda(\XX)$, $\lambda(\hat\rho)$ and $\XX(\hat\rho)$
do not involve small nor large parameters.
Therefore, in addition to the density ${\cal M}^4 \sim \bar\rho_{\rm de0}$,
these models automatically introduce another density scale
$\rho_{\alpha}$,
\be
\rho_{\alpha} = \frac{\gamma{\cal M}^4}{\alpha} \sim \frac{\gamma\bar\rho_{\rm de0}}{\alpha}
\gtrsim 10^6 \; \gamma \bar\rho_{\rm de0} ,
\;\;\; \hat\rho = \frac{\rho}{\rho_{\alpha}} .
\label{rho-alpha-def}
\ee
This implies that, from the point of view of the coupling function $\ln A$,
the low-redshift mean density of the Universe is within its very low density
regime if $\gamma \gg \alpha$.
Moreover, as we shall check below, there is a cosmological transition between low-density
and high-density regimes at the redshift $z_{\alpha}$ where $\bar\rho \sim \rho_{\alpha}$,
which corresponds to
\beq
a_{\alpha} \sim (\alpha/\gamma)^{1/3} \lesssim 0.01 \gamma^{-1/3} , \;\;\;
z_{\alpha} \sim (\alpha/\gamma)^{-1/3} \gtrsim 100 \gamma^{1/3} .
\label{zalpha-def}
\eeq
In the numerical computations presented in this paper, we will consider the case
$\alpha=10^{-6}$ with $\gamma \sim 1, 10^{-3}$, and $10^{-6}$, to investigate scenarios
where the transition redshift $z_{\alpha}$ goes from 100 to 0.

\subsection{Coupling function}
\label{sec:coupling}

To define explicit models we simply need to give the explicit expression of the coupling
function $A(\tilde\chi)$, or of the rescaled function $\lambda(\XX)$ of
Eq.(\ref{lambda-def}). From Eq.(\ref{rho-hat-def}), we derive the properties
\beq
\frac{d\lambda}{d\XX} > 0 , \;\;\;
\left. \frac{d\lambda}{d\XX} \right|_{\rho=0} = +\infty , \;\;\;
\left. \frac{d\lambda}{d\XX} \right|_{\rho=\infty} = 0.
\label{lambda-constraints-general}
\eeq
We also require an accelerated expansion in the late Universe, corresponding to the
dark energy era. From Eqs.(\ref{rho-de-def}) and (\ref{rho-phi-p-phi-def}) this means
that $\bar{p}_{\varphi}(z=0) < 0$ and $\bar{\tilde\chi}(z=0) < 1$.
To have a unique and well-defined mapping
$\rho\leftrightarrow \XX$, the function
$d\lambda/d\XX$ must be monotonic. Therefore, we have two choices,
where $\lambda(\XX)$ is either concave or convex.

\subsubsection{Concave coupling function $\lambda(\XX)$}
\label{sec:concave}

\paragraph{General concave case}
\label{sec:general-concave}
\mbox{}\newline

If $\lambda(\XX)$ is concave, the first derivative $d\lambda/d\XX$
is a decreasing function of $\XX$, which implies that $\XX$ is an increasing
function of $\rho$, and we have the asymptotic behaviors
\beq
\rho \to 0 : \;\;\; \tilde\chi \to 0 , \;\; \XX \to 0 , \;\; \frac{d\lambda}{d\XX} \to +\infty ,
\label{rho-0-chi-concave}
\eeq
\beq
\rho \to \infty : \;\;\; \tilde\chi \to \gamma , \;\; \XX \to 1, \;\; \frac{d\lambda}{d\XX} \to 0 .
\label{rho-infty-chi-concave}
\eeq
In the following we consider power-law behaviors at these boundaries,
\beqa
\XX \to 0 :  && \;\;\; \frac{d\lambda}{d\XX} \sim \XX^{\mu_-  -1}  \;\;\;
\mbox{with} \;\;\; 0 < \mu_- < 1 , \nonumber \\
\hat\rho \ll 1 : && \;\;\; \XX \sim \hat\rho^{1/(1-\mu_-)} ,
\label{mu-m-def-concave}
\eeqa
and
\beqa
\XX \to 1 : && \;\;\; \frac{d\lambda}{d\XX} \sim (1-\XX)^{\mu_+ -1}  \;\;\;
\mbox{with} \;\;\; \mu_+ > 1 , \nonumber \\
\hat\rho \gg 1 : && \;\;\; 1 - \XX \sim \hat\rho^{-1/(\mu_+ -1)} .
\label{mu-p-def-concave}
\eeqa

From Eqs.(\ref{rho-phi-p-phi-def}) and (\ref{rho-de-def}), we find that at early and late times
the dark energy density behaves as
\beq
z \gg z_{\alpha} : \;\;\; \bar{\tilde\chi} \to \gamma , \;\;\;
\bar\rho_{\rm de} \to (1-\gamma) {\cal M}^4 ,
\eeq
\beq
z \ll z_{\alpha} : \;\;\; \bar{\tilde\chi} \to 0 , \;\;\; \bar\rho_{\rm de} \to {\cal M}^4 .
\label{M4-concave}
\eeq
Thus, the dark energy density grows with time.
If $\gamma \gg \alpha$ we have seen that the transition redshift $z_{\alpha}$ is very large,
so that Eq.(\ref{M4-concave}) implies ${\cal M}^4= \bar\rho_{\rm de0}$.
If $\gamma \lesssim \alpha \ll 1$ the dark energy density is almost constant and very close
to ${\cal M}^4$. Therefore, in all cases we obtain
\beq
\mbox{concave:} \;\;\; {\cal M}^4 = \bar\rho_{\rm de0} , \;\;\;
\hat\rho = \frac{\alpha\rho}{\gamma \bar\rho_{\rm de0}} ,
\label{M4-rhode0-concave}
\eeq
and the dark energy density is almost constant and positive at low $z$.
This sets the energy scale ${\cal M}$ in terms of the dark energy density today.

To study the evolution of linear matter perturbations in Eq.(\ref{DL}) we need
the factor $\epsilon_1(\tau)$. From the definitions (\ref{beta-n-def}) and
(\ref{eps1-def}) we have
\beq
\beta_1 = \frac{\alpha}{\gamma} \frac{d\lambda}{d\XX} , \;\;\;
\beta_2 = \frac{\alpha}{\gamma^2} \frac{d^2\lambda}{d\XX^2} , \;\;
\epsilon_1 = \alpha \frac{ ( d\lambda/d\XX)^2 }{ d^2\lambda/d\XX^2 } .
\eeq
Like the coupling function $\ln A$, the amplitude $\epsilon_1$
is proportional to $\alpha$ and does not depend on $\gamma$.
Thus, the small value of the parameter $\alpha$ simultaneously ensures that the
constraints (\ref{dlnA-small}) and (\ref{epsilon1-bound1}) are satisfied.
For a concave coupling function $\lambda$, we have
\beq
\lambda \;\; \mbox{concave:} \;\;\;  \beta_1 > 0 , \;\; \beta_2 < 0 , \;\; \epsilon_1 < 0 .
\eeq
This means that concave coupling functions imply a slower growth of large-scale
structures than the $\Lambda$-CDM cosmology.
At high redshift we obtain for the power-law models (\ref{mu-p-def-concave})
\beq
z \gg z_{\alpha} : \;\;\; \epsilon_1 \sim - \alpha (1-\bar{\XX})^{\mu_+} \sim
- \alpha (a/a_{\alpha})^{3\mu_+/(\mu_+ -1)} ,
\label{epsilon1-high-z-concave}
\eeq
while at low redshift the power-law models (\ref{mu-m-def-concave}) give
\beq
z \ll z_{\alpha} : \;\;\; \epsilon_1 \sim - \alpha \bar{\XX}^{\mu_-} \sim
- \alpha (a/a_{\alpha})^{-3\mu_-/(1-\mu_-)} .
\label{epsilon1-low-z-concave}
\eeq
Thus, the amplitude $| \epsilon_1(\tau) |$ peaks at the transition redshift
$z_{\alpha}$, with power-law decays at higher and lower redshifts,
and
\beq
\epsilon_1(z_{\alpha}) \sim - \max(| \epsilon_1 |) \sim - \alpha .
\label{epsilon1-max-concave}
\eeq

\paragraph{A simple explicit concave model}
\label{sec:explicit-concave}
\mbox{}\newline

For illustration, in the numerical computations presented in this paper we choose the
simple coupling function
\beq
0 \leq \XX \leq 1 : \;\;\; \lambda(\XX) = \sqrt{\XX (2-\XX)} ,
\label{lambda-concave-explicit}
\eeq
which we display in Fig.~\ref{fig_lambda_chiI}.
This corresponds to the exponents
\beq
\mu_- = 1/2 , \;\;\; \mu_+ = 2 ,
\eeq
the density and the scalar-field kinetic term are related by
\beq
\hat\rho = \frac{\sqrt{\XX (2-\XX)}}{1-\XX} , \;\;\;
\XX = 1 - \frac{1}{\sqrt{\hat\rho^2+1}} ,
\eeq
which gives
\beqa
&& \lambda = \frac{\hat\rho}{\sqrt{\hat\rho^2+1}} , \;\;\;
\frac{d\lambda}{d\ln\rho} = \frac{\hat\rho}{(\hat\rho^2+1)^{3/2}} , \nonumber\\
&& \epsilon_1 = - \alpha \frac{\bar{\hat\rho}}{(\bar{\hat\rho}^2+1)^{3/2}} .
\label{lambda-rho-concave-explicit}
\eeqa

\subsubsection{Convex coupling function $\lambda(\tilde\chi)$}
\label{sec:convex}

\paragraph{General convex case}
\label{sec:general-convex}
\mbox{}\newline

If $\lambda(\XX)$ is convex, the first derivative $d\lambda/d\XX$
is an increasing function of $\XX$, which implies that $\XX$ is a decreasing
function of $\rho$, and we have the asymptotic behaviors
\beq
\rho \to 0 : \;\;\; \tilde\chi \to \gamma , \;\; \XX \to 1, \;\; \frac{d\lambda}{d\XX} \to +\infty ,
\label{rho-0-chi-convex}
\eeq
\beq
\rho \to \infty : \;\;\; \tilde\chi \to 0 , \;\; \XX \to 0 , \;\; \frac{d\lambda}{d\XX} \to 0 .
\label{rho-infty-chi-convex}
\eeq
We again focus on power-law behaviors at these boundaries,
\beqa
\XX \to 0 : && \;\;\; \frac{d\lambda}{d\XX} \sim \XX^{\mu_+  -1}  \;\;\;
\mbox{with} \;\;\; \mu_+ > 1 , \nonumber \\
\hat\rho \gg 1 : && \;\;\; \XX \sim \hat\rho^{-1/(\mu_+-1)} ,
\label{mu-p-def-convex}
\eeqa
and
\beqa
\XX \to 1 : && \;\;\; \frac{d\lambda}{d\XX} \sim (1 -\XX)^{\mu_- -1}  \;\;\;
\mbox{with} \;\;\; 0 < \mu_- < 1 , \nonumber \\
\hat\rho \ll 1 : && \;\;\; 1 - \XX \sim \hat\rho^{1/(1-\mu_-)} .
\label{mu-m-def-convex}
\eeqa

From Eqs.(\ref{rho-phi-p-phi-def}) and (\ref{rho-de-def}), we find that at early and late times
the dark energy density behaves as
\beq
z \gg z_{\alpha} : \;\;\; \bar{\tilde\chi} \to 0 , \;\;\; \bar\rho_{\rm de} \to {\cal M}^4 ,
\eeq
\beq
z \ll z_{\alpha} : \;\;\; \bar{\tilde\chi} \to \gamma , \;\;\; \bar\rho_{\rm de} \to
(1-\gamma) {\cal M}^4 .
\label{M4-convex}
\eeq
Thus, the dark energy density decreases with time.
If $\gamma \gg \alpha$ the transition redshift $z_{\alpha}$ is very large
and Eq.(\ref{M4-convex}) implies ${\cal M}^4= \bar\rho_{\rm de0}/(1-\gamma)$.
If $\gamma \lesssim \alpha \ll 1$ the dark energy density is almost constant and very close
to ${\cal M}^4 \simeq (1-\gamma) {\cal M}^4$. Therefore, in all cases we obtain
\beq
\mbox{convex:} \;\;\;  {\cal M}^4 = \frac{\bar\rho_{\rm de0}}{1-\gamma} , \;\;\;
\hat\rho = \frac{\alpha (1-\gamma) \rho}{\gamma \bar\rho_{\rm de0}} .
\label{M4-rhode0-convex}
\eeq
In particular, for the convex models we obtain an upper bound on the parameter $\gamma$,
\beq
\lambda \;\; \mbox{convex} : \;\;\; \gamma < 1 ,
\label{gamma-lt-1-convex}
\eeq
so that the dark energy density today is positive.
Then, the dark energy density is almost constant and positive at low $z$.

For a convex coupling function $\lambda$, we have
\beq
\lambda \;\; \mbox{convex:} \;\;\;  \beta_1 > 0 , \;\; \beta_2 > 0 , \;\; \epsilon_1 > 0 .
\eeq
This means that convex coupling functions imply a faster growth of large-scale
structures than the $\Lambda$-CDM cosmology.
At high redshift we obtain for the power-law models (\ref{mu-p-def-convex})
\beq
z \gg z_{\alpha} : \;\;\; \epsilon_1 \sim \alpha \bar{\XX}^{\mu_+} \sim
\alpha (a/a_{\alpha})^{3\mu_+/(\mu_+ -1)} ,
\label{epsilon1-high-z-convex}
\eeq
while at low redshift the power-law models (\ref{mu-m-def-convex}) give
\beq
z \ll z_{\alpha} : \;\;\; \epsilon_1 \sim \alpha (1 - \bar{\XX})^{\mu_-} \sim
\alpha (a/a_{\alpha})^{-3\mu_-/(1-\mu_-)} .
\label{epsilon1-low-z-convex}
\eeq
Thus, $\epsilon_1(\tau)$ again peaks at the transition redshift
$z_{\alpha}$, with power-law decays at higher and lower redshifts, and
\beq
\epsilon_1(z_{\alpha}) \sim \max(\epsilon_1) \sim \alpha .
\label{epsilon1-max-convex}
\eeq

\paragraph{A simple explicit convex model}
\label{sec:explicit-convex}
\mbox{}\newline

\begin{figure}
\begin{center}
\epsfxsize=8. cm \epsfysize=5.5 cm {\epsfbox{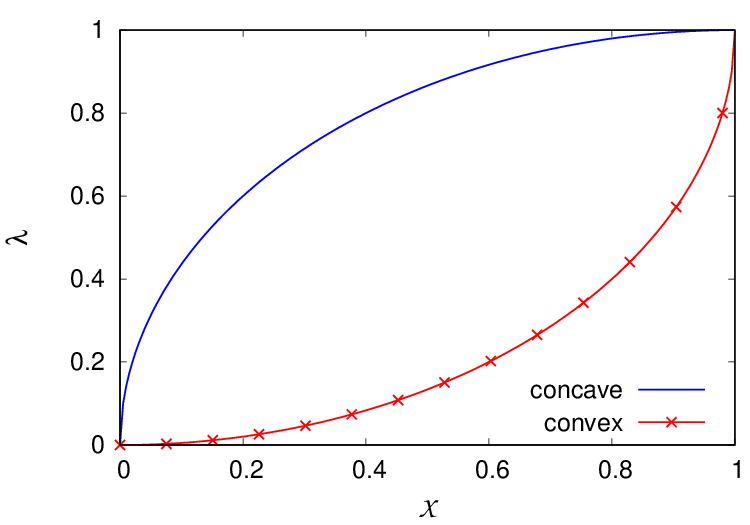}}
\end{center}
\caption{
Normalized coupling functions $\lambda(\XX)$ for the concave and
convex models of Eqs.(\ref{lambda-concave-explicit}) and (\ref{lambda-convex-explicit})
used in the numerical computations.
}
\label{fig_lambda_chiI}
\end{figure}

For the numerical computations we choose the simple coupling function
\beq
0 \leq \XX \leq 1 : \;\;\; \lambda(\XX) = 1 - \sqrt{1 - \XX^2} ,
\label{lambda-convex-explicit}
\eeq
which we display in Fig.~\ref{fig_lambda_chiI}.
This again corresponds to the exponents
\beq
\mu_- = 1/2 , \;\;\; \mu_+ = 2 ,
\eeq
the density and the scalar-field kinetic term are related by
\beq
\hat\rho = \frac{\sqrt{1 - \XX^2}}{\XX} , \;\;\;
\XX = \frac{1}{\sqrt{\hat\rho^2+1}} ,
\eeq
which gives
\beqa
&& \lambda = 1- \frac{\hat\rho}{\sqrt{\hat\rho^2+1}} , \;\;\;
\frac{d\lambda}{d\ln\rho} = - \frac{\hat\rho}{(\hat\rho^2+1)^{3/2}} , \nonumber \\
&& \epsilon_1 =  \alpha \frac{\bar{\hat\rho}}{(\bar{\hat\rho}^2+1)^{3/2}} .
\label{lambda-rho-convex-explicit}
\eeqa

\subsubsection{Common high-density and low-density behaviors}
\label{sec:common-behaviors}

At high densities, which corresponds to
\beq
\rho \gg \rho_{\alpha} , \;\;\; \hat\rho \gg 1, \;\;\; \frac{d\lambda}{d\XX} \ll 1 ,
\eeq
in Eq.(\ref{rho-hat-def}), both the concave and convex models introduced
in Eqs.(\ref{mu-p-def-concave}) and (\ref{mu-p-def-convex}) satisfy
\beq
\frac{d\lambda}{d\XX} \sim | \XX - \XX_{+} |^{\mu_+ -1} ,
\;\;\; \mu_+ > 1 , \;\;\;  | \XX - \XX_{+} | \ll 1,
\eeq
with $\XX_{+} = 1$ or $0$. This gives
\beq
| \XX - \XX_{+} | \sim \hat\rho^{\, -1/(\mu_+ -1)} , \;\;\;
\left | \frac{d\lambda}{d\ln\rho} \right | \sim \hat\rho^{\, -\mu_+/(\mu_+ -1)} .
\label{lambda-high-density-common}
\eeq

At low densities, which corresponds to
\beq
\rho \ll \rho_{\alpha} , \;\;\; \hat\rho \ll 1, \;\;\; \frac{d\lambda}{d\XX} \gg 1 ,
\eeq
in Eq.(\ref{rho-hat-def}), both the concave and convex models introduced
in Eqs.(\ref{mu-m-def-concave}) and (\ref{mu-m-def-convex}) satisfy
\beq
\frac{d\lambda}{d\XX} \sim | \XX - \XX_{-} |^{\mu_- -1} ,
\;\;\; 0 < \mu_- < 1 , \;\;\;  | \XX - \XX_{-} | \ll 1,
\eeq
with $\XX_{-} = 0$ or $1$. This gives
\beq
| \XX - \XX_- | \sim \hat\rho^{\, 1/(1-\mu_-)} , \;\;\;
\left | \frac{d\lambda}{d\ln\rho} \right | \sim \hat\rho^{\, \mu_-/(1-\mu_-)} .
\label{lambda-low-density-common}
\eeq

\section{Stability of the scalar-field solution}
\label{sec:stability}

As we explained in section~\ref{sec:Scalar-field}, we focus on the solution (\ref{K-G-2-def})
of the scalar-field equation of motion (\ref{K-G-1-def}). We now check that this solution is
stable, for the cosmological background.
We consider a small perturbation $\delta\varphi$ around the solution $\bar\varphi$ of
Eq.(\ref{K-G-2-def}),
\beq
\varphi = \bar\varphi + \delta\varphi, \;\;\;
\tilde\chi = \bar{\tilde\chi} + \delta\tilde\chi ,
\eeq
with, at linear order,
\beq
\bar{\tilde\chi} = \frac{1}{2{\cal M}^4 \tilde{a}^2} \left( \frac{d\bar\varphi}{d\tau} \right)^2 , \;\;\;
\delta\tilde\chi = \frac{1}{{\cal M}^4 \tilde{a}^2} \frac{d\bar\varphi}{d\tau}
\frac{\partial\delta\varphi}{\partial\tau} .
\eeq
Substituting into Eq.(\ref{K-G-1-def}) gives at linear order
\beq
a^3 \sqrt{\bar{\XX}} \left[ \frac{d\bar{\XX}}{d\bar{\lambda}}
\frac{d^2\bar{\lambda}}{d\bar{\XX}^2}
+ 4 \alpha \frac{d\bar{\lambda}}{d\bar{\XX}} \right] \delta\XX =  {\rm constant} ,
\label{stability-eq}
\eeq
where we used $\bar{A} \simeq 1$.

Let us first consider the evolution of $\delta\XX$ at high redshift, in the high-density
regime $\bar\rho \gg \rho_{\alpha}$ and $z \gg z_{\alpha}$.
From Eq.(\ref{lambda-high-density-common}), we obtain
\beq
| \bar{\XX} - \XX_{+} | \sim \bar{\hat\rho}^{\, -1/(\mu_+ -1)}
\sim (a/a_{\alpha})^{3/(\mu_+ -1)} ,
\eeq
which grows with time but remains small in the high-density regime.
Then, Eq.(\ref{stability-eq}) yields
\beq
\delta\XX \propto \frac{ | \bar{\XX} - \XX_+ | }{a^3 \sqrt{\bar{\XX}}} .
\eeq
If $\XX_+ \neq 0$ we have $\sqrt{\bar{\XX}} \simeq \sqrt{\XX_+}$, whereas if
$\XX_+ = 0$ the factor $\sqrt{\bar{\XX}}$ grows with time.
Thus, $\delta\XX$ becomes increasingly small as compared with
$ | \bar{\XX} - \XX_+ |$ as time grows.
Therefore, in the high-density era, $\bar\rho \gg \rho_{\alpha}$,
the perturbation to the solution (\ref{K-G-2-def}) decays and that solution is stable.

At low redshift, in the low-density regime $\bar\rho \ll \rho_{\alpha}$ and
$z \ll z_{\alpha}$, we obtain from Eq.(\ref{lambda-low-density-common})
\beq
| \bar{\XX} - \XX_- | \sim \bar{\hat\rho}^{\, 1/(1-\mu_-)}
\sim (a/a_{\alpha})^{-3/(1-\mu_-)} ,
\label{chi-chi*-a-low}
\eeq
which is small and decreases with time.
Then, Eq.(\ref{stability-eq}) yields
\beq
\delta\XX \propto \frac{ | \bar{\XX} - \XX_- | }{a^3 \sqrt{\bar{\XX}}} .
\eeq
If $\XX_- \neq 0$ we have $\sqrt{\bar{\XX}} \simeq \sqrt{\XX_-}$,
and $\delta\XX$ becomes increasingly small as compared with
$ | \bar{\XX} - \XX_- |$ as time grows.
If $\XX_- = 0$, which can only happen for the concave models, the perturbation
only becomes negligible if $a^3  \sqrt{\bar{\XX}}$ grows with time.
Using Eq.(\ref{chi-chi*-a-low}) this gives the condition
\beq
\mbox{concave models with} \;\; \XX_- = 0 : \;\;\; \mu_- < \frac{1}{2} .
\label{stability-mu-m}
\eeq

Thus, we find that the solution (\ref{K-G-2-def}) of the scalar-field equation of motion
(\ref{K-G-1-def}) is stable at high redshift, $z > z_{\alpha}$.
At low redshift, $z < z_{\alpha}$, it is stable if the low-density limit of $\tilde\chi(\rho)$ is
$\tilde\chi_- > 0$, or if the exponent $\mu_- < 1/2$.
In practice, because the low-redshift era has a finite span until today, and the perturbation
has had time to become negligible during the high-redshift era, we can slightly relax
the condition (\ref{stability-mu-m}). In particular, we can keep the marginal case
$\mu_- = 1/2$ in the model (\ref{mu-m-def-concave}), which corresponds to a perturbation
that remains small and constant in relative terms in the low-redshift era.
In all these cases, the solution (\ref{K-G-2-def}) is physical.

\section{Linear perturbations}
\label{sec:linear}

\subsection{Cosmological background and factor $\epsilon_1$}
\label{sec:background}

\begin{figure}
\begin{center}
\epsfxsize=8. cm \epsfysize=5. cm {\epsfbox{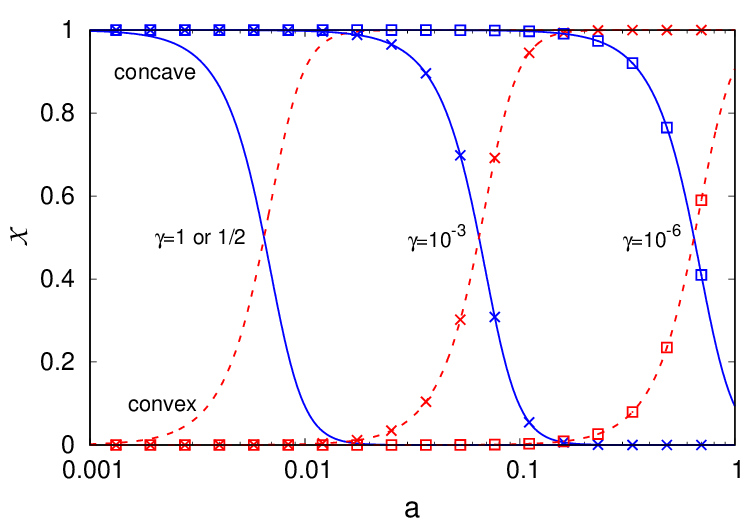}}
\epsfxsize=8. cm \epsfysize=5. cm {\epsfbox{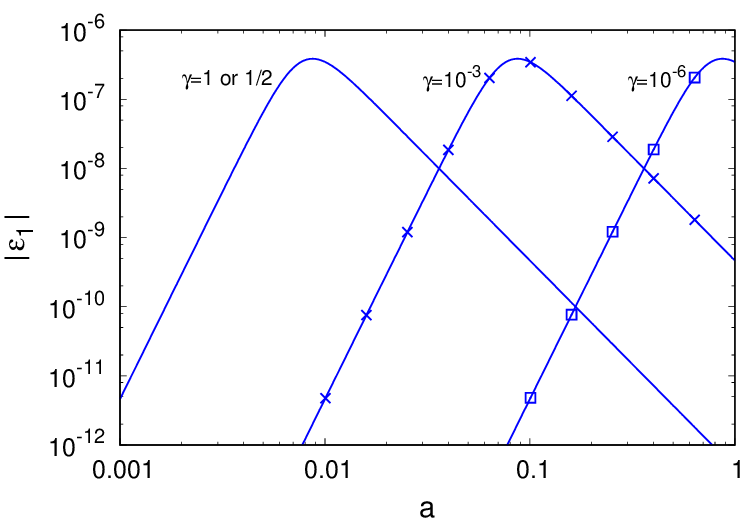}}
\end{center}
\caption{
{\it Upper panel:} normalized background scalar-field kinetic term $\bar{\XX}$ as a function
of the scale factor for the concave (solid lines) and convex (dashed lines) models
(\ref{lambda-concave-explicit}) and (\ref{lambda-convex-explicit}).
We show the results for $\gamma=1, 10^{-3}, 10^{-6}$ for the concave models,
and $\gamma=1/2, 10^{-3}, 10^{-6}$ for the convex models.
{\it Lower panel:} amplitude $|\epsilon_1(a)|$ of the fifth-force factor that enters the evolution
of the linear growing mode, for the models shown in the upper panel.
The concave and convex models have opposite signs but the same amplitude
(exactly for $\gamma= \{ 1,1/2 \}$ and approximately for $\gamma=10^{-3}$ and
$\gamma=10^{-6}$).
}
\label{fig_eps1}
\end{figure}

As $A(\tilde\chi)$ remains very close to unity, and the factor $\epsilon_2$ is very small,
see Eqs.(\ref{dlnA-small}) and (\ref{epsilon2-small}), the Einstein and Jordan frames
are very close and the Jordan-frame Planck mass in Eq.(\ref{Planck-J}) is constant
within an accuracy of $10^{-6}$.
The scalar-field energy density and pressure are almost constant in the dark energy era,
evolving more slowly than the Hubble rate by a factor $| \epsilon_2 | \lesssim 10^{-6}$
from Eq.(\ref{chi-tau-epsilon2}), while they are subdominant at high redshift
(they are actually bounded for the models introduced in
section~\ref{sec:coupling} whereas the matter and radiation densities keep increasing
with redshift).
Therefore, we recover the $\Lambda$-CDM background cosmology within an accuracy
of $10^{-6}$.

For the numerical computations presented in this paper we take
$\alpha = 10^{-6}$, $\gamma = \{ 1, 10^{-3} , 10^{-6} \}$ for the concave model
(\ref{lambda-concave-explicit}), and $\gamma = \{ 1/2, 10^{-3} , 10^{-6} \}$
for the convex model (\ref{lambda-convex-explicit}).
We choose $\gamma=1/2$ instead of $1$ for the convex model because of the condition
(\ref{gamma-lt-1-convex}).

We display in the upper panel in Fig.~\ref{fig_eps1} the normalized background
kinetic term $\bar{\XX}$.
We can check that it shows a transition at the scale factor
$a_{\alpha} = (\alpha/\gamma)^{1/3}$ between the high-density and low-density regimes.
For $\gamma \gg \alpha$, $z_{\alpha} \gg 1$ and both $\bar{\tilde\chi}$ and
$\bar\rho_{\rm de}$ are very close to a constant at low $z$.
For $\gamma=\alpha=10^{-6}$, the transition occurs at $z \sim 1$, but the dark energy
density still remains close to a constant because
$\bar\rho_{\rm de} \simeq {\cal M}^4 (1-\bar{\tilde\chi})$ and
$0 \leq  \bar{\tilde\chi} \leq \gamma \ll 1$.
Therefore, in all cases we recover the $\Lambda$-CDM background cosmology.

We display in the lower panel in Fig.~\ref{fig_eps1} the factor $\epsilon_1(a)$
of Eq.(\ref{eps1-def})
[hence $\epsilon_2(a)$ also as $\epsilon_2=3\epsilon_1$].
It remains very small, with a peak of order $10^{-6}$ at $a_{\alpha}$, in agreement
with Eqs.(\ref{epsilon1-high-z-concave})-(\ref{epsilon1-max-concave}) and
(\ref{epsilon1-high-z-convex})-(\ref{epsilon1-max-convex}).
This again means that $\bar{A}$ remains very close to unity, from the definition
(\ref{H-J-E}).

The curves $|\epsilon_1(a)|$ are the same for the concave and convex models
used in our numerical computations. Indeed, in terms of the rescaled density
$\hat\rho$, the functions $\epsilon_1(\hat\rho)$ of Eqs.(\ref{lambda-rho-concave-explicit})
and (\ref{lambda-rho-convex-explicit}) happen to have exactly the same amplitude,
and opposite signs. Then, comparing Eqs.(\ref{M4-rhode0-concave}) and
(\ref{M4-rhode0-convex}), we find that $\gamma_{\rm concave}=1$ and
$\gamma_{\rm convex}=1/2$ give the same normalized density $\hat\rho$ for
a given physical density $\rho$. For $\gamma=10^{-3}$ and $\gamma=10^{-6}$,
we obtain approximately the same value $\hat\rho$, within an accuracy of
$10^{-3}$ or $10^{-6}$, as $\gamma \ll 1$.
This means that, for a given matter background, whether the cosmological
background or the density profile of virialized halos, the three pairs of concave and convex
models computed in this paper have the same amplitude for the fifth force
but opposite signs.

From Eqs.(\ref{DL}) and (\ref{eps-def}), the factor $\epsilon_1$ measures the deviation
of the linear growing mode from the $\Lambda$-CDM prediction.
In all cases, it decays at high redshifts, which means that we recover the standard
$\Lambda$-CDM cosmology in the early universe and we can use the same initial conditions
for the matter power spectrum, on cosmological scales.
This is because at early time the matter density is very high, $\bar\rho \gg \rho_{\alpha}$,
and the field $\XX$ and the conformal factor $A$ converge to their boundary values
$\XX_+$ and $A_+$. Then, because a constant coupling $A$ gives back General Relativity,
as the two metrics in Eq.(\ref{g-Jordan-def}) become identical up to a constant factor,
there is no fifth force and the gravitational instability proceeds in the standard fashion.
This will be clearly seen in the next sections where we study the linear growing mode
$D_+(k,a)$.

\subsection{Concave models}
\label{sec:linear-concave}

\begin{figure}
\begin{center}
\epsfxsize=8. cm \epsfysize=5.5 cm {\epsfbox{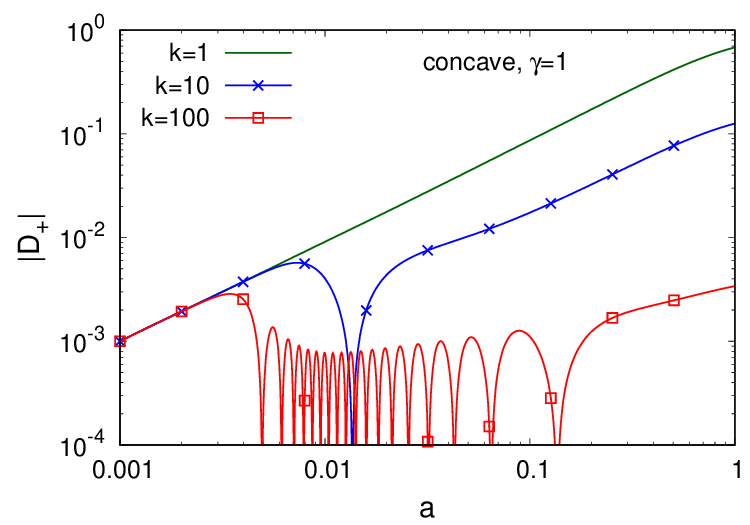}}\\
\epsfxsize=8. cm \epsfysize=5.5 cm {\epsfbox{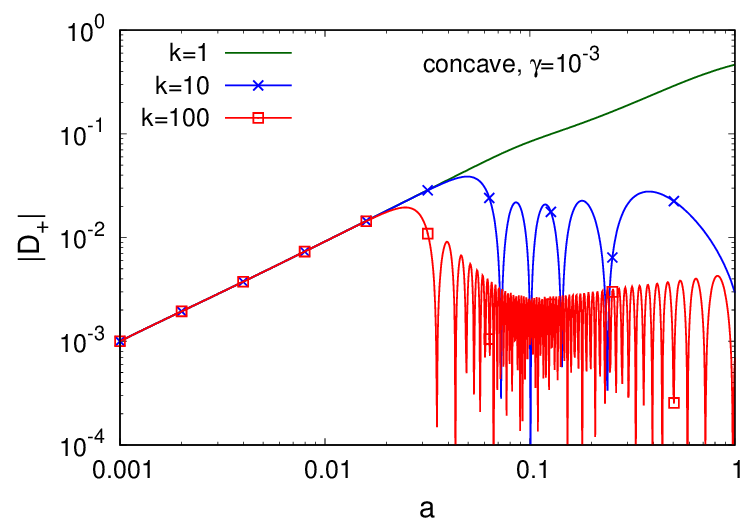}}\\
\epsfxsize=8. cm \epsfysize=5.5 cm {\epsfbox{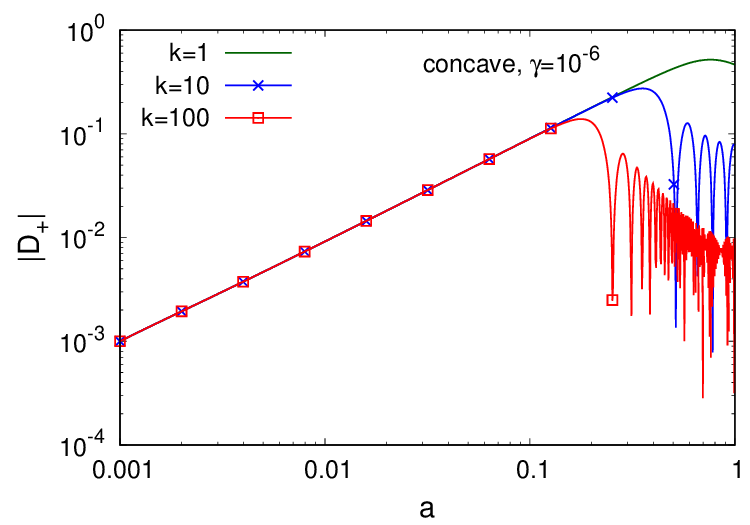}}
\end{center}
\caption{
Linear growing mode $D_+(k,a)$ as a function of the scale factor
for $k=1, 10$ and $100 h\rm{Mpc}^{-1}$, from top to bottom.
The different panels are the concave models with $\gamma=1, 10^{-3}$, and $10^{-6}$,
from top to bottom.
}
\label{fig_Dp_concave}
\end{figure}

\begin{figure}
\begin{center}
\epsfxsize=8. cm \epsfysize=6 cm {\epsfbox{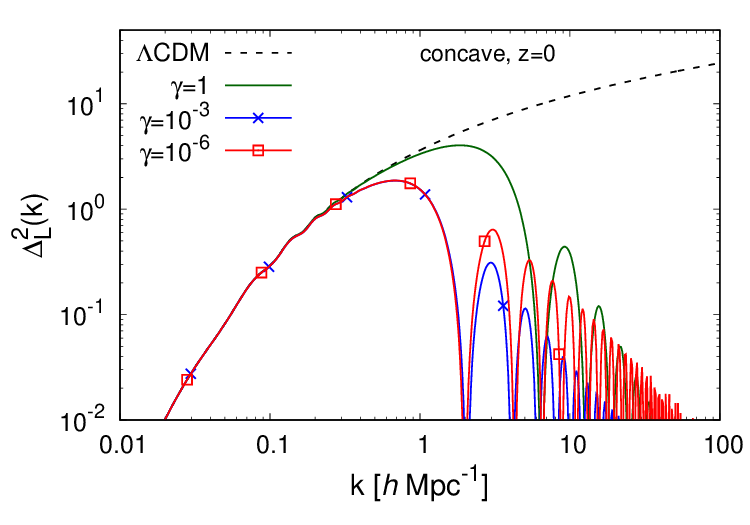}}
\end{center}
\caption{
Logarithmic linear power spectrum $\Delta^2_L(k,a) = 4 \pi k^3 P_L(k,a)$
for the concave models at redshift $z=0$.
}
\label{fig_DeltaL_k_concave}
\end{figure}

We first consider the case of concave coupling functions, introduced in
section~\ref{sec:concave}. This leads to a negative $\epsilon_1$ and the fifth force
decreases the Newtonian gravity.
The linear modes $D_{\pm}(k,a)$ of the matter density contrast obey the evolution equation
(\ref{DL}), where the departure from the $\Lambda$-CDM cosmology only comes from the
factor $\epsilon(k,a)$.
Because $|\epsilon_1| \lesssim \alpha \ll 1$, the factor $1$ in Eq.(\ref{eps-def}) gives a
negligible contribution to $(1+\epsilon)$ and we can write
\beq
\epsilon(k,a) = \epsilon_1(a) \frac{2}{3\Omega_{\rm m}} \left( \frac{ck}{aH} \right)^2 .
\label{eps-k-a}
\eeq
On Hubble scales we have $\epsilon \sim \epsilon_1$, hence $|\epsilon| \lesssim \alpha \ll 1$
and we recover the $\Lambda$-CDM growth of structures.
On smaller scales $|\epsilon(k,a)|$ grows as $k^2$ and it reaches unity at a wave number
\beq
k_{\alpha}(a) \simeq \frac{aH}{c\sqrt{|\epsilon_1|}}
\simeq \frac{3\times 10^{-4}}{\sqrt{a|\epsilon_1|}} h \, \rm{Mpc}^{-1} ,
\label{k-alpha-def}
\eeq
where we used $H^2 \propto a^{-3}$ in the matter era.
From Eqs.(\ref{epsilon1-high-z-concave})-(\ref{epsilon1-low-z-concave})
we have
\beq
a \ll a_{\alpha} : \;\;\; k_{\alpha}(a) \simeq \frac{H_0}{c} \alpha^{-2/3} \gamma^{1/6}
\left( \frac{a}{a_{\alpha}} \right)^{\! -(4 \mu_+ -1)/[2(\mu_+-1)]} ,
\label{kalpha-a-low-a}
\eeq
\beq
a \gg a_{\alpha} : \;\;\; k_{\alpha}(a) \simeq \frac{H_0}{c} \alpha^{-2/3} \gamma^{1/6}
\left( \frac{a}{a_{\alpha}} \right)^{\! (4 \mu_- -1)/[2(1-\mu_-)]} .
\label{kalpha-a-high-a}
\eeq
Thus, $k_{\alpha}(a)$ decreases with time in the high-redshift era $a<a_{\alpha}$,
which means that increasingly large comoving scales enter the regime where
the fifth force dominates over Newtonian gravity and stops the growth of matter
perturbations.
In the low-redshift era, $a>a_{\alpha}$, $k_{\alpha}(a)$ increases with time
if $\mu_->1/4$ and decreases if $\mu_-<1/4$. Thus, we find that the minimum value
of $k_{\alpha}$ until today is given by
\beq
\mu_- > \frac{1}{4} : \;\;\; k_{\alpha}^{\min} = k_{\alpha}(a_{\alpha})
= \frac{H_0}{c} \alpha^{-2/3} \gamma^{1/6} ,
\label{kalpha-min-mum-high}
\eeq
\beq
\mu_- < \frac{1}{4} : \;\;\; k_{\alpha}^{\min} = k_{\alpha}(1)
= \frac{H_0}{c} \alpha^{-1/[2(1-\mu_-)]} \gamma^{\mu_- /[2(1-\mu_-)]} .
\label{kalpha-min-mum-low}
\eeq
Wave numbers below $k_{\alpha}^{\min}$, i.e. comoving scales greater than
$1/k_{\alpha}^{\min}$, have never entered the fifth-force regime yet.
Higher wave numbers have entered the fifth-force regime at the scale factor
$a_-(k)$ given by
\beq
a_-(k) = \alpha^{1/(4\mu_+ -1)} \gamma^{-\mu_+/(4\mu_+ -1)}
\left( \frac{ck}{H_0} \right)^{\! -2(\mu_+ -1)/(4\mu_+ -1)} ,
\label{am-k-def}
\eeq
if $\mu_- >1/4$ or $k>k_{\alpha}(a_{\alpha})$,
and by
\beq
a_-(k) = \alpha^{-1/(1-4\mu_-)} \gamma^{\mu_-/(1 - 4\mu_-)}
 \left( \frac{ck}{H_0} \right)^{\! -2(1-\mu_-)/(1-4\mu_-)} ,
\label{am-k-def-2}
\eeq
if $\mu_- <1/4$ and $k_{\alpha}(1) < k < k_{\alpha}(a_{\alpha})$.
If $\mu_->1/4$ they have left the fifth-force regime at the scale factor
$a_+(k)$ given by (if $a_+ < 1$)
\beqa
\mu_- > \frac{1}{4} : \;\;
a_+(k) & = & \alpha^{1/(4\mu_- -1)} \gamma^{-\mu_-/(4\mu_- -1)} \nonumber \\
&& \times \left( \frac{ck}{H_0} \right)^{2(1-\mu_-)/(4\mu_- -1)} , \;\;\;
\label{ap-k-def-mum-high}
\eeqa
while if $\mu_- < 1/4$ they have never left the fifth-force regime yet,
$a_+(k) > 1$.
In the time interval $[a_-,a_+]$, the factor $(1+\epsilon)$ in the linear evolution equation
(\ref{DL}) is dominated by $\epsilon$ and becomes negative.
This means that the density fluctuations no longer feel an attractive gravity but a pressure-like
force. Then, the linear growing mode $D_+(k,a)$ stops growing and develops
an oscillatory behavior. As the fifth-force regime starts earlier and ends later
for higher wave numbers, $D_+(k,a)$ is more strongly suppressed for higher wave numbers
as compared with the $\Lambda$-CDM growing mode. At low wave numbers,
$k < k_{\alpha}^{\min}$, the fifth force was never relevant and we recover the
$\Lambda$-CDM prediction.
We refer to \cite{Brax:2016vpd} for a more detailed analysis, for the case $\gamma=1$.

This agrees with the results shown in Fig.~\ref{fig_Dp_concave}.
For $k \ll 10 h {\rm Mpc}^{-1}$ we recover the
$\Lambda$-CDM steady growth of matter perturbations, while for
$k \gg 10 h {\rm Mpc}^{-1}$  we have an oscillatory phase around $a_{\alpha}$.
For smaller $\gamma$, the transition scale factor $a_{\alpha}$ increases and the
oscillatory phase occurs at lower redshifts.

These behaviors lead to a sharp falloff in the logarithmic linear power spectrum
$\Delta_L(k)$ at high $k$, shown in Fig.~\ref{fig_DeltaL_k_concave}.
The dependence of the power spectrum at $z=0$ on $\gamma$ is not simple,
because of the interplay between the mean redshift and the duration
of the oscillatory phase.
We find that $\gamma = 10^{-3}$ and $10^{-6}$ give similar cutoffs, which appear
at somewhat larger scales than for $\gamma=1$.
The larger-scale cutoff for $\gamma = 10^{-3}$ as compared with $\gamma=1$ agrees
with Eq.(\ref{kalpha-min-mum-high}).
The cutoff does not decrease further for $\gamma=10^{-6}$ because the fifth-force
factor $|\epsilon_1|$ has just peaked and the Hubble rate in Eq.(\ref{k-alpha-def})
is somewhat higher in the dark-energy era than predicted by the matter-era scaling
that we used here.

\subsection{Convex models}
\label{sec:linear-convex}

\begin{figure}
\begin{center}
\epsfxsize=8. cm \epsfysize=5.5 cm {\epsfbox{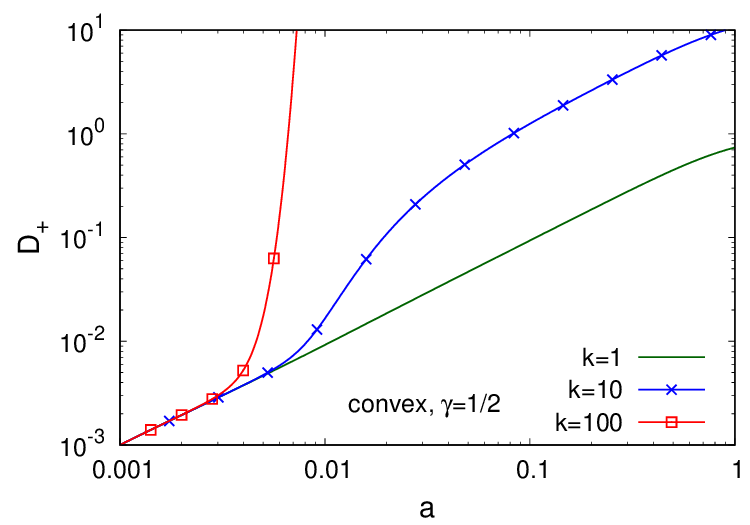}}\\
\epsfxsize=8. cm \epsfysize=5.5 cm {\epsfbox{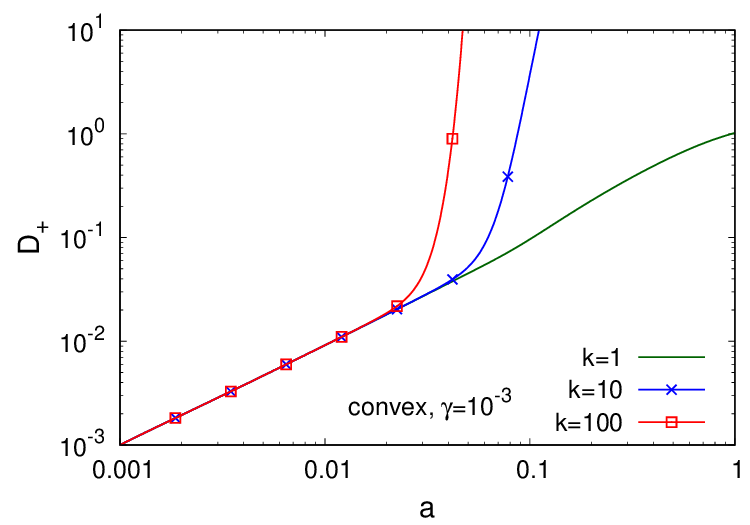}}\\
\epsfxsize=8. cm \epsfysize=5.5 cm {\epsfbox{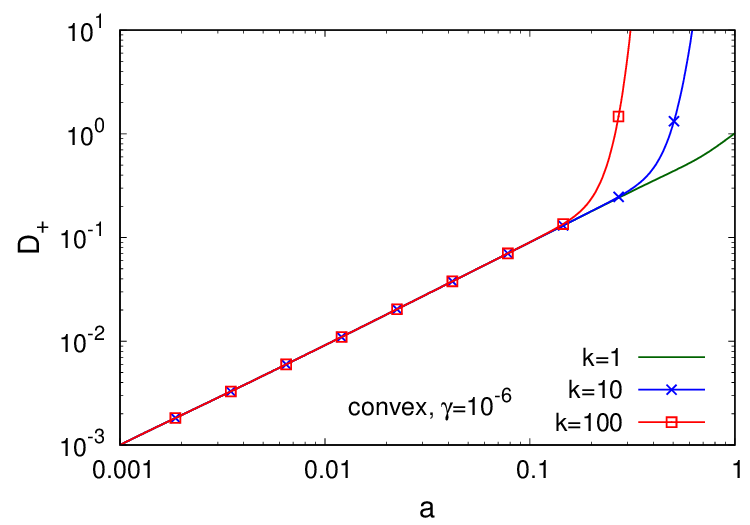}}
\end{center}
\caption{
Linear growing mode $D_+(k,a)$ as a function of the scale factor
for $k=1, 10$ and $100 h\rm{Mpc}^{-1}$, from bottom to top.
The different panels are the convex models with $\gamma=1/2, 10^{-3}$, and $10^{-6}$,
from top to bottom.
}
\label{fig_Dp_convex}
\end{figure}

\begin{figure}
\begin{center}
\epsfxsize=8. cm \epsfysize=6 cm {\epsfbox{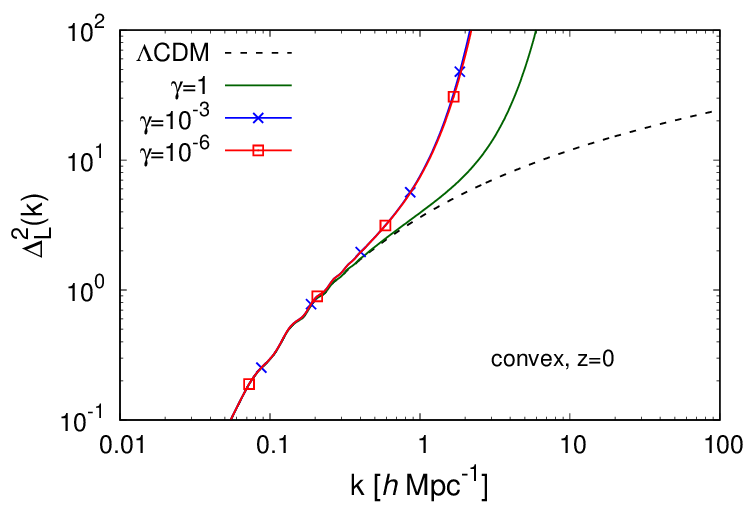}}
\end{center}
\caption{
Logarithmic linear power spectrum $\Delta^2_L(k,a)= 4 \pi k^3 P_L(k,a)$
for the convex models at redshift $z=0$.
}
\label{fig_DeltaL_k_convex}
\end{figure}

We now consider the case of convex coupling functions, introduced in
section~\ref{sec:convex}. This leads to a positive $\epsilon_1$ and the fifth force
increases the Newtonian gravity.
We can again use the approximation (\ref{eps-k-a}) for $\epsilon(k,a)$,
define the transition wave number $k_{\alpha}(a)$ to the fifth-force regime as in
Eq.(\ref{k-alpha-def}), with the asymptotic behaviors
(\ref{kalpha-a-low-a})-(\ref{kalpha-a-high-a}), and the minimum values
(\ref{kalpha-min-mum-high})-(\ref{kalpha-min-mum-low}).
For a given comoving wave number $k$, the fifth-force regime extends over the time
range $[a_-,a_+]$ given by Eqs.(\ref{am-k-def})-(\ref{ap-k-def-mum-high}).

In the time interval $[a_-,a_+]$, the factor $(1+\epsilon)$ in the linear evolution equation
(\ref{DL}) is dominated by $\epsilon$ and becomes large and positive.
This means that the linear growing mode $D_+(k,a)$ grows much faster than in the
$\Lambda$-CDM scenario, and increasingly so for higher $k$ because of the
$k^2$ dependence of $\epsilon(k,a)$ in Eq.(\ref{eps-k-a}).
Moreover, this fifth-force regime starts earlier and ends later
for higher wave numbers.
Therefore, $D_+(k,a)$ is more strongly amplified for higher wave numbers.
At low wave numbers, $k < k_{\alpha}^{\min}$, the fifth force was never relevant
and we again recover the $\Lambda$-CDM prediction.
We refer to \cite{Brax:2016vpd} for a more detailed analysis, for the case $\gamma=1$.

This agrees with the results shown in Fig.~\ref{fig_Dp_convex}.
For $k \ll 10 h {\rm Mpc}^{-1}$ we recover the slow
$\Lambda$-CDM growth of matter perturbations, while for $k \gg 10 h {\rm Mpc}^{-1}$
we have a very fast amplification phase around $a_{\alpha}$.
This leads to a sharp increase in the logarithmic linear power spectrum $\Delta_L(k)$
at high $k$, shown in Fig.~\ref{fig_DeltaL_k_convex}.
In a fashion similar to the concave case, $\gamma = 10^{-3}$ and $10^{-6}$ give similar
departures from the $\Lambda$-CDM power, with now a strong amplification that appears
at somewhat larger scales than for $\gamma=1$.

\section{Spherical collapse}
\label{sec:Spherical-collapse}

\subsection{Equation of motion}
\label{sec:eq-spherical}

As can be seen from Eq.\eqref{Euler-J}, on large scales where the
baryonic pressure is negligible the particle trajectories $\vr(t)$ are given by
\beq
\frac{d^2 \vr}{d t^2}  - \frac{1}{a} \frac{d^2 a}{d t^2} \vr = - \nabla_{\vr} \left( \Psi_{\rm N}
+ \Psi_A \right) ,
\label{trajectory-Jordan}
\eeq
where $\vr=a\vx$ is the physical coordinate, $\nabla_{\vr}=\nabla/a$ the gradient
operator in physical coordinates, and $\Psi_A= c^2 \ln A$ is the fifth force contribution
to the metric potential $\Phi$ in Eq.(\ref{Phi-J-Psi-J-PsiN}).
As in \cite{Brax:2016vpd}, to study the spherical collapse before shell crossing
it is convenient to label each shell by its Lagrangian radius $q$ or enclosed mass $M$,
and to introduce its normalized radius $y(t)$ by
\beq
y(t) = \frac{r(t)}{a(t) q} \;\;\; \mbox{with} \;\;\;
q = \left(\frac{3M}{4\pi\bar\rho_0}\right)^{1/3} , \;\;\; y(t=0) = 1 .
\label{y-def-Jordan}
\eeq
In particular, the matter density contrast within radius $r(t)$ reads as
$1+ \delta_{<}(r) = y(t)^{-3}$.
Then, the equation of motion of the normalized radius $y$ reads as
\cite{Brax:2016vpd}
\beqa
&& \frac{d^2 y}{d(\ln a)^2} + \left( 2+\frac{1}{H^2} \frac{d H}{d t} \right)
\frac{d y}{d\ln a} + \frac{\Omega_{\rm m}}{2} y (y^{-3} - 1) = \nonumber \\
&& - y \left( \frac{c}{Hr} \right)^2 \frac{d\ln A}{d\ln\rho}
\frac{r}{1+\delta} \frac{\partial\delta}{\partial r} .
\label{y-lna-1}
\eeqa
In contrast with the $\Lambda$-CDM case, where the dynamics of different shells are
decoupled before shell crossing, the fifth force introduces a coupling as it depends on the
density profile, through the local density contrast $\delta(r)$ and its first derivative
$\partial\delta/\partial r$.
As in \cite{Brax:2016vpd}, to bypass this difficulty we decouple the motion of the shell
of mass $M$ of interest by assuming a constant shape for the density profile,
which is then fully parameterized by the density contrast $\delta_<$ within the mass shell
$M$ [more precisely, we choose the typical profile associated with a Gaussian field
of power spectrum $P_L(k)$].

\subsection{Concave models}
\label{sec:spherical-concave}

\begin{figure}
\begin{center}
\epsfxsize=8. cm \epsfysize=5.5 cm {\epsfbox{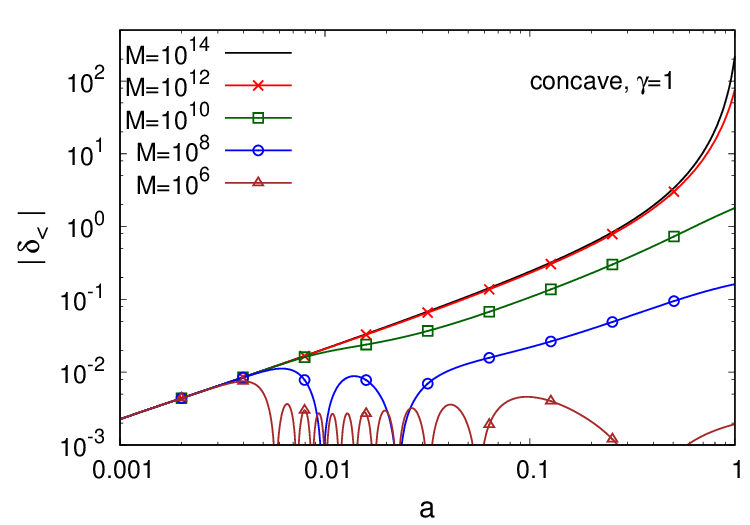}}\\
\epsfxsize=8. cm \epsfysize=5.5 cm {\epsfbox{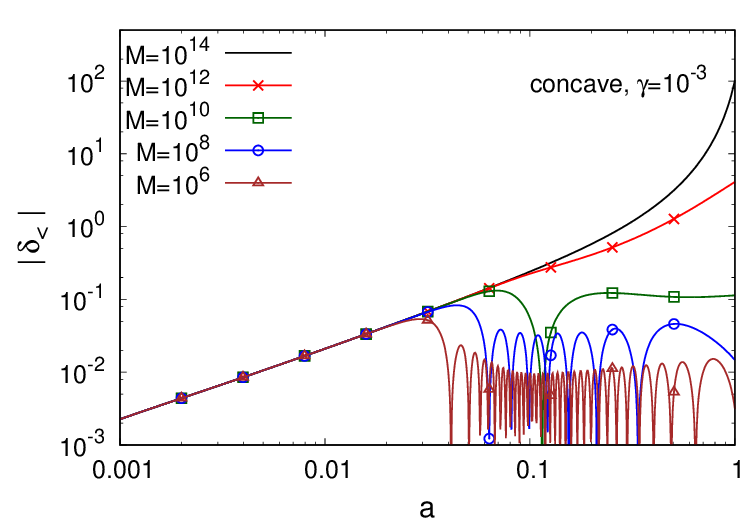}}\\
\epsfxsize=8. cm \epsfysize=5.5 cm {\epsfbox{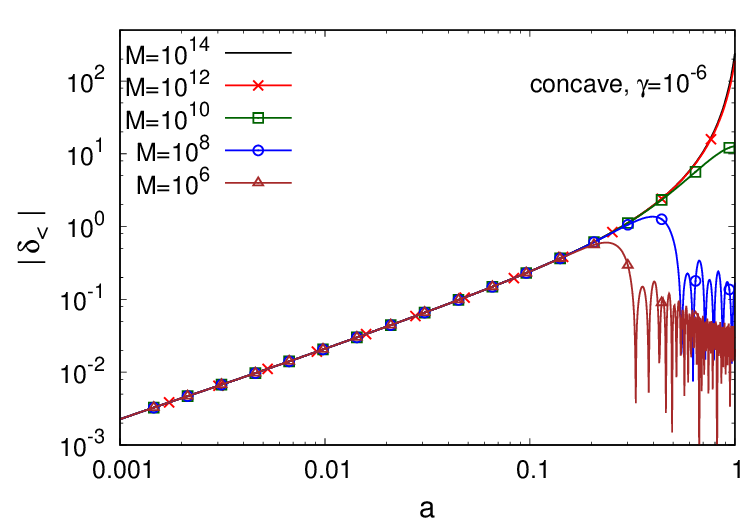}}\end{center}
\caption{
Evolution with time of the nonlinear density contrast $\delta_{<}(a)$
for several masses, from $M=10^{14}$ to $10^6 h^{-1} M_{\odot}$ from top to bottom,
with the same initial condition that corresponds to the $\Lambda$-CM linear density
threshold today $\delta_{<L}^{\Lambda \rm -CDM}=1.6$.
The different panels are the concave models with $\gamma=1, 10^{-3}$, and $10^{-6}$,
from top to bottom.
}
\label{fig_delta_concave}
\end{figure}

\begin{figure}
\begin{center}
\epsfxsize=8. cm \epsfysize=5.5 cm {\epsfbox{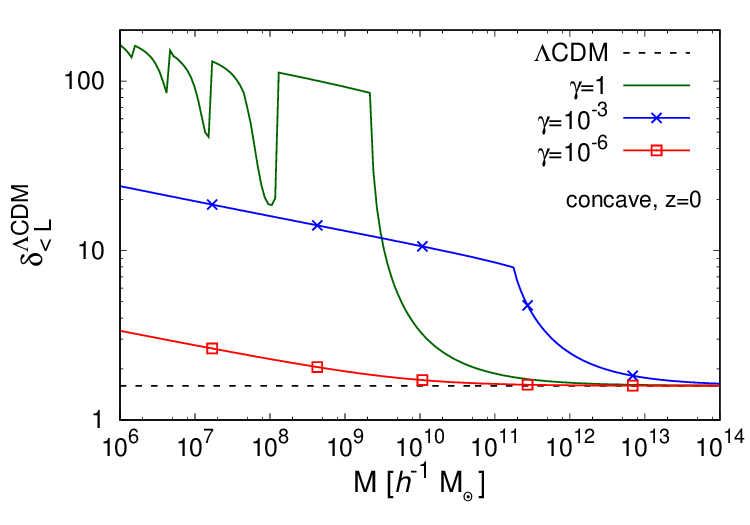}}
\end{center}
\caption{
Initial linear density contrast, as measured by
$\delta_{<L}^{\Lambda \rm -CDM}$, that gives rise to a nonlinear density contrast
$\delta_{<}=200$ at $z=0$, as a function of the halo mass $M$.
}
\label{fig_delta_threshold_concave}
\end{figure}

We show in Fig.~\ref{fig_delta_concave} the evolution of the matter
density contrast $\delta_<$ within a spherical shell $M$, given by the spherical collapse
dynamics. We consider several mass scales $M$. All curves follow the same behavior
at very high redshift, before the fifth-force becomes important, as we choose the same
initial conditions, which corresponds to a linear density contrast today of
$\delta_{<L}^{\Lambda \rm -CDM}=1.6$ in the $\Lambda$-CDM cosmology.
In agreement with the results found in section~\ref{sec:linear-concave}, large masses,
associated with large scales, follow the $\Lambda$-CDM behavior, whereas small masses
show an oscillatory phase around $a_{\alpha}$.
This significantly delays the collapse of small scales.
Moreover, because of the change of signs due to the oscillations, an initial overdensity
may turn underdense at the end of the oscillatory phase, which means that it will never
collapse after the fifth force becomes negligible (within the spherical no-shell-crossing
approximation).
We refer to \cite{Brax:2016vpd} for a more detailed analysis of the case $\gamma=1$,
where we also study the spherical dynamics for different values of the initial density
contrast.

We show in Fig.~\ref{fig_delta_threshold_concave} the linear density contrast threshold,
measured by $\delta_{<L}^{\Lambda \rm -CDM}$ (i.e., the extrapolation up to $z=0$
of the initial density contrast by the $\Lambda$-CDM growth rate), required to reach
a nonlinear density contrast $\delta_{<}=200$ today.
Again, at large mass we recover the $\Lambda$-CDM linear density threshold,
$\delta_{<L}^{\Lambda \rm -CDM} \simeq 1.6$, whereas at small mass we obtain
a much greater linear density threshold $\delta_{<L}^{\Lambda \rm -CDM} \sim 100$.
This is required to compensate the slower growth at low mass found
in Fig.~\ref{fig_delta_concave}.

For $\gamma=1$, we find a non-monotonic curve, which is due to the oscillation phase and
the complex behavior found in Fig.~\ref{fig_delta_concave}.
Indeed, as explained above, some overdensities become underdensities at the end
of the oscillation phase and never collapse.
This implies that such initial conditions cannot contribute to the curve
$\delta_{<L}^{\Lambda \rm -CDM}$ of the linear density threshold, and that only initial
conditions associated with an even number of oscillations are relevant.
These jumps from one even number to the next, skipping the odd number in-between,
lead to the jumps found for $\gamma=1$, see also the discussion in
Ref.~\cite{Brax:2016vpd}.
For $\gamma=10^{-3}$ and $10^{-6}$ these jumps or steps are beyond the mass range
shown in the figure.

In any case, the formation of low mass halos is strongly suppressed as compared with
the $\Lambda$-CDM scenario.
In fact, rather than forming in the usual bottom-up hierarchical fashion of CDM models,
low-mass halos may form later in a top-down fashion, by fragmentation of larger-mass halos,
as in Warm Dark Matter (WDM) scenarios.
For $\gamma=10^{-6}$ the oscillatory phase only takes place at low redshift.
Then, for relevant cosmological masses, $M \gtrsim 10^6 h^{-1} M_{\odot}$,
the threshold $\delta_{<L}^{\Lambda \rm -CDM}$ is not so much higher than in the
$\Lambda$-CDM cosmology, as it is sufficient that the collapse was already
sufficiently advanced at $z \sim 1$.
As compared with the linear power spectrum shown in Fig.~\ref{fig_DeltaL_k_concave},
where the cases $\gamma=10^{-3}$ and $10^{-6}$ gave similar results,
they now give very different results for $\delta_{<L}^{\Lambda \rm -CDM}$.
Moreoever, at $M \gtrsim 10^{10} h^{-1} M_{\odot}$,
the curve obtained for $\gamma=1$ falls in between the results obtained
for $\gamma=10^{-3}$ and $\gamma=10^{-6}$.
This is due to the complex behavior of the fifth force, associated with these oscillatory
phases and their different redshifts, that lead to different orderings between the models
depending on the measured quantity.

\subsection{Convex models}
\label{sec:spherical-convex}

\begin{figure}
\begin{center}
\epsfxsize=8. cm \epsfysize=5.5 cm {\epsfbox{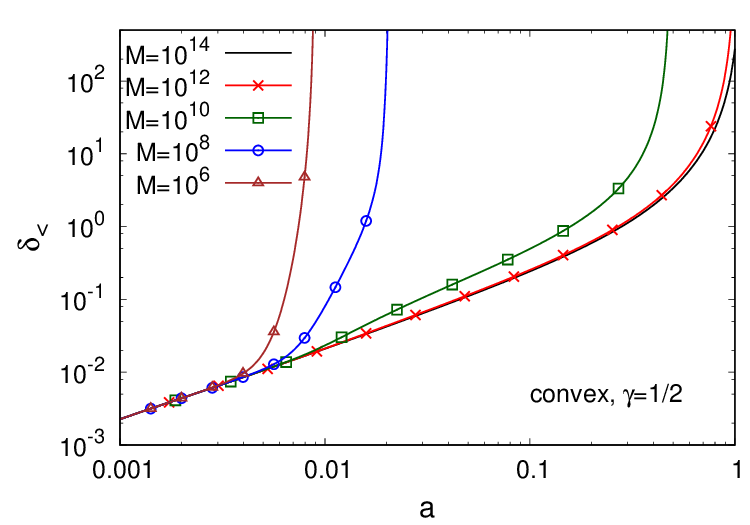}}\\
\epsfxsize=8. cm \epsfysize=5.5 cm {\epsfbox{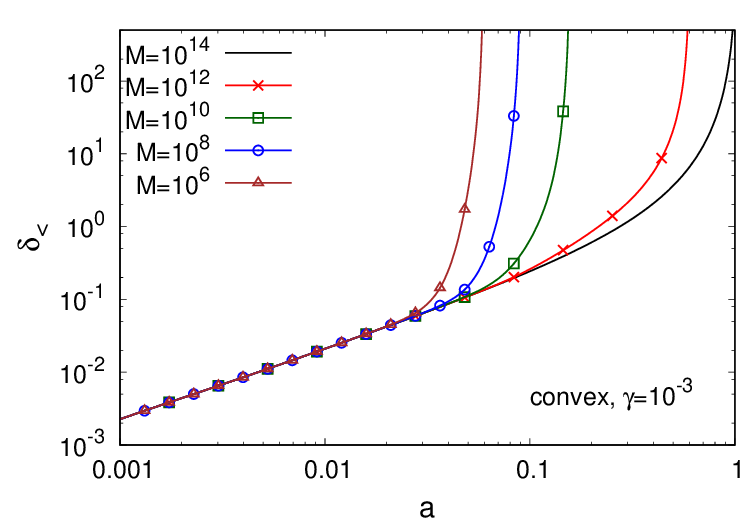}}\\
\epsfxsize=8. cm \epsfysize=5.5 cm {\epsfbox{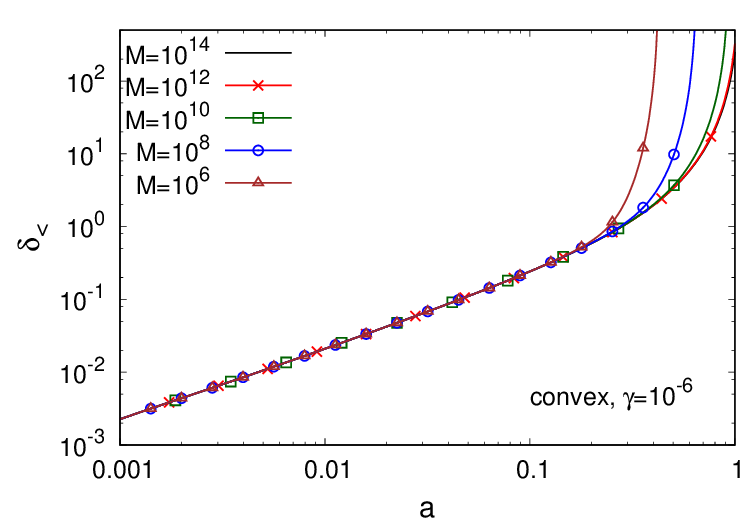}}\end{center}
\caption{
Evolution with time of the nonlinear density contrast $\delta_{<}(a)$
for several masses, from $M=10^{14}$ to $10^6 h^{-1} M_{\odot}$ from bottom to top,
with the same initial condition that corresponds to the $\Lambda$-CM linear density
threshold today $\delta_{<L}^{\Lambda \rm -CDM}=1.6$.
The different panels are the convex models with $\gamma=1/2, 10^{-3}$, and $10^{-6}$,
from top to bottom.
}
\label{fig_delta_convex}
\end{figure}

\begin{figure}
\begin{center}
\epsfxsize=8. cm \epsfysize=5.5 cm {\epsfbox{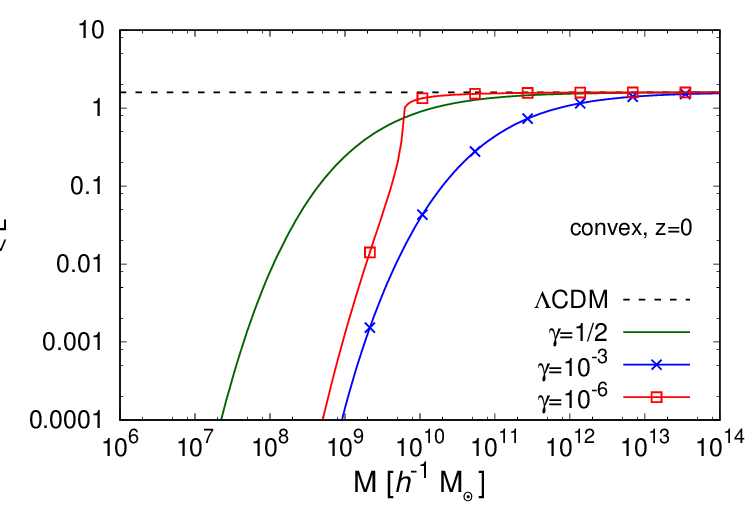}}
\end{center}
\caption{
Initial linear density contrast, as measured by
$\delta_{<L}^{\Lambda \rm -CDM}$, that gives rise to a nonlinear density contrast
$\delta_{<}=200$ at $z=0$, as a function of the halo mass $M$.
}
\label{fig_delta_threshold_convex}
\end{figure}

We show in Fig.~\ref{fig_delta_convex} the evolution of the matter
density contrast $\delta_<$ given by the spherical collapse dynamics for the convex
models, as was done in Fig.~\ref{fig_delta_concave} for the concave models.
In agreement with the results found in section~\ref{sec:linear-convex}, large masses
again follow the $\Lambda$-CDM behavior, whereas small masses
show a very fast amplification phase around $a_{\alpha}$.
This significantly accelerates the collapse of small scales.

The initial condition, measured by the linear density contrast threshold
$\delta_{<L}^{\Lambda \rm -CDM}$, required to reach a nonlinear density
contrast $\delta_{<}=200$ today, is shown in Fig.~\ref{fig_delta_threshold_convex}.

At large mass we recover the $\Lambda$-CDM linear density threshold, whereas at small
mass we obtain a much smaller linear density threshold.
This is required to compensate the faster growth found at low mass in
Fig.~\ref{fig_delta_convex}.
Again, the dependence on $\gamma$ is not simple and for
$M \gtrsim 10^{10} h^{-1} M_{\odot}$, the curve obtained for $\gamma=1$ falls in between
the results obtained for $\gamma=10^{-3}$ and $\gamma=10^{-6}$.
As compared with the linear power spectrum shown in Fig.~\ref{fig_DeltaL_k_convex},
where the cases $\gamma=10^{-3}$ and $10^{-6}$ gave similar results,
this again shows that different measured quantities can lead to a different ordering
between the models.

\subsection{Halo mass function}
\label{sec:mass-function}

\begin{figure}
\begin{center}
\epsfxsize=8. cm \epsfysize=5.5 cm {\epsfbox{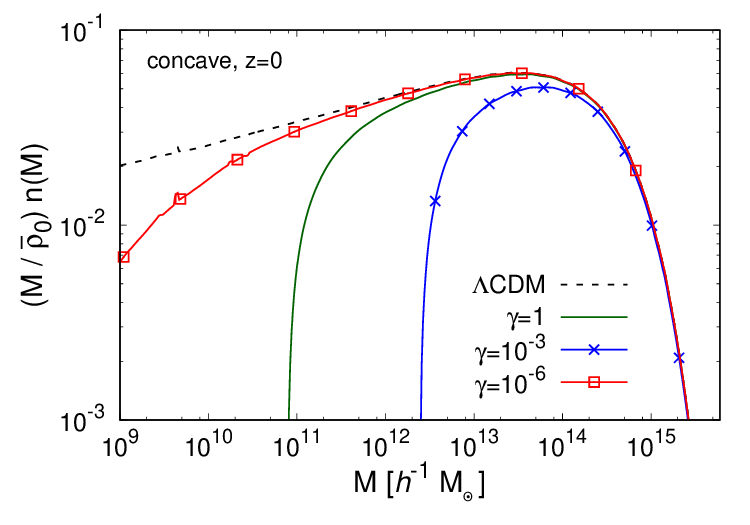}}\\
\epsfxsize=8. cm \epsfysize=5.5 cm {\epsfbox{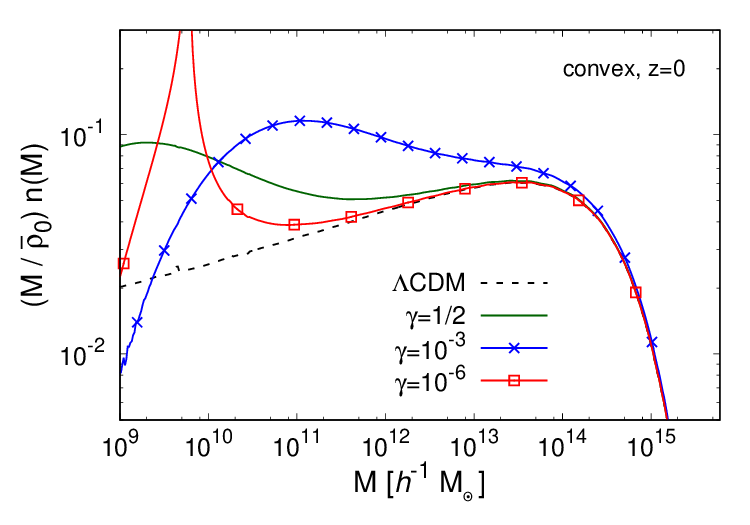}}
\end{center}
\caption{
Halo mass function at $z=0$ for the concave and convex models, and for the
$\Lambda$-CDM reference.
}
\label{fig_nM}
\end{figure}

Following a Press-Schechter approach \cite{Press1974}, we write the comoving halo mass
function as
\beq
n(M) \frac{d M}{M} = \frac{\bar\rho_0}{M} f(\nu) \frac{d\nu}{\nu} ,
\label{nM-def}
\eeq
where the scaling variable $\nu(M)$ is defined as
\beq
\nu(M) = \frac{\delta_{<L}^{\Lambda \rm CDM}(M)}{\sigma(M)} ,
\label{nu-def}
\eeq
and $\delta_{<L}^{\Lambda \rm CDM}(M)$ is again the initial linear density contrast
(extrapolated up to $z=0$ by the $\Lambda$-CDM linear growth factor) that is required
to build a collapsed halo (which we define here by a nonlinear density contrast of 200 with
respect to the mean density of the Universe).
The variable $\nu$ measures whether such an initial condition corresponds to a  rare and
very high overdensity in the initial Gaussian field ($\nu \gg 1$) or to a typical fluctuation
($\nu \lesssim 1$).
In the Press-Schechter approach, we have $f(\nu) = \sqrt{2/\pi} \nu e^{-\nu^2/2}$
\cite{Press1974}. Here we use the same function as in \cite{Valageas2009}.
Then, the impact of the modified gravity only arises through the linear threshold
$\delta_{<L}^{\Lambda \rm CDM}(M)$, as we assume the same initial matter density
power spectrum as for the $\Lambda$-CDM reference at high redshift.
As noticed in Fig.~\ref {fig_eps1} and sections \ref{sec:linear-concave} to
\ref{sec:spherical-convex}, this is legitimate because at high redshifts the fifth
force vanishes and we recover the standard $\Lambda$-CDM cosmology (because the 
conformal coupling essentially converges to a constant).

We display the halo mass function in Fig.~\ref{fig_nM}.
For both concave and convex models we recover the $\Lambda$-CDM
large-mass tail, in agreement with the convergence to the $\Lambda$-CDM
threshold in Figs.~\ref{fig_delta_threshold_concave} and \ref{fig_delta_threshold_convex}.
At low mass, the halo mass function is smaller/greater for the concave/convex models
than the $\Lambda$-CDM reference, in agreement with the higher/lower thresholds
in Figs.~\ref{fig_delta_threshold_concave} and \ref{fig_delta_threshold_convex}.
Again, the behavior is not monotonic with the parameter $\gamma$, and from
Fig.~\ref{fig_nM} we can conclude that $\gamma \sim 10^{-3}$ is probably excluded
by observations, as it yields large deviations for masses $10^{12}-10^{13} h^{-1} M_{\odot}$
that correspond to massive objects (large galaxies or small groups), whose properties
should not suffer too much from poorly constrained baryonic effects.

For the concave model, the mass function defined by Eq.(\ref{nM-def}) actually becomes
negative below $10^{11} h^{-1} M_{\odot}$ for $\gamma=1$,
and at a higher or lower mass for $\gamma=10^{-3}$ and $10^{-6}$.
This is due to the fast increase of $\delta_{<L}^{\Lambda \rm CDM}(M)$ at low mass,
which makes $\nu$ increase at lower mass (in this range), instead of decreasing,
and gives a change of sign for $d\nu/d M$.
Of course, this means that the prediction (\ref{nM-def}) is not physical at low mass.
This is because at low mass structure formation no longer follows the standard
hierarchical picture.
Because of the damping of density perturbations on small scales, found in
Figs.~\ref{fig_Dp_concave} and \ref{fig_delta_concave}, small scales do not
collapse first. Then, smaller-mass objects would form (if they do) in a top-down scenario,
where larger masses, $\gtrsim 10^{11} h^{-1} M_{\odot}$, collapse first and next
fragment into smaller substructures, as in a hot dark matter scenario.
In contrast, for the convex model the hierarchical pattern of structure formation
is even stronger than in the $\Lambda$-CDM cosmology, because the growth
of matter perturbations is increasingly fast on smaller scales.
The spike at $5\times 10^9 h^{-1} M_{\odot}$ for the convex model with
$\gamma=10^{-6}$ is due to the sudden change of $\delta_{<L}^{\Lambda \rm CDM}(M)$
at this mass scale found in Fig.~\ref{fig_delta_threshold_convex}.
In practice, we do not expect the Press-Schechter picture, based on the spherical collapse,
to provide a reliable prediction of such sharp features, and this peak is probably
smoothened by perturbations from the spherical collapse and mergers. However,
we can expect a greater abundance of halos around this mass scale than for the
$\Lambda$-CDM cosmology.

The spherical collapse dynamics and the Press-Schechter approach described above
assume smooth density fields without strong effects from the nonlinearities of the
scalar field sector on smaller scales (i.e., substructures).
In the concave case, where the fifth force damps the formation of small-scale structures,
this is legitimate and should be an even better approximation than for the
$\Lambda$-CDM cosmology. In the convex case, where the formation of small-scale structures
is enhanced, a catastrophic amplification of small-scale fluctuations could lead
to strongly inhomogeneous systems where the analysis described above breaks down.
We shall investigate this issue in section~\ref{sec:nonlinear-analysis-convex} below.
Using a thermodynamical approach, we find that such strong inhomogeneities do not
develop during the early cosmological evolution (because the temperature or velocity
dispersion is sufficient to stabilize the system) and the standard spherical analysis
described above should apply.

\section{Newtonian and fifth-force regimes}
\label{sec:newtonian}

\subsection{Clusters and galaxies}
\label{sec:clusters}

We study in this section how the fifth force compares to Newtonian gravity on the scales
of large cosmological structures such as galaxies and clusters.
As in \cite{Brax:2016vpd}, we can associate to the Newtonian force the circular
velocity $v_{\rm N}^2$ at radius $r$,
\beq
F_{\rm N} = - \frac{v_{\rm N}^2(r)}{r} , \;\;\; v_{\rm N}^2 =   \frac{{\cal G}_{\rm N} M(<r)}{r} ,
\label{FN-v2}
\eeq
and to the fifth force a characteristic velocity $c_s^2$ with
\beq
F_A = \pm \frac{c_s^2}{r} , \;\;\; c_s^2 = c^2
\left | \frac{d\ln A}{d\ln\rho} \frac{d\ln\rho}{d\ln r} \right | ,
\label{FA-cs2}
\eeq
so that the force ratio $\eta$ reads as
\beq
| \eta | = \left | \frac{F_A}{F_{\rm N}} \right | = \frac{c_s^2}{v_{\rm N}^2} .
\label{eta-cs-vN}
\eeq
To keep the ratio $\eta$ below unity in typical astrophysical and cosmological structures,
we need $c_s^2 \lesssim v_{\rm N}^2 \lesssim 10^{-6} c^2$. This implies again that
the coupling function in Eq.(\ref{FA-cs2}) must be small, $| \delta \ln A | \lesssim 10^{-6}$,
and we recover the constraint (\ref{dlnA-small}).

Let us now consider typical cosmological structures, such as clusters and galaxies,
in the low-redshift Universe at $z \simeq 0$.
In the low-density regime, we have from Eq.(\ref{lambda-low-density-common})
\beq
\rho \ll \rho_{\alpha} : \;\; c_s^2 \sim c^2 \alpha \hat\rho^{\, \mu_-/(1-\mu_-)} ,
\label{cs-low-density}
\eeq
and
\beq
z \simeq 0 : \;\;\; v_{\rm N}^2 \sim (H_0 R)^2 \gamma \hat\rho / \alpha ,
\eeq
for objects of rescaled density $\hat\rho$ and radius $R$.
This gives
\beq
\eta \sim \left( \frac{\alpha c}{\sqrt{\gamma} H_0 R} \right)^2 \hat\rho^{\, -(1-2\mu_-)/(1-\mu_-)} .
\label{eta-cluster-mum}
\eeq
For the particular case $\mu_-=1/2$, this yields
\beq
\mu_- = \frac{1}{2} : \;\;\;
\eta \sim \frac{1}{\gamma} \, \left( \frac{\alpha}{10^{-6}} \right)^{\! 2} \,
\left( \frac{3 h^{-1} {\rm kpc}}{R} \right)^{\! 2} .
\label{eta-cluster}
\eeq
Thus, at low $z$ for $\alpha=10^{-6}$ and $\gamma=1$
 the fifth force is negligible on extragalactic scales, such as cluster sizes
($1 h^{-1} {\rm Mpc}$) and beyond, while it becomes of order of the Newtonian force
on galactic scales, around $3 h^{-1} {\rm kpc}$.
For $\gamma=10^{-6}$, Eq.(\ref{eta-cluster}) would suggest that
the fifth force would be important on Mpc scales, associated with clusters,
but this is not the case because for such a value of $\gamma$ virialized structures
with a density contrast greater than 200 are no longer in the low-density regime
(\ref{cs-low-density}). We give a more detailed analysis in
Fig.~\ref{fig_eta_z0} below.

The scale dependence $R^{-2}$, which amplifies the fifth force on small scales,
is related to the $k^2$ dependence in the factor $\epsilon(k,a)$ in
Eq.(\ref{eps-k-a}), which governs the evolution of linear cosmological matter
perturbations.
This is due to the fact that the fifth-force potential depends on the density,
$\rho \propto \nabla^2 \Psi_{\rm N}$, instead of the gravitational potential $\Psi_{\rm N}$.

\subsection{Spherical halos}
\label{sec:spherical-halos}

\begin{figure}
\begin{center}
\epsfxsize=8. cm \epsfysize=5.5 cm {\epsfbox{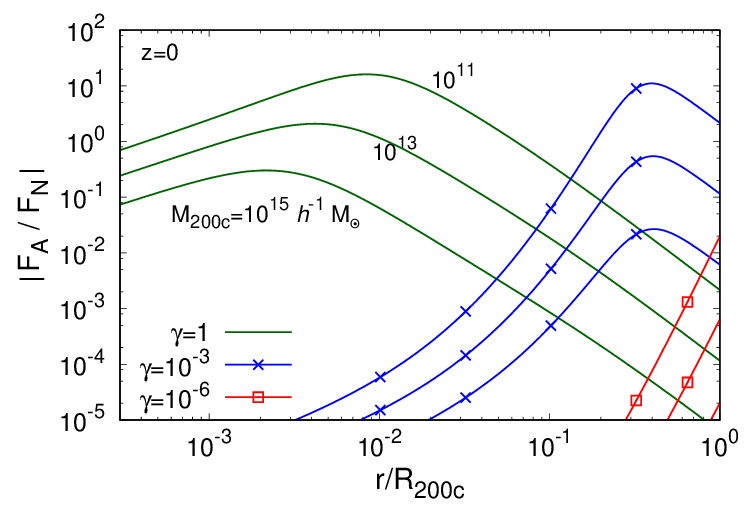}}
\end{center}
\caption{
Absolute value of the ratio $\eta=F_A/F_{\rm N}$, as a function of the radius $r$,
within spherical halos.
We display the halo masses $M_{\rm 200c}= 10^{15}$, $10^{13}$ and
$10^{11} h^{-1} M_{\odot}$, from bottom to top, at $z=0$.
The concave and convex models have the same amplitude $|\eta|$ but opposite signs,
for $\gamma=\{1,10^{-3},10^{-6}\}$ and $\gamma=\{1/2,10^{-3},10^{-6}\}$.
}
\label{fig_eta_z0}
\end{figure}

Let us now study the behavior of the fifth force inside spherical halos.
For simplicity we consider power-law density profiles,
\beq
\rho(r) \sim \rho_{\alpha} \left( \frac{r}{r_{\alpha}} \right)^{-\gamma_p} .
\label{gamma-def}
\eeq
For a Navarro-Frenk-White (NFW) density profile \cite{Navarro:1996}, we have
$\gamma_p=1$ in the inner parts of the halos.
In the high-density regime, we have from Eq.(\ref{lambda-high-density-common})
\beq
\rho \gg \rho_{\alpha} : \;\; c_s^2 \sim c^2 \alpha \hat\rho^{\, -\mu_+ /(\mu_+ -1)} ,
\eeq
while $v_{\rm N}^2 \propto \rho r^2$.
Then, we find that in the power-law density profile (\ref{gamma-def}) the ratio
$\eta$ behaves as
\beq
r \ll r_{\alpha} : \;\;\; \eta \propto r^{-2+\gamma_p (2\mu_+ -1)/(\mu_+ -1)} .
\label{eta-gamma}
\eeq
In particular, the fifth force becomes negligible at the center of the halo if we have
\beq
\eta \to 0 \;\; \mbox{for} \;\; r \to 0 \;\; \mbox{if} \;\;
\gamma_p > 1 - \frac{1}{2\mu_+ -1} .
\eeq
Thus, the fifth force always becomes negligible inside NFW halos, for any exponent
$\mu_+$.
This screening is due to the fact that at high densities the scalar-field kinetic term
goes to the constant $\tilde\chi_+$, for the models introduced in
section~\ref{sec:coupling}, which means that the coupling function also goes to
the constant $A_+=A(\tilde\chi_+)$.
Because a constant coupling function does not generate any fifth force
(which is proportional to $\nabla \ln A$), this leads to a suppression of the fifth force
at high densities. Within a halo with a radial density profile, the efficiency of this
damping depends on the growth rate of the density at smaller radii.
Thus, $\eta$ goes faster to zero at small radii in Eq.(\ref{eta-gamma}) for a larger
slope $\gamma_p$.

We show the amplitude of the force ratio $\eta$ in Fig.~\ref{fig_eta_z0}.
As for $\epsilon_1(a)$ shown in Fig.~\ref{fig_eps1}, the concave
and convex models with $\gamma=\{1,10^{-3},10^{-6}\}$ and
$\gamma=\{1/2,10^{-3},10^{-6}\}$ give the same amplitude and opposite signs
for $\eta(r)$ within a given halo.
This is because the functions $d\lambda/d\ln\rho$ of
Eqs.(\ref{lambda-rho-concave-explicit}) and (\ref{lambda-rho-convex-explicit})
have the same amplitude and opposite signs while the normalized densities
$\hat\rho$ are exactly or almost equal.

At large radii, where $\rho \ll \rho_{\alpha}$, we have the $r^{-2}$ relative
decrease of the fifth force as in Eq.(\ref{eta-cluster}), whereas at small radii,
we have the decrease given by Eq.(\ref{eta-gamma}).
At the transition between the small-radii and large-radii regimes, we have
$\hat\rho \simeq 1$ and $r \simeq r_{\alpha}$. From Eq.(\ref{eta-cluster-mum})
we obtain
\beq
r_{\alpha} \propto (\gamma/\alpha)^{-1/\gamma_p} \;\;\; \mbox{and} \;\;\;
\eta(r_{\alpha}) \propto \alpha^{2-2/\gamma_p} \gamma^{2/\gamma_p-1} ,
\label{r-apha-eta-alpha}
\eeq
where we assumed that the power-law (\ref{gamma-def}) holds up to the
halo radius $R_{200 \rm c}$.
For $\gamma_p=2$, which corresponds to the typical slope at intermediate radii,
this yields $\eta(r_{\alpha}) \propto \alpha$.
This agrees with Fig.~\ref{fig_eta_z0}, where we find that the amplitude
of the peak of $\eta$, at the transition between the small- and large-radii regimes,
does not depend much on $\gamma$, while its location $r_{\alpha}$ grows
roughly as $1/\sqrt{\gamma}$ for smaller $\gamma$.

In agreement with the analysis of Eq.(\ref{eta-cluster}) above, the fifth force is
negligible for massive halos such as clusters of galaxies, but becomes important
for smaller galactic halos.
In particular, we find that the fifth force becomes significantly greater than the
Newtonian gravity at intermediate radii for halos of mass
$\lesssim 10^{11} h^{-1} M_{\odot}$.
This suggests that these models, with $\gamma \gtrsim 10^{-4}$, are actually ruled out
by observations, which are consistent with Newtonian gravity on galactic scales.
This constraint disappears for models with $\gamma \lesssim 10^{-4}$, because
the peak of $|\eta|$ at $r_{\alpha}$ does not exist as it would be pushed outside of
the halo in the figure. This is because virialized halos have a large overdensity,
of order $200/\Omega_{\rm m}$ at the virial radius, so that they entirely fall in the
high-density regime of the coupling function $\ln A$ (and even more so at higher redshift).
Then, the peak associated with the transition between the high-density and low-density
regimes is never reached.

For the convex models with $\gamma \gtrsim 10^{-4}$, where the fifth force amplifies
the formation of structures,
this problem may be circumvented by nonlinearities. Indeed, the fifth force
could lead to the fragmentation of the system into smaller and denser substructures,
where the fifth force would next become negligible because of the screening
mechanism described in section~\ref{sec:screening} below, due to the locality
of the fifth force.
This would not change the spherically-averaged properties of the halo but
inhomogeneities would be significant and would invalidate the mean-field
analysis used in Eq.(\ref{eta-gamma}) and Fig.~\ref{fig_eta_z0}, because of the
local and nonlinear character of the fifth force, which is not self-averaging.

For the concave models, this nonlinear process is unlikely to solve the problem found
in Fig.~\ref{fig_eta_z0} at low mass for $\gamma \gtrsim 10^{-4}$.
Indeed, because the fifth force is now repulsive and
acts like a standard polytropic pressure, instead of promoting the fragmentation of the
system and the building of inhomogeneous nonlinear structures, it prevents
the collapse and tends to smooth density profiles. Therefore, a mean-field analysis,
where we neglect substructures, should be a good approximation, and we can expect
the problem found in Fig.~\ref{fig_eta_z0} to be real.
A related issue is that the large repulsive fifth force found in Fig.~\ref{fig_eta_z0}
suggests that halos of mass $\lesssim 10^{11} h^{-1} M_{\odot}$
and moderate densities cannot form in the first place, as the collapse is stopped by
the fifth force.
This means that concave models with $\gamma \gtrsim 10^{-4}$ are probably ruled out
by observations.

\subsection{Fifth-force dominated regime}
\label{sec:fifth-force-regime}

We now study more generally which regions in the planes $(\rho,R)$ or $(M,R)$,
for matter structures of global density $\rho$, radius $R$, and mass $M$, are dominated
by the fifth force.
Writing $M=4\pi\rho R^3/3$, we obtain from Eq.(\ref{eta-cs-vN})
\beq
| \eta | \sim \frac{2}{\Omega_{\rm m 0}} \frac{\bar\rho_0}{\rho}
\left( \frac{c}{R H_0} \right)^{\!2} \left| \frac{d\ln A}{d\ln\rho} \right| .
\label{eta-R-rho}
\eeq
Then, the fifth force is greater than Newtonian gravity if we have
\beq
|\eta| \geq 1 : \;\;\; R^2 \leq \left( \frac{c}{H_0} \right)^{\!2}  \frac{2}{\Omega_{\rm m 0}}
\frac{\bar\rho_0}{\rho} \left| \frac{d\ln A}{d\ln\rho} \right| .
\label{R-rho}
\eeq
Although for convenience we write the right-hand side in terms of the cosmological quantities
$H_0$, $\bar\rho_0$ and $\Omega_{\rm m0}$ at $z=0$, this expression does not depend
on redshift nor on cosmology. Moreover, it is only a function of the density $\rho$.
Then, in a density-radius plane $(\rho,R)$, the domain where $|\eta|\geq 1$ is given by the
area under the curve $R_{\eta}(\rho)$, where $R_\eta(\rho)$ is the density-dependent
radius defined by the right-hand side in Eq.(\ref{R-rho}).

\begin{figure}
\begin{center}
\epsfxsize=9 cm \epsfysize=7. cm {\epsfbox{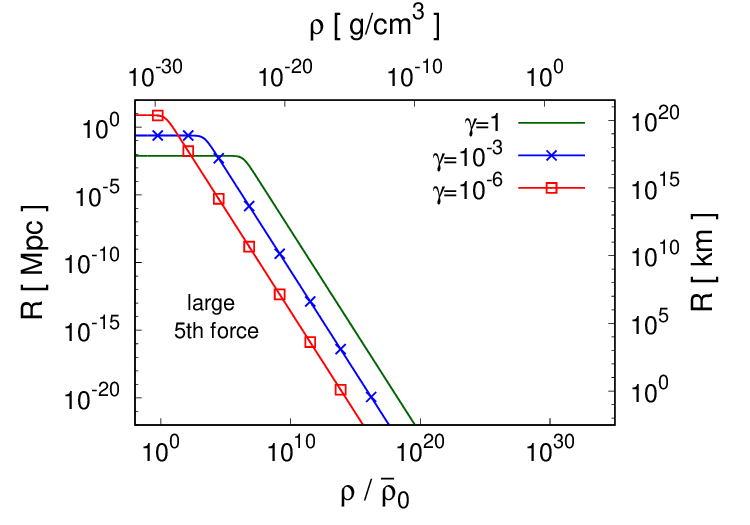}}
\end{center}
\caption{
Domain in the density-radius plane where the fifth force is greater than Newtonian gravity.
This domain is identical for the concave and convex models
with $\gamma=\{1,10^{-3},10^{-6}\}$ and $\gamma=\{1/2,10^{-3},10^{-6}\}$.
The horizontal axis is the typical density of the structure, $\rho$, given in units of the mean
matter cosmological density today, $\bar\rho_0$, in the bottom-border scale, and in units
of ${\rm g . cm}^{-3}$ in the top-border scale.
The vertical axis is the typical radius of the structure, $R$, given in Mpc in the left-border
scale and in km in the right-border scale.
}
\label{fig_eta_R_rho}
\end{figure}

At low densities, we obtain from Eq.(\ref{lambda-low-density-common})
\beq
\rho \ll \rho_{\alpha} : \;\;\; R_{\eta}(\rho) \sim R_{\alpha}
\left( \frac{\rho}{\rho_{\alpha}} \right)^{(2\mu_- -1)/(2-2\mu_-)} ,
\label{Reta-low-density}
\eeq
where we defined
\beq
R_{\alpha} = \frac{\alpha c}{\sqrt{\gamma}H_0} , \;\;\;
R_{\alpha} \sim 10 \,{\rm kpc} \;\; \mbox{for} \;\; \alpha = 10^{-6} , \; \gamma = 1 .
\label{Ralpha-def}
\eeq
Thus, if $\mu_- = 1/2$, at low densities we obtain a constant radius threshold, of order
$R_{\alpha}$.
If $\mu_- < 1/2$ the threshold $R_{\eta}(\rho)$ increases at lower densities.
At high densities, we obtain from Eq.(\ref{lambda-high-density-common})
\beq
\rho \gg \rho_{\alpha} : \;\;\; R_{\eta}(\rho) \sim R_{\alpha}
\left( \frac{\rho}{\rho_{\alpha}} \right)^{-(2\mu_+ -1)/(2\mu_+ -2)} .
\label{Reta-high-density}
\eeq
Thus, the threshold $R_{\eta}(\rho)$ decreases at high densities.
As analysed for the behavior (\ref{eta-gamma}) of the fifth force inside spherical halos,
this decrease of the fifth-force regime at high densities is due to the convergence
to a finite value $A_+=A(\tilde\chi_+)$ of the coupling function, as a constant coupling
function $A$ does not generate a fifth force.
This analysis agrees with the results displayed in Fig.~\ref{fig_eta_R_rho}.
The transition radius and density scale as $R_{\alpha} \propto \alpha/\sqrt{\gamma}$
and $\rho_{\alpha} \propto \gamma/\alpha$.

\begin{figure}
\begin{center}
\epsfxsize=9 cm \epsfysize=7. cm {\epsfbox{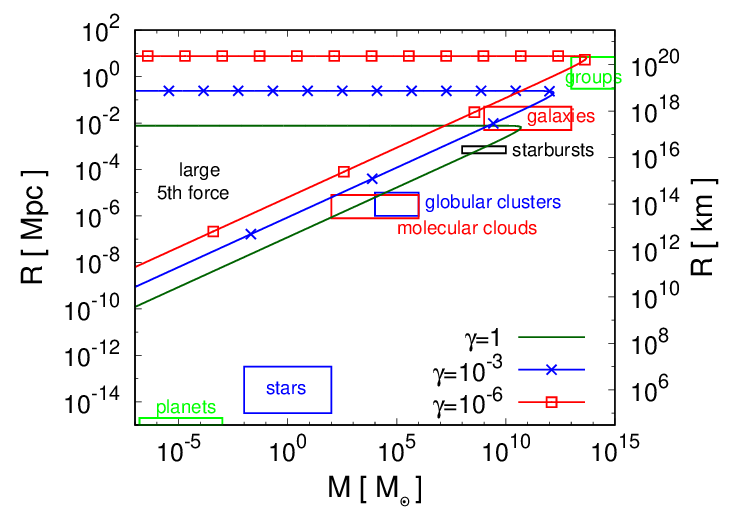}}
\end{center}
\caption{
Domain in the mass-radius plane where the fifth force is greater than Newtonian gravity.
This domain is identical for the concave and convex models
with $\gamma=\{1,10^{-3},10^{-6}\}$ and $\gamma=\{1/2,10^{-3},10^{-6}\}$.
The horizontal axis is the typical mass of the structure, $M$, given in units of the solar mass.
The vertical axis is the typical radius of the structure, $R$, given in Mpc in the left-border
scale and in km in the right-border scale.
The rectangles show the typical scales of various astrophysical structures.
}
\label{fig_eta_R_M}
\end{figure}

To simplify the comparison with astrophysical structures, it is convenient to display
the fifth-force domain (\ref{R-rho}) in the mass-radius plane $(M,R)$.
The low-density branch (\ref{Reta-low-density}) writes as
\beq
\rho \ll \rho_{\alpha} : \;\;\; R_{\eta}(M) \sim R_{\alpha}
\left( \frac{M}{M_{\alpha}} \right)^{-(1-2\mu_-)/(4\mu_- -1)} ,
\label{RetaM-low-density}
\eeq
where we defined
\beq
M_{\alpha} = \rho_{\alpha} R_{\alpha}^3 , \;\;\;
M_{\alpha} \simeq \left( \frac{\alpha}{10^{-6}} \right)^2 \gamma^{-1/2} \;
10^{10} M_{\odot} ,
\label{Malpha-def}
\eeq
while the high-density branch (\ref{Reta-high-density}) reads as
\beq
\rho \gg \rho_{\alpha} : \;\;\; R_{\eta}(M) \sim R_{\alpha}
\left( \frac{M}{M_{\alpha}} \right)^{(2\mu_+ -1)/(4\mu_+ -1)} .
\label{RetaM-high-density}
\eeq
If $\mu_->1/4$, the fifth-force domain is a triangle in the $(\ln M,\ln R)$
plane that extends down to $M \to 0$ and has a right corner at the maximum mass
$M_{\alpha}$. Then, objects more massive than $M_{\alpha}$ are not dominated
by the fifth force, whatever their size.
We display this diagram in Fig.~\ref{fig_eta_R_M}, along with typical cosmological
and astrophysical structures.

In agreement with section~\ref{sec:clusters}, the fifth force is negligible for clusters and
groups (at their global scale), while it is of the same order as Newtonian gravity for galaxies
if $\gamma \gtrsim 10^{-4}$.
It is interesting to note that various galactic structures, from the molecular clouds and
extended starburst regions, to the overall extent of low-mass galaxies, as well as the small
old globular clusters, all lie close to the boundary of the fifth-force region if
$\gamma \gtrsim 10^{-3}$.
Thus, in agreement with Fig.~\ref{fig_eta_z0},
these objects may provide strong constraints on the models considered in this paper.

\subsection{Screening on astrophysical scales and in the laboratory}
\label{sec:screening}

\subsubsection{Screening mechanism}
\label{sec:screening-mechanism}

Alternative theories to General Relativity are strongly constrained,
or even ruled out, by Solar System tests, based on the trajectories of planets around the Sun
(measurements by the Cassini satellite  \cite{Bertotti:2003rm})
or the motion of the Moon around the Earth
(Lunar Laser Ranging experiment \cite{Williams:2004qba}).
To remain consistent with these data, modified-gravity scenarios often involve nonlinear
screening mechanisms that ensure convergence to General Relativity in small-scale
and high-density environments (typically by suppressing the gradients of the scalar field
or its coupling to matter).
In our case, as for the ultralocal models described in \cite{Brax:2016vpd} that obey
the same equations of motion, if we consider stars, planets and moons as isolated objects
in the vacuum, the screening is provided by the definition of the model itself and is
$100\%$ efficient.
Indeed, because the fifth force is exactly local, as
${\bf F}_A = -c^2 \nabla \ln A(\rho)$ only depends on the local density and its gradient,
the impact of the Sun onto the motion of the Earth through the fifth force is exactly zero.
Therefore, the Sun is completely ``screened'' as viewed from the Earth by the fifth force,
as well as all planets and moons of the Solar System.
Then, the trajectories of astrophysical objects in the Solar System are exactly given
by the usual Newtonian gravity, or more accurately General Relativity,
and all Solar Systems tests of gravity are satisfied, to the same accuracy as
General Relativity.

This screening mechanism, which is the one of the ultralocal models \cite{Brax:2016vpd},
is different from the screening mechanisms of usual conformal coupling
scenarios, leading to Yukawa interactions with matter.
The four usual mechanisms are the chameleon mechanism
\cite{Khoury:2003aq,Khoury:2003rn,Brax:2004qh},
where the field is almost constant and short-ranged in high-density environments
because its mass grows with the matter density,
the Damour-Polyakov mechanism \cite{Damour:1994zq},
where the coupling to matter vanishes in high-density environments,
the K-mouflage \cite{Babichev:2009ee,Brax:2014b} and
Vainshtein \cite{Vainshtein:1972sx} mechanisms, where gradients of the scalar field
are suppressed by the nonlinearities of the kinetic term (i.e., the Lagrangian
includes terms of higher order than quadratic over the field derivatives,
that involve $\partial \varphi$ or $\Box\varphi$).
In contrast, the ultralocal screening does not rely on the suppression of the scalar field
gradients, but on their local character. Because of the lack of propagation, the
fifth force is essentially short-ranged and insensitive to distant masses.
This may be seen as an extreme limit of the chameleon mechanism.

In the case of the models with kinetic conformal coupling that we study in this paper,
the scalar-field Lagrangian includes a kinetic term and the equation of motion
(\ref{K-G-1-def}) includes derivative terms and looks like a modified Klein-Gordon equation.
However, it admits the ultralocal non-propagating solution (\ref{K-G-2-def}),
which recovers the ultralocal screening mechanism.

\subsubsection{Fifth-force pressure}
\label{sec:pressure}

Even though the fifth force on the Earth is not significantly influenced by the Sun and
other planets, it does not vanish as it is sensitive to the local gradient of the matter density.
Then, we must check that this local force is small enough to have avoided detection
in the laboratory or on the Earth (e.g., at its surface or in the atmosphere).
As seen in Eq.(\ref{Euler-1-J}), the local nature of the scalar field
(more precisely of its kinetic term $\tilde\chi$) makes the fifth
force appear as a polytropic pressure $p_A(\rho)$, given by Eq.(\ref{pA-def}), where
$\rho$ is now the baryonic matter density as the dark matter density and its gradient
can be neglected.
Since ${\cal M}^4 \sim \bar\rho_{\rm de 0}$, and $\tilde\chi \leq \gamma$ for the models
considered in this paper, we obtain for a typical density of 1 g/cm$^{3}$,
\beq
\rho \sim 1 \; {\rm g . cm}^{-3} : \;\;\;
\frac{p_A}{\rho} \sim 3 \times 10^{-13} \, \gamma \; ({\rm m/s})^2 .
\label{pA-rho}
\eeq
This corresponds to very small velocities and motions, for any $\gamma \leq 1$.
To compare this pressure with the thermal motions found on the Earth or in the laboratory,
we write Eq.(\ref{pA-rho}) as a temperature,
\beq
\frac{m_p p_A}{\rho k_B} \sim 3 \times 10^{-17} \, \gamma \; {\rm K} ,
\label{p-T}
\eeq
where again we chose $\rho \sim 1 \; {\rm g/cm}^{3}$, $m_p$ is the proton mass and
$k_B$ the Boltzmann constant.
This gives a very low temperature that is much smaller than the temperature reached by
cold-atoms experiments in the laboratory, $T \sim 10^{-7}$ K.
Thus, the fifth force can be neglected in the laboratory and on the Earth,
and in other astrophysical objects.

\section{Nonlinear analysis of the convex models}
\label{sec:nonlinear-analysis-convex}

As we noticed in section~\ref{sec:spherical-halos}, within a mean-field
(i.e., spherically averaged) analysis the fifth force can be greater than Newtonian
gravity at intermediate radii in low-mass halos, $M \lesssim 10^{11} h^{-1} M_{\odot}$,
if $\gamma \gtrsim 10^{-4}$.
As in \cite{Brax:2016vpd}, we pointed out that for the convex models this problem may be
circumvented by nonlinearities, which could lead to a fragmentation of the system
and a screening of the fifth force, due to its locality, as described in
section~\ref{sec:screening}.
On the other hand, these nonlinearities may also invalidate the analysis of cosmological
perturbations presented in section~\ref{sec:linear}. Indeed, this linear analysis
assumes that the fifth-force gradients on large cosmological scales are set by the
large-scale density gradients. This may no longer hold if strong small-scale inhomogeneities
develop, which make the fifth force independent of large-scale density gradients.

Following \cite{Brax:2016vpd}, in this section we first obtain the scale associated
with the cosmological nonlinear transition. Then, to go beyond perturbation theory
and spherically-symmetric approximations, we use a thermodynamical approach
to obtain the phase diagram associated with the fifth force.
Next, we compare this phase diagram with the trajectories associated with the
cosmological nonlinear transition, and with the density profiles inside halos.

\subsection{Evolution of the cosmological nonlinear transition for the convex model}
\label{sec:cosmo-trajectory}

\begin{figure}
\begin{center}
\epsfxsize=8. cm \epsfysize=5.5 cm {\epsfbox{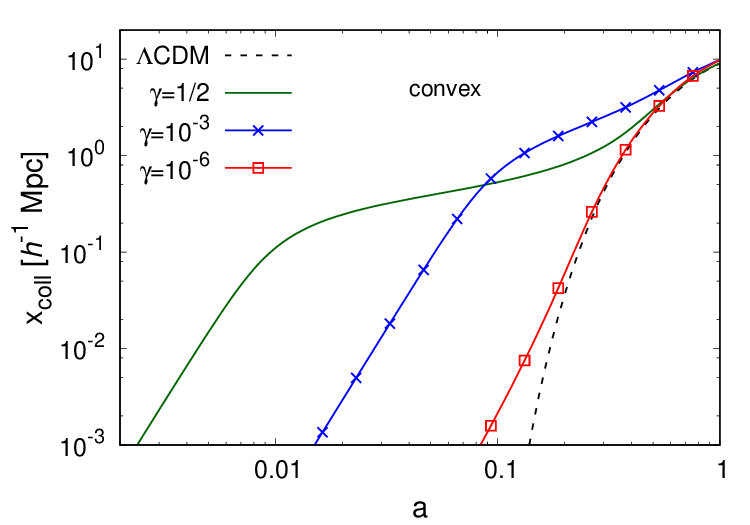}}
\end{center}
\caption{
Nonlinear transition scale $x_{\rm coll}(a)$ (in comoving coordinates)
as a function of the scale factor $a$ for the convex models.
}
\label{fig:r_c-convex}
\end{figure}

\begin{figure}
\begin{center}
\epsfxsize=8. cm \epsfysize=5.5 cm {\epsfbox{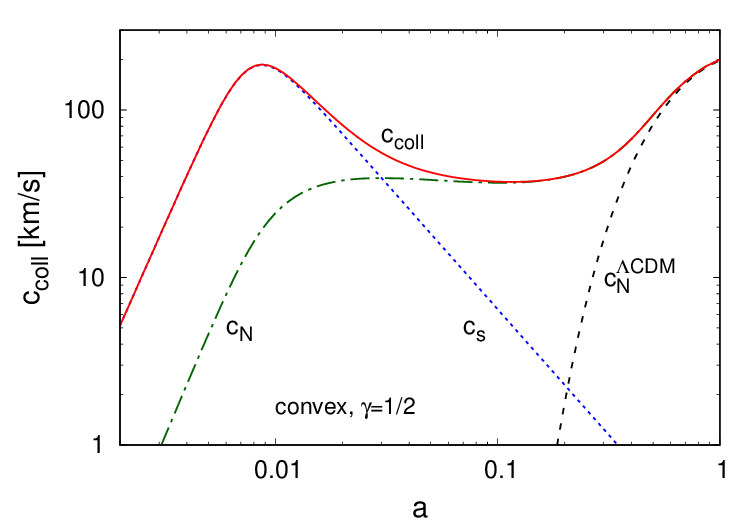}}\\
\epsfxsize=8. cm \epsfysize=5.5 cm {\epsfbox{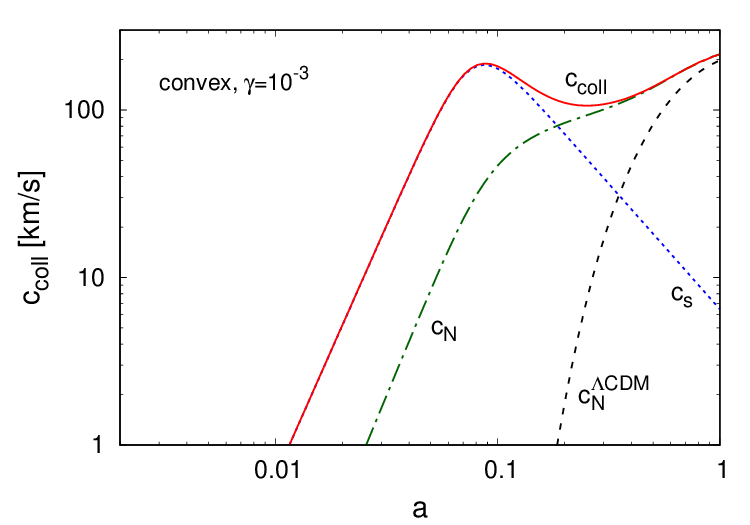}}\\
\epsfxsize=8. cm \epsfysize=5.5 cm {\epsfbox{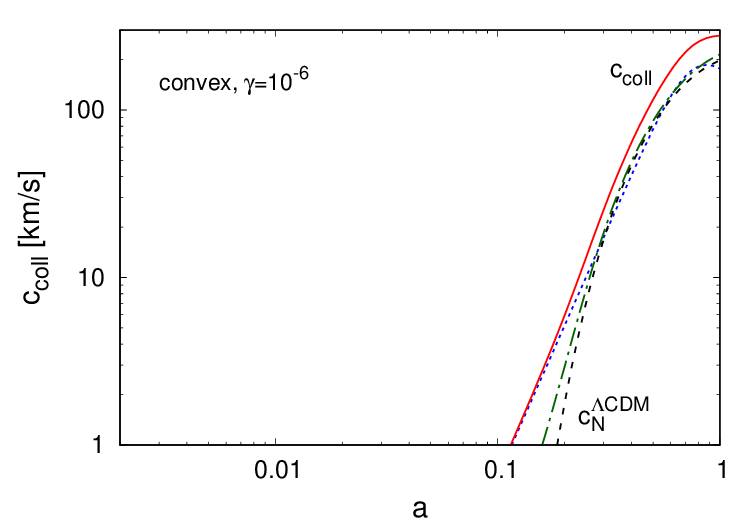}}
\end{center}
\caption{
Collapse velocity scale $c_{\rm coll}(a)$ (solid line) for the convex models.
The dotted and dot-dashed lines are $c_s$ and $c_{\rm N}$, whereas the dashed line
on the right is the result $c_{\rm coll}^{\Lambda \rm -CDM} = c_{\rm N}^{\Lambda \rm -CDM}$
in the case of the $\Lambda$-CDM cosmology.
}
\label{fig:v_c-convex}
\end{figure}

We show in Fig.~\ref{fig:r_c-convex} the comoving cosmological nonlinear
transition scale $x_{\rm coll}(a)$, defined by
\beq
\Delta^2_L(\pi/x_{\rm coll},z) = 1.5
\label{condition-collapse-radius}
\eeq
Because of the faster growth of density fluctuations on small scales, analysed in
sections~\ref{sec:linear-convex} and \ref{sec:mass-function}, the comoving
scale $x_{\rm coll}$ is much greater than the $\Lambda$-CDM prediction at high redshift.
It shows a steady rise for $a< a_{\alpha}$, when the characteristic fifth-force factor
$\epsilon_1(a)$ keeps growing with cosmic time as seen in the lower panel in
Fig.~\ref{fig_eps1}.
For the case $\gamma = 1/2$, in the time range $a_{\alpha} \lesssim a \lesssim 0.3$,
$x_{\rm coll}$ is roughly constant, which means that the hierarchical formation process
stops as no larger structures collapse.
This is due to the fast decrease of $\epsilon_1(a)$.
It means that the fifth force decays with time, so that it cannot generate the collapse
of greater structures, and we need to wait until $a \simeq 0.3$ for Newtonian gravity
to take the lead and generate the collapse of larger structures.
Then, at low redshift we recover the $\Lambda$-CDM behavior.
For $\gamma=10^{-6}$ this plateau does not exist because the fifth-force peak occurs
at low redshift where Newtonian gravity also becomes dominant.
The case $\gamma=10^{-3}$ gives an intermediate configuration.

For the thermodynamic analysis presented in section~\ref{sec:thermo-equilibrium}
below, we need the initial kinetic energy or typical velocity of the collapsing region.
From the evolution equation (\ref{delta-evol}) of the linear density, we define an effective
velocity scale $c_{\rm coll}$ by
\beq
c_{\rm coll}^2(a) = c_s^2 + c_{\rm N}^2 ,
\label{ceff-def}
\eeq
with
\beq
c_s^2 = \epsilon_1 c^2 , \;\;\;\;
c_{\rm N}^2 = (1+\epsilon_1) \frac{3\Omega_{\rm m}}{2\pi^2}
\left( H a x_{\rm coll} \right)^2 .
\label{cs-cN-def}
\eeq
The factor $c_s^2$ comes from the pressure-like term $\epsilon_1 c^2 \nabla^2\delta$
in Eq.(\ref{delta-evol}), while the term $c_{\rm N}^2$ comes from the right-hand
side, associated with the usual gravitational force (amplified by the negligible
factor $\epsilon_1$).
We show our results in Fig.~\ref{fig:v_c-convex}.
We also display the case of the $\Lambda$-CDM cosmology where
$c_{\rm coll}^{\Lambda \rm -CDM} = c_{\rm N}^{\Lambda \rm -CDM}$
as there is no pressure-like term.

Let us first consider the upper panel with $\gamma=1/2$.
The component $c_s$, due to the pressure-like term associated with the
fifth-force, dominates at high redshift. Its amplitude follows the rise and
fall of $\epsilon_1(a)$ displayed in Fig.~\ref{fig_eps1}. This also explains the
rise until $a_{\alpha}$ and next the stop of $x_{\rm coll}(a)$.
The component $c_{\rm N}$, associated with the Newtonian gravity, explicitly depends
on the scale $x_{\rm coll}(a)$. It grows with time, along with $x_{\rm coll}(a)$, and
dominates at late times, $a \gtrsim 0.03$. The plateau for $0.01 \lesssim a \lesssim 0.3$
follows from the very slow growth of $x_{\rm coll}(a)$ found in
Fig.~\ref{fig:r_c-convex} in this redshift range.
At late times we recover the standard $\Lambda$-CDM behavior.
Thus, we can distinguish three regimes from Fig.~\ref{fig:v_c-convex}.
At early times, $a<0.01$, the fifth force dominates and increasingly large scales
enter the nonlinear regime. This is the period when the thermodynamic analysis of
section~\ref{sec:thermo-equilibrium} below applies and allows us to estimate the behavior
of the system in the nonlinear regime.
For $0.01 < a < 0.3$, the hierarchical process of structure formation stops,
as the fifth force decreases and becomes subdominant with respect to Newtonian
gravity, which is still weak on these scales.
Finally, for $0.3 < a < 1$, we recover the $\Lambda$-CDM behavior, as Newtonian gravity
is dominant and strong enough to generate the collapse of new greater scales,
which have never been strongly modified by earlier fifth-force effects.

In the lower panel with $\gamma=10^{-6}$, we find that the deviation from the
$\Lambda$-CDM velocity scale is much smaller, as the fifth force is negligible at
high redshift and only becomes significant at low $z$, where however it is does not
become greater than Newtonian gravity on cosmological scales.
This is because Newtonian gravity already gives $c_{\rm N} \sim 300$ km/s
for the cosmological scales that turn nonlinear today, which is of the same order
as the maximum value $\max(c_s^2) = \max(\epsilon_1) c^2 \simeq \alpha c^2$.
The case $\gamma=10^{-3}$ is closer to the case $\gamma=1/2$, as there is
a distinct fifth-force era around $a_{\alpha}=0.1$.

We can note that in the case $\gamma = 1/2$
this history singles out a characteristic mass and velocity scale, associated
with the plateau found in Fig.~\ref{fig:v_c-convex},
\beqa
\gamma = 1/2: && x_* \sim 0.355 \; h^{-1} {\rm Mpc} , \;\;\; c_* \sim 50 \; {\rm km/s} ,
\nonumber \\
&& M_* \sim 2 \times 10^{10} \; h^{-1} M_{\odot}  .
\label{x*-M*-c*-def}
\eeqa
As in Fig.~\ref{fig_eta_R_M}, we recover the scales associated with small galaxies.
However, it is not clear whether this could alleviate or worsen some of the problems
encountered on galaxy scales by the standard $\Lambda$-CDM scenario.
This would require detailed numerical studies that are beyond the scope of this
paper.

\subsection{Thermodynamic equilibrium in the fifth-force regime for the convex model}
\label{sec:thermo-equilibrium}

So far we have implicitly assumed that during the initial phase
$a<a_{\alpha}$ of structure formation, governed by the fifth force in the cases
$\gamma=1/2$ and $10^{-3}$, the density field remains smooth on cosmological scales,
so that a standard linear analysis of matter cosmological perturbations can be applied.
This is not obvious because small scales, $x\leq x_{\rm coll}(a_{\alpha})$, have already
turned nonlinear at high redshift, $z > z_{\alpha}$.
Then, the density field could have become strongly inhomogeneous, and the gradient
of the fifth-force potential $\nabla\ln A$ at a given location in space would be unrelated
with the gradient of the density field smoothed on cosmological scales.
This strong sensitivity to the small-scale distribution of the density
field does not arise for the Newtonian gravitational force, because the Newtonian potential
is given by the Poisson equation, $\Psi_{\rm N} \propto \nabla^{-2} \rho$, which regularizes
the density field, whereas the fifth force potential $\ln A$ is a direct function of the local
density through Eq.(\ref{K-G-J-pert}).
This issue only arises in the first stage $a<a_{\alpha}$ found in
Fig.~\ref{fig:r_c-convex}, where new scales enter the nonlinear regime and are
dominated by the fifth force.

To address this question we need to go beyond perturbation theory and spherical
dynamics, as this is a highly nonlinear and inhomogeneous problem.
Following \cite{Brax:2016vpd}, we use a thermodynamic analysis, which provides
a simple analytic framework, see the Appendix~\ref{sec:Phase} for more details.
Assuming that the scales that turn nonlinear because of the fifth force at high redshift
reach a statistical equilibrium through the rapidly changing effects of the fluctuating
potential, in a fashion somewhat similar to the violent relaxation that takes
place for gravitational systems \cite{Lynden-Bell1967},
we investigate the properties of this thermodynamic equilibrium.
This first requires the study of the phase transitions and of the phase
diagram associated with the potential $\ln A(\rho)$ that defines our model.
Because this issue arises from the behavior of the fifth force in the regime where it
dominates over Newtonian gravity, we can neglect the latter to investigate this point.
Note that contrary to the usual gravitational case, the
potential $\ln A$ is both bounded and short-ranged , so that we cannot build infinitely
large negative (or positive) potential energies and a stable thermodynamic equilibrium
always exists, and it is possible to work with either micro-canonical, canonical or
grand-canonical ensembles.
In this respect, a thermodynamic analysis is better suited for such systems than for
standard 3D gravitational systems, where the potential energy is unbounded
from below and stable equilibria do not always exist, and different statistical
ensembles are not equivalent \cite{Padmanabhan1990}.

\begin{figure}
\begin{center}
\epsfxsize=8. cm \epsfysize=5.5 cm {\epsfbox{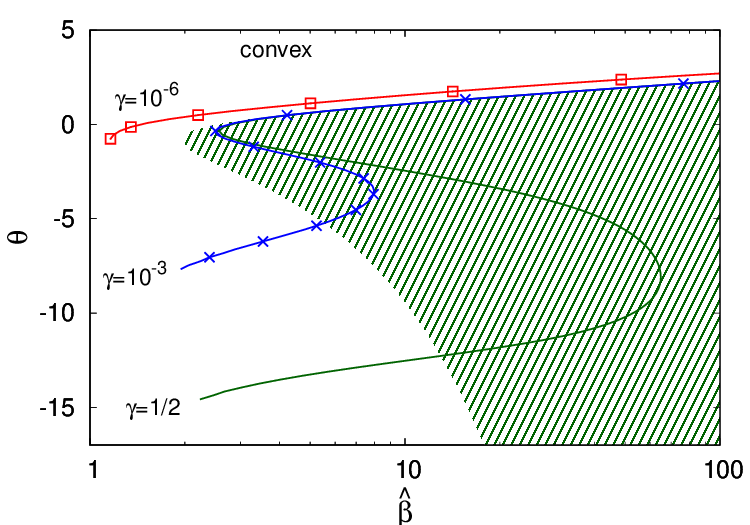}}
\end{center}
\caption{
Thermodynamic phase diagram of the convex models. The shaded area is the region
of initial inverse temperature $\hat\beta$ and density $\theta$ where the system
reaches an inhomogeneous thermodynamic equilibrium. The white area corresponds to
the homogeneous phase.
The solid lines are the cosmological trajectories $(\hat\beta_{\rm coll}(a),\theta_{\rm coll}(a))$,
for $\gamma=1/2, 10^{-3}$, and $10^{-6}$ from bottom to top.
}
\label{fig:cosmo-tra-convex}
\end{figure}

We refer to \cite{Brax:2016vpd} for the derivation of the thermodynamical diagram
associated with the fifth-force potential $\ln A$, see also the Appendix~\ref{sec:Phase}.
The main result is that we have
a first-order phase transition, between a high-temperature homogeneous phase
and a low-temperature inhomogeneous phase.
Defining the dimensionless inverse temperature $\hat\beta$ and density $\theta$,
\beq
\hat\beta = \frac{\alpha c^2}{k_B T} , \;\;\; \theta = \ln \hat\rho ,
\label{beta-def}
\eeq
the inhomogeneous phase is shown by the shaded area in Fig.~\ref{fig:cosmo-tra-convex},
while the white area is the homogeneous phase.
In terms of the normalized inverse temperature $\hat\beta$ and density $\theta$,
the phase diagram no longer depends on the parameters $\alpha$ and $\gamma$,
so that all three convex models computed in this paper have the same phase diagram
as displayed in Fig.~\ref{fig:cosmo-tra-convex}.
For a given normalized density $\theta$, the inhomogeneous phase extends to the right
up to infinite $\hat\beta$, i.e., down to zero temperature $T$.
The leftmost point of the inhomogeneous phase defines the critical temperature $T_c$,
which for the convex model (\ref{lambda-convex-explicit}) is given by
\beq
\hat\beta_c = \frac{15 (11-\sqrt{105}) \sqrt{15+\sqrt{105}}}{4 (\sqrt{105}-3)} \simeq 1.96
\label{betac-def}
\eeq
At low temperature the high- and low-density boundaries of the inhomogeneous phase
obey the asymptotic behaviors
\beq
\hat\beta \rightarrow \infty : \;\;\; \theta_+ \sim \frac{1}{2} \ln(\hat\beta) , \;\;\;
\theta_- \sim - \hat\beta .
\label{mus-large-beta}
\eeq
Then, if we consider a region of space with average initial temperature and
density, $(1/\hat\beta,\theta)$, which fall outside of the shaded region, the system remains
in the homogeneous phase.
If the initial condition falls inside the shaded region, the system becomes inhomogeneous
and splits over domains with density $\theta_-$ or $\theta_+$, with a proportion such that
the total mass over the full volume is conserved.

\subsection{Cosmological trajectory in the phase diagram}
\label{sec:thermo-diagram-cosmo}

The solid curves in Fig.~\ref{fig:cosmo-tra-convex} give the cosmological
trajectories in the phase diagram associated with the nonlinear transition scale,
$x_{\rm coll}(a)$ of Eq.(\ref{condition-collapse-radius}) and Fig.~\ref{fig:r_c-convex},
for $\gamma=1/2, 10^{-3}$, and $10^{-6}$.
To this scale we associate the mean density of the Universe,
$\rho_{\rm coll}(a) = \bar\rho(a)$, as the transition corresponds to density contrasts
of order unity, hence
\beq
\rho_{\rm coll}(a) = \bar\rho(a) , \;\;\;
\theta_{\rm coll}(a) = \ln \left[ \frac{\alpha\bar\rho(a)}{\gamma {\cal M}^4} \right]  ,
\label{theta-coll-def}
\eeq
and the inverse temperature
\beq
\beta_{\rm coll}(a) = \frac{1}{c^2_{\rm coll}(a)} \;\;\; \mbox{hence} \;\;\;
\hat\beta_{\rm coll}(a) = \frac{\alpha c^2}{c^2_{\rm coll}(a)} .
\label{beta-coll-def}
\eeq
As cosmic time grows and the density $\rho_{\rm coll}(a)$ decreases the cosmological
trajectory runs downwards in Fig.~\ref{fig:cosmo-tra-convex}.

Let us first consider the case $\gamma=1/2$.
In agreement with Fig.~\ref{fig:v_c-convex},
the inverse temperature $\hat\beta_{\rm coll}$ first decreases until $a \simeq 0.01$,
as the velocity $c_{\rm coll}$ grows.
Next, $\hat\beta_{\rm coll}$ increases while $c_{\rm coll}(z)$ decreases until
$a \simeq 0.2$, when we recover the $\Lambda\rm -CDM$ behavior,
and $\hat\beta_{\rm coll}$ decreases again thereafter.
We are interested in the first era, $a<a_{\alpha}$, when the hierarchical process of
structure formation is governed by the fifth force, and we find that the cosmological
trajectory is almost indistinguishable from the upper boundary $\theta_+(\hat\beta)$
of the inhomogeneous thermodynamic phase.
Indeed, from Eq.(\ref{ceff-def}) and Fig.~\ref{fig:v_c-convex} we have at early times
$c_{\rm coll} \simeq c_s$, hence $\hat\beta_{\rm coll} \simeq \alpha/\epsilon_1$.
Using Eq.(\ref{lambda-rho-convex-explicit}), we have at high densities, which also
correspond to $a<a_{\alpha}$,
$\epsilon_1 \simeq \alpha/\hat\rho^2 = \alpha e^{-2\theta}$,
hence
\beq
a \ll a_{\alpha} : \;\;\; \theta_{\rm coll} \sim \frac{1}{2} \ln( \hat\beta_{\rm coll}) ,
\label{theta-beta-coll-II}
\eeq
and we recover the asymptote (\ref{mus-large-beta}) of $\theta_+(\hat\beta)$.
This means that, according to the thermodynamic analysis, the cosmological density
field does not develop strong inhomogeneities that are set by the cutoff scale
of the theory when it enters the fifth-force nonlinear regime.
Therefore, density gradients remain set by the large-scale cosmological density
gradients and the analysis of the linear growing modes in
section~\ref{sec:linear-convex} and of the spherical collapse
in section~\ref{sec:spherical-convex} are valid.
Of course, on small nonlinear scales and at late times, where Newtonian gravity
becomes dominant, we recover the usual gravitational instability that we neglected in
this analysis and structure formation proceeds as in the standard $\Lambda$-CDM
case.

For the case $\gamma=10^{-3}$, we recover the same S-shape for the cosmological
trajectory, but the low-density part is shorter, because the redshift $z_{\alpha}$
is lower, and corresponds to a higher normalized density because of the factor
$\gamma$ in the definition of $\theta$ in Eq.(\ref{theta-coll-def}).
We again have the high-density behavior (\ref{theta-beta-coll-II}), which does not
depend on $\gamma$, and we recover the same conclusions as for $\gamma=1/2$.

For the case $\gamma=10^{-6}$, the cosmological trajectory no longer shows the
S-shape. This is because $z_{\alpha} \sim 1$ and there is no longer an intermediate
era where structure formation stops as the fifth force declines while Newtonian
gravity increases but remains small, at the scale $x_{\rm coll}$, in agreement
with Figs.~\ref{fig:r_c-convex} and \ref{fig:v_c-convex}.
Then, the cosmological trajectory always remains in the homogeneous phase.

Therefore, in all cases the analysis of the linear growing modes in
section~\ref{sec:linear-convex} and of the spherical dynamics
in section~\ref{sec:spherical-convex} are valid.

\subsection{Halo profiles}
\label{sec:thermo-halo-centers}

\begin{figure}
\begin{center}
\epsfxsize=8. cm \epsfysize=5.5 cm {\epsfbox{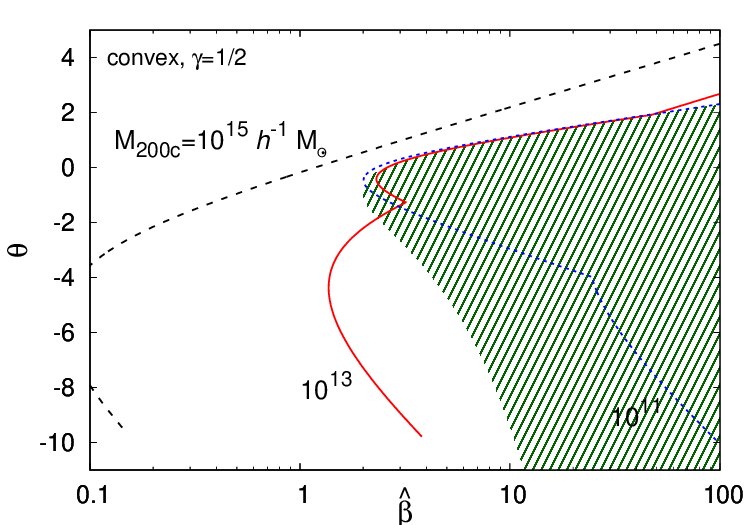}}\\
\epsfxsize=8. cm \epsfysize=5.5 cm {\epsfbox{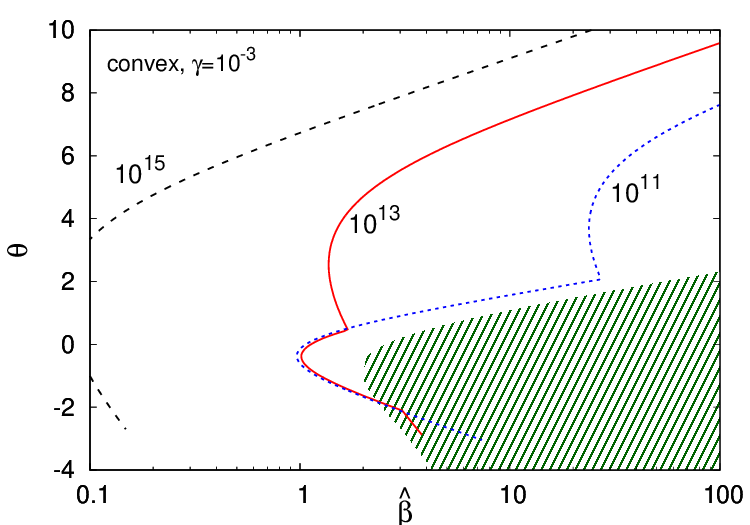}}\\
\epsfxsize=8. cm \epsfysize=5.5 cm {\epsfbox{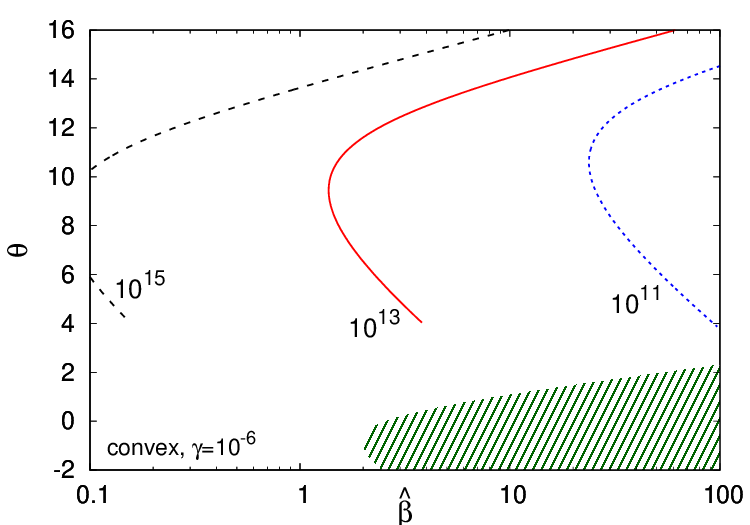}}\end{center}
\caption{
Radial trajectories $(\hat\beta_r,\theta_r)$ over the thermodynamic phase diagram
inside halos of mass $M_{\rm 200c}= 10^{15}$, $10^{13}$ and
$10^{11} h^{-1} M_{\odot}$, at $z=0$ for the convex models.
}
\label{fig:diagram_Halo_convex}
\end{figure}

We show in Fig.~\ref{fig:diagram_Halo_convex} the trajectories over the phase diagram
associated with density profiles in spherical halos, with a NFW profile \cite{Navarro:1996}.
Within a radius $r$ inside the halo, the averaged reduced density is
\beq
\theta_r = \ln \left[ \frac{\alpha \rho(<r)}{\gamma {\cal M}^4} \right]
= \ln\left[ \frac{\alpha 3 M(<r)}{\gamma 4\pi r^3 {\cal M}^4} \right] ,
\label{theta-halo-def}
\eeq
and we write the reduced inverse temperature as
\beq
\hat\beta_r = \frac{\alpha c^2}{{\rm Max}(c_s^2,v_{\rm N}^2)} ,
\label{beta-halo-def}
\eeq
where $v_{\rm N}$ is the circular velocity (\ref{FN-v2}) associated with the Newtonian gravity
while $c_s$ is the velocity scale (\ref{FA-cs2}) associated with the fifth force.
We choose the non-analytic interpolation ${\rm Max}(c_s^2,v_{\rm N}^2)$
instead of the smooth interpolation $c_s^2+v_{\rm N}^2$, which we used
in Eq.(\ref{ceff-def}) for the cosmological analysis, for illustrative convenience.
Indeed, the discontinuous changes of slope in Fig.~\ref{fig:diagram_Halo_convex}
show at once the location of the transition $|\eta|=1$ between the
fifth-force and the Newtonian gravity regimes.
As we move inside the halo, towards smaller radii $r$, the density $\theta_r$ grows
and we move upward in Fig.~\ref{fig:diagram_Halo_convex}.

Let us first consider the case $\gamma=1/2$ of the upper panel.
The turn-around of $\hat\beta_r$ at $\theta_r \simeq -4$ corresponds to the NFW
radius $r_s$ where the local slope of the density goes through $\gamma=2$
and the circular velocity $v_{\rm N}$ is maximum.
In agreement with Fig.~\ref{fig_eta_z0}, for high-mass halos,
$M \gtrsim 10^{13} h^{-1} M_{\odot}$, Newtonian gravity dominates at all radii and
there is no discontinuity in the radial trajectory as the maximum in Eq.(\ref{beta-halo-def})
is always $v_{\rm N}^2$.
Moreover, we are always in the homogeneous phase, so that the subdominant fifth force
does not lead to further fragmentation of the system.
For low-mass halos, $M \lesssim 10^{13} h^{-1} M_{\odot}$, the fifth force
becomes dominant at intermediate radii, between the two discontinuous changes
of slope of the radial trajectory. Moreover, a significant part of the outer halo
falls in the inhomogeneous thermodynamic phase.
This suggests a fragmentation of the system in the outer regions into smaller
substructures, which could give rise to some observational signatures.
This nonlinear process would next lead to a screening of the fifth force,
as discussed for the Solar System and the Earth in section~\ref{sec:screening},
because of the ultralocal character of the fifth force and the
disappearance of large-scale collective effects.

In the case $\gamma=10^{-3}$, as compared with $\gamma=1/2$ the radial trajectories
move upward with respect to the phase diagram in the normalized coordinates
$(\hat\beta,\theta)$ because of the factor $\gamma$ in Eq.(\ref{theta-halo-def}).
Then, even for $M \sim 10^{11} h^{-1} M_{\odot}$ most of the radial
trajectory is within the homogeneous phase. This suggests that nonlinearities
may not lead to a fragmentation of the system and that the spherically-averaged
analysis would be valid. Then, because we found in Fig.~\ref{fig_eta_z0} that
for $M \sim 10^{11} h^{-1} M_{\odot}$ this analysis gives a fifth force that is significantly
greater than Newtonian gravity these models are probably ruled out by observations.

In the case $\gamma=10^{-6}$ shown in the lower panel, the full radial trajectories
are within the homogeneous phase, so that the spherically-averaged analysis is
valid for all masses $M \gtrsim 10^9 h^{-1} M_{\odot}$.
Because we found in Fig.~\ref{fig_eta_z0} that for these masses
this analysis gives a fifth force that is smaller than Newtonian gravity these models are
consistent with observations.

\section{Conclusion}
\label{sec:Conclusion}

We have investigated the low-energy regime of particle physics models where the
standard model particles, e.g. the fermions such as the electron, as well as the dark matter,
are coupled to a hidden
sector comprising a scalar field charged under an Abelian symmetry. At energies below
the symmetry breaking scale, which is taken here to be of the $\rm meV$ order of magnitude,
the remaining Goldstone mode couples derivatively to the standard model fields.
In particular, it couples to Cold Dark Matter and, on coarse-grained scales larger than 1 mm,
the presence of this coupling can have an influence both on the cosmology of the late-time
Universe and the astrophysics of large-scale structures.

On the cosmological side, the potential energy in the Goldstone sector can serve as the
vacuum energy leading to the late cosmic acceleration. Moreover, the existence of a massless
scalar field coupled to matter is not in contradiction with local tests of gravity as the coupling
is ultralocal and the fifth force between isolated compact objects vanishes.
On the other hand, within extended structures with large-scale density
gradients, the coupling between matter and the scalar field can lead to several effects.

First of all, the linear growth of structures on very large scales, larger than 10 Mpc, is either
increased if the coupling function $\ln[A(\tilde \chi)]$, where $\chi$ is the normalized kinetic
term of the Goldstone mode, is convex, or decreased if the function is concave.
Thus, this class of models can either amplify or damp the formation of large-scale
structures. This can be contrasted to many modified-gravity models, such as
chameleon or K-mouflage scenarios, where the fifth force is always attractive and
amplifies Newtonian gravity (because  stability constraints restrict the sign of
the coupling parameters).

In the convex case, this could lead to more structure and more clumped objects
on nonlinear scales smaller than 10 Mpc. We use a thermodynamical analysis to show
that there is no catastrophic formation of very small high-density clumps at high redshifts.
In the concave case, fewer small structures are formed in a way more akin to Warm Dark
Matter scenarios.
Depending on the value of the model parameters, the dynamics inside halos can also
be significantly modified.
In the convex case where structure formation is enhanced,
in the outer shells of massive halos a thermodynamic phase transition could
take place with the formation of two phases containing two different populations of objects.
Such a mixing phenomenon is beyond the reach of this article and should be analyzed with
numerical simulations. This could provide interesting signatures for these models and
we hope to come back to these issues in the near future.
In the concave case, the repulsiveness of the fifth force is expected to damp nonlinearities,
but if it is too large as compared with Newtonian gravity it would modify the velocity
and density profiles of galactic halos.

Thus, as for the ultralocal models, these kinetic conformal coupling models
are very close to the $\Lambda$-CDM background but can generate significant deviations
on nonlinear cosmological scales, which increase on smaller scales.
In particular, for realistic models these deviations peak on galactic scales, which
could be interesting as the $\Lambda$-CDM cosmology encounters some tension with
observations on these scales.
Both models share the same screening mechanism on small astrophysical scales,
due to the ultralocal behavior of the fifth force, which does not propagate.
This screening mechanism is different from the usual chameleon, Damour-Polyakov,
K-mouflage and Vainshtein screening mechanisms of conformal coupling scenarios
with Yukawa interactions to matter.

As compared with the ultralocal models, the kinetic conformal coupling models
show two main theoretical and observational differences:
a) they can be associated with an Abelian symmetry and Goldstone modes,
which gives an additional motivation for the form of the Lagrangian,
and b) they can lead to both an attractive or a repulsive fifth force, depending on the
choice of the coupling function.
For the ultralocal models the case of a repulsive force is theoretically disfavored because
the model becomes unstable with respect to the addition of a small positive kinetic term,
which is expected in realistic scenarios as it is not forbidden by some symmetry.
For the kinetic conformal coupling models studied in this paper this issue does not
arise and the sign of the fifth force is not constrained by such stability requirements.

In summary, the Goldstone models that we have presented here offer a new mechanism 
to screen local effects of a scalar degree of freedom in the Solar System while allowing 
deviations from the $\Lambda$-CDM predictions on astrophysical and cosmological scales. 
As detailed above, the main observational effect of the new scalar interaction mediated 
in space-dependent density environments is an increase or decrease of clustering 
on small scales. 
In particular, models of the concave type where growth is depleted compared 
to $\Lambda$-CDM are promising, as they are very different from other scalar-tensor theories 
such as $f(R)$ models where an increase is always guaranteed.
We have shown that the impact of the fifth force is most important on galactic scales.
Therefore, these models should be mostly constrained by the observational properties
of galaxies (profiles, abundance of small galaxies, satellites and substructures), rather than by
larger-scale probes such as the CMB or clusters of galaxies. Lyman-$\alpha$ forest
clouds are likely to provide strong constraints as well, especially since they are
quasi-linear objects and better understood than galaxies, so that it would be easier
to distinguish fifth-force effects from possible baryonic effects (e.g., poorly understood
stellar and AGN outflows that can modify galaxy properties).
This would require numerical simulations, to go beyond the analytical treatment 
that we have given here.

In terms of constraints, the main one would be on the magnitude of the factor 
$\epsilon_1$, which measures the potential deviations from $\Lambda$-CDM
of the linear growing mode.
In this paper, we have used a conservative bound, $|\epsilon_1 | \lesssim 10^{-7}$,
associated with the bound $| A -1 | \lesssim 10^{-6}$ on the range of the conformal factor.
In terms of the main two parameters of these models, $\alpha$ and $\gamma$,
which describe the amplitude and the width of the coupling function $A$,
see Eqs.(\ref{lambda-def}) and (\ref{gamma-chi-def}), this gives $\alpha \lesssim 10^{-6}$.
(The third parameter, the energy scale ${\cal M}$, is set by the dark energy density today,
${\cal M}^4 \simeq \bar\rho_{\rm de0}$.)
Models with $\alpha \ll 10^{-6}$ or $\gamma \ll \alpha$ are identical to $\Lambda$-CDM
for practical purposes. 
Therefore, the interesting regime is $10^{-7} \lesssim \alpha \lesssim 10^{-6}$ and 
$\alpha \lesssim \gamma \lesssim 1$.
(For the concave case $\gamma > 1$ is allowed but for the convex case we have
$\gamma  <1$ for theoretical consistency.)
In this paper, we have considered the cases $\alpha = 10^{-6}$ and 
$\gamma \sim 1, 10^{-3}, 10^{-6}$.

We have seen that the linear growing mode, as a function of scale and time, 
is strongly sensitive to both $\alpha$ and $\gamma$, but the linear power spectrum today
is mostly sensitive to $\alpha$. Therefore, probes at different redshifts would provide
useful complementary constraints, such as galaxies at $z \lesssim 1$ and Lyman-$\alpha$
clouds at $z \sim 3$. In particular, in contrast with some other modified-gravity scenarios,
these models predict significant deviations from $\Lambda$-CDM at high redshift, $z \gtrsim 2$,
before the dark energy governs the background (provided one can probe small scales).

The low-mass tail of the halo mass function is sensitive to $\gamma$ and shows
a non-monotonic behavior. In particular, the deviations from $\Lambda$-CDM
are largest for $\gamma \sim 10^{-3}$ and this case is probably ruled out by the data,
as it would give too many or too few objects of mass $10^{12}-10^{13} h^{-1} M_{\odot}$.
We leave a more detailed study to further works.

From the analysis of the fifth force within spherical halos of $10^{11} h^{-1} M_{\odot}$,
we concluded that concave models with $\gamma \gtrsim 10^{-4}$ are ruled out,
as the fifth force is greater than Newtonian gravity at some intermediate radius.
For the convex models we could not conclude as small-scale nonlinearities of the scalar
sector may play a significant role in such objects and invalidate our mean-field spherical
approach. Numerical simulations and a detailed comparison with small galaxies,
such as dwarfs, would provide better constraints.

Thus, the realistic regime corresponds to $10^{-7} \lesssim \alpha \lesssim 10^{-6}$
and $10^{-7} \lesssim \gamma \lesssim 10^{-4}$. Smaller values converge to 
the $\Lambda$-CDM predictions, and the model has no more practical interest,
while higher values lead to deviations that are too large for 
$10^{11}-10^{13} h^{-1} M_{\odot}$ objects.
As explained above, better constraints will require numerical simulations and more detailed
comparison with data, especially from galaxy, small group and Lyman-$\alpha$ forest
probes.
This will require much more work and is left for the future.

\begin{acknowledgments}

This work is supported in part by the French Agence Nationale de la Recherche under Grant ANR-12-BS05-0002.

\end{acknowledgments}

\appendix

\section{Change of frame and fermions}
\label{sec:frames-fermions}

We consider massive Dirac fermions $\psi$ in the Jordan frame with the action
\be
S_\psi=-\int d^4x \sqrt{-g} \; ( \ii \bar \psi \gamma^\mu \nabla_\mu \psi + m_\psi \bar \psi \psi)
\ee
and the change of frame
\be
g_{\mu\nu}= A^2(\varphi,\tilde \chi) \, \tilde g_{\mu\nu} ,
\ee
where the conformal change of metric is an arbitrary function of $\varphi$ and
$\tilde \chi= -\tilde g^{\mu\nu} \partial_\mu \varphi \partial_\nu \varphi/(2{\cal M}^4)$.
The covariant derivative involves the spin connection and reads
\be
\nabla_\mu = \partial_\mu +\frac{1}{4} \gamma^{ab} \omega_{ab\mu} ,
\ee
where $a,b$ are Lorentz flat indices raised and lowered with the Minkowski metric
$\eta_{ab}$ while $\mu$ is a curved index.
The $\gamma^a$ matrices are such that
\be
\{\gamma^a,\gamma^b\}=2 \eta^{ab}
\ee
and we have introduced the commutator
\be
\gamma^{ab}=\frac{1}{2}[\gamma^a,\gamma^b].
\ee
The spin connection is most easily calculated using the Cartan formalism. Defining the vielbeins such that
\be
g_{\mu\nu}= \eta_{ab} e^a_\mu e^b_\nu
\ee
and the one-form $e^a= e^a_\mu dx^\mu$, the torsion-free condition implies that
\be
de^a + \omega^a_b \wedge e^b=0 ,
\label{spin}
\ee
and the one-form $\omega^a_b$ defines the spin connection
\be
\omega^a_b= \omega^a_{b\mu}dx^\mu.
\ee
Upon changing frame the vielbeins transform as
\be
 e^a_\mu= A \tilde  e^a_\mu,
\ee
the Dirac matrices become
\be
\gamma^\mu= A^{-1} \tilde \gamma^\mu ,
\ee
where $\gamma^\mu= e^\mu_a \gamma^a$.
We also redefine the Dirac fields as
\be
\psi= A^{-3/2} \tilde \psi ,
\ee
which gives
\be
\partial_\mu \psi= A^{-3/2} \left( \partial_\mu \tilde \psi - \frac{3}{2} \frac{\partial_\mu A}{A}
\tilde \psi \right) .
\ee
Using (\ref{spin}) and defining
\be
d\tilde e^a + \tilde \omega^a_b \wedge\tilde e^b = 0 ,
\ee
we find that
\be
\omega_{ab}=  \tilde \omega_{ab} + \frac{\partial_\mu A}{A}
(\tilde e^\mu_b \tilde e_a -\tilde e^\mu_a \tilde e_b) ,
\ee
which is antisymmetric in $(ab)$. As a result we find that
\beqa
\gamma^\mu\nabla_\mu \psi & = & A^{-5/2} \tilde \gamma^\mu \left( \tilde \nabla _\mu \tilde \psi -\frac{3}{2}\frac{\partial_\mu A}{A}\tilde \psi \right. \nonumber \\
&& \left. + \frac{1}{2} \frac{\partial_\nu A}{A} \tilde e^\nu_b \tilde e_{a\mu} \gamma^{ab}
\tilde \psi \right) ,
\eeqa
and upon using
$\tilde \gamma^\mu ( \tilde \gamma_\mu \tilde \gamma^\nu - \gamma^\nu \tilde \gamma_\mu ) = 6 \tilde \gamma^\nu$
we find that
\be
\gamma^\mu\nabla_\mu \psi=  A^{-5/2} \tilde \gamma^\mu \tilde \nabla _\mu \tilde \psi .
\ee
As a result the fermionic action becomes
\be
 S_\psi= -\int d^4x \sqrt{-\tilde g} \; ( \ii \bar{\tilde \psi} \tilde\gamma^\mu \tilde\nabla_\mu
 \tilde \psi + A(\tilde\chi) m_\psi \bar{\tilde\psi}\tilde \psi ) .
\ee
In particular, the mass in the Einstein frame can be identified with
\be
\tilde{m}_\psi = A \, m_\psi .
\ee

\section{Thermodynamics and Phase Transition}
\label{sec:Phase}

In this appendix, we recall some of the essential ingredients needed in the main text,
see \cite{Brax:2016vpd} for more details.

In the continuum limit, where the mass $m$ of the dark matter particles goes to zero,
we describe the system by the smooth phase-space distribution function
$f(\vx,\vv)$. The mass $M$, the energy $E$ and the entropy $S$ of the system are given by
\beqa
M & = &\int d^3 x d^3 v \; f(\vx,\vv) ,
\label{mass-thermo} \\
E & = & \int d^3 x d^3 v \; f(\vx,\vv) \left( \frac{v^2}{2} + c^2 \ln A[\rho(\vx)] \right) ,
\hspace{0.5cm}
\label{energy-thermo} \\
S & = & -\int d^3 x d^3 v \; f(\vx,\vv) \; \ln \frac{f(\vx,\vv)}{f_0} ,
\label{entropy-thermo}
\eeqa
where $f_0$ is a normalization constant and we used the fact that the potential
$\ln A$ is a function of the local density.
In the grand-canonical ensemble, the statistical equilibrium is obtained by minimizing
the grand-canonical potential $\Omega$, which is given by
\beq
\Omega = E - S/\beta - \mu M ,
\label{grand-potential}
\eeq
where $\beta$ and $\mu$ are the inverse temperature and the chemical potential.
The equilibrium phase-space distribution is given by the minimum of the grand potential,
${\cal D} \Omega / {\cal D} f = 0$, which implies that
\beqa
f(\vx,\vv) = f_0 \; e^{-\beta \left(v^2/2 + c^2 \ln A + c^2 d \ln A / d \ln \rho \right)
+ \beta\mu - 1 } .
\label{phase-space-eq}
\eeqa
We recover a Maxwellian distribution over velocities, i.e.
$f(\vx,\vv) \propto \rho(\vx) e^{-\beta v^2/2}$.
Integrating over velocities gives
\beq
f(\vx,\vv) = \left( \frac{\beta}{2\pi} \right)^{3/2} \rho(\vx) \; e^{-\beta v^2/2} ,
\label{f-rho-beta}
\ee
and Eq.(\ref{phase-space-eq}) yields
\beq
\rho(\vx) = f_0  \left( \frac{2\pi}{\beta} \right)^{3/2}
e^{-\beta c^2 \left( \ln A + d \ln A / d \ln \rho \right)+ \beta\mu - 1 } .
\label{rho-lnA-eq}
\eeq

It is convenient to introduce the rescaled
dimensionless potential and density $\lambda$ and $\hat\rho$, from
Eqs.(\ref{lambda-def}) and (\ref{rho-hat-def}).
Defining also the rescaled dimensionless inverse temperature $\hat\beta$ and
chemical potential $\hat\mu$,
\beq
\hat\beta = \alpha c^2 \beta ,
\label{beta-hat-def}
\eeq
\beq
\hat\mu = \ln \left[ \frac{\alpha f_0 c^3}{{\cal M}^4}
\left( \frac{2\pi}{\beta c^2} \right)^{3/2} \right] + \beta\mu -1 ,
 \label{mu-hat-def}
\eeq
the equilibrium condition (\ref{rho-lnA-eq}) reads as
\beq
\hat\mu = \theta + \hat\beta \, \nu(\theta) ,
\label{mu-theta-nu-eq}
\eeq
where we introduced
\beq
\theta = \ln\hat\rho , \;\;\; \nu(\theta) = \lambda + \frac{d\lambda}{d\theta} .
\label{theta-nu-def}
\eeq
For a given value of the rescaled inverse temperature $\hat\beta$ and chemical
potential $\hat\mu$, this gives the equilibrium density $\theta$ as the solution
of the implicit equation (\ref{mu-theta-nu-eq}).
In terms of these dimensionless variables, the grand-canonical potential
(\ref{grand-potential}) reads as
\beq
\Omega = \frac{{\cal M}^4 c^2 V}{\hat{\beta}} \hat\Omega \;\;\; \mbox{with} \;\;\;
\hat\Omega = e^{\theta} \left[ \hat\beta \lambda - \hat\mu-1 + \theta \right] ,
\label{hat-Omega-def}
\eeq
where $V$ is the total volume of the system. Thus, the equilibrium equation
(\ref{mu-theta-nu-eq}) is the condition $d\hat\Omega/d\theta = 0$, as the
thermodynamic equilibrium corresponds to the minimization of the grand-potential.

We analyse the system at a fixed temperature, which corresponds to
a given initial velocity dispersion, as a function of the chemical potential $\hat\mu$
or of the density $\hat\rho$, seen as conjugate variables.
At high temperature, $\hat\beta\rightarrow 0$, Eq.(\ref{mu-theta-nu-eq}) becomes
$\hat\mu=\theta$ and there is a unique density for each $\hat\mu$. This corresponds
to the high-temperature homogeneous phase where we recover a perfect gas as
the potential energy is negligible.
At low temperature, $\hat\beta\rightarrow \infty$, the right-hand side of
Eq.(\ref{mu-theta-nu-eq}) can become non-monotonic so that for some values of
the chemical potential $\hat\mu$ there are several solutions $\theta_i$.
This corresponds to the inhomogeneous phase, where the system splits over several
regions of different densities $\theta_i$, with an admixture such that the mean
density over the large scale $x=V^{1/3}$ is the initial density $\bar\rho$,
see \cite{Balian2007} for an analysis of such phase transitions.

More precisely, the function $\hat\mu(\theta)$ becomes non-monotonic below a temperature
$1/\hat\beta_c$.
 As explained above, for low $\hat\beta$ (i.e., high temperature)
the function $\hat\mu(\theta)$ is monotonic while for high $\hat\beta$
(i.e., low temperature) it is non-monotonic over some range of densities,
with a first-order phase transition at $\hat\beta_c$.
Then, for $\hat\beta < \hat\beta_c$, we always have a single solution $\theta(\hat\mu)$
for any chemical potential $\hat\mu$.
For $\hat\beta > \hat\beta_c$, in a finite range $[\hat\mu_1,\hat\mu_2]$ and
$[\theta_1,\theta_2]$, we have three solutions, $\theta_- < \theta_{\rm m} < \theta_+$,
for a given chemical potential $\hat\mu$.
Both $\theta_-$ and $\theta_+$ are local minima of the grand-potential $\hat\Omega$
whereas $\theta_{\rm m}$ is a local maximum. Then, the physical solution
$\theta(\hat\mu)$ is the global minimum among $\{\theta_-,\theta_+\}$ (i.e., the deepest
minimum).
There is a critical value $\hat\mu_s$ where
there is a  transition from $\theta_-$ to $\theta_+$. This happens at the crossing
of their values of the grand-potential, when
$\hat\Omega(\theta_-;\hat\mu_s) = \hat\Omega(\theta_+;\mu_s)$
\cite{Balian2007}.
Thus, we have a first-order phase transition, as the density of the system jumps from
$\theta_-(\hat\mu_s)$ to $\theta_+(\hat\mu_s)$ when the chemical potential goes
through $\hat\mu_s$. At $\hat\mu_s$, where $\hat\Omega_-=\hat\Omega_+$,
there is a coexistence of the two phases. One part of the volume $V$ is at the low
density $\theta_-$ and the other part at the high density $\theta_+$.
The relative fraction between the two phases is set by the mean density $\bar\theta$
of the full volume, $\theta_- \leq \bar\theta \leq \theta_+$, which is given by the initial
condition of the system (the constraint on the average density of the full system).

The thermodynamic phase diagram of the system, in the inverse temperature - density
plane, is shown by the shaded area in Fig.~\ref{fig:cosmo-tra-convex}.
This domain is limited at low $\hat\beta$ by the critical temperature
$\hat\beta_c$. The lower and upper limits of the domain are the curves
$\theta_-(\hat\beta) \equiv \theta_-(\hat\mu_s(\hat\beta),\hat\beta)$ and
$\theta_+(\hat\beta) \equiv \theta_+(\hat\mu_s(\hat\beta),\hat\beta)$.
The meaning of the diagram in Fig.~\ref{fig:cosmo-tra-convex} is the following.
If the average initial temperature and density, $(1/\hat\beta,\theta)$, fall outside of
the shaded region, the system remains in the homogeneous phase.
If the initial condition falls inside the shaded region, the system becomes inhomogeneous
and splits over domains with density $\theta_-$ or $\theta_+$, with a proportion such that
the total mass over the full volume is conserved.

\bibliography{ref1}   % name your BibTeX data base

\end{document}